\documentclass[onecolumn,floatfix,superscriptaddress,showpacs,showkeys,nofootinbib,preprint]{revtex4}

\usepackage{epsfig}
\usepackage{amssymb,latexsym,amsmath}
\newcommand{\eq}[1]{\begin{align} #1 \end{align}}

\begin{document}

\title{Multiplicity Distributions in\\
 Canonical and Microcanonical Statistical Ensembles}
\author{M. Hauer}
\affiliation{Helmholtz Research School, University of Frankfurt, Frankfurt,
  Germany}
\affiliation{University of Cape Town, South Africa}
\author{V.V. Begun}
\affiliation{ Bogolyubov Institute for Theoretical Physics, Kiev,
Ukraine}
\author{M.I. Gorenstein}
\affiliation{ Bogolyubov Institute for Theoretical Physics, Kiev, Ukraine}
\affiliation{Frankfurt Institute for Advanced Studies, Frankfurt,
Germany}

\begin{abstract}
The aim of this paper is to introduce a new technique for calculation of
observables, in particular multiplicity distributions, in various statistical
ensembles at finite volume. The method is based on Fourier analysis of the
grand canonical partition function. Taylor expansion of the generating
function is used to separate contributions to the partition function in
their power in volume. We employ Laplace's asymptotic expansion to show that
any equilibrium distribution of multiplicity, charge, energy, etc. tends to a
multivariate normal distribution  in the thermodynamic
limit. Gram-Charlier expansion allows additionally for calculation of finite
volume corrections. Analytical formulas are presented for inclusion of
resonance decay and finite acceptance effects directly into the system
partition function. This paper consolidates and extends previously
published results of current investigation into properties of statistical
ensembles.  
\end{abstract}


\maketitle
\tableofcontents
\section{Introduction}
\label{Introduction}

During the last couple of decades the statistical hadronization model, first
introduced by Fermi \cite{Fermi} and Hagedorn \cite{hagedorn}, has been
surprisingly successful in describing fundamental properties of systems
created in heavy ion collisions, cosmic rays, and elementary particle
reactions. In the context of heavy ion collisions this model has been applied
to an extensive set of data on hadron production, ranging form the center of
mass energies of the experiments at the SIS, AGS, SPS, and most recently, RHIC
facilities. For reviews see \cite{GSIfits,AGSfits,SPSfits,RHICfits}. A
systematic evolution of thermodynamic parameters as collision energy (and size
of colliding ions) is changed \cite{FreezeOut} has allowed to establish the
`chemical freeze-out line', which is now a vital part in our
understanding of the phase diagram of strongly interacting matter.
More controversially this model has also been applied to a range
of elementary collisions \cite{ElemetaryFits}, where only few
particles are produced, and the picture of a gas of hadrons can
hardly be suitable. The remarkable ability of the statistical
model to explain these data has lead to the suggestion
\cite{ElemetaryFits,SHM} that thermal particle production is a general
property of the hadronization process itself, rather than the result of a long
sequence of microscopic interactions. In this work we will not 
argue about possible physical interpretations \cite{Meaning} of
the partition function of statistical mechanics. However we note that
in order to apply a semi-classical approximation a volume of
$\mathcal{O}(10fm^3)$ seems to be sufficient \cite{QFTMCE}.

One of the answers still outstanding in high energy physics is the
one of a possible formation of a  deconfined state of matter,
where degrees of freedom are quarks and gluons, rather than
hadrons, and the nature of the phase transition between these two
phases. The growing interest in the study of event-by-event
fluctuations in strong interactions  is thus motivated by
expectations of anomalies in the vicinity of the onset of 
deconfinement \cite{OnsetOfDecon} and in the case when the
expanding system goes through the transition line between
quark-gluon plasma and hadron gas \cite{PhaseTrans}. In
particular, a critical point of strongly interacting matter may be
accompanied by a characteristic power-law pattern in fluctuations
\cite{CriticalPoint}. Multiplicity and charge fluctuations have
been indeed proposed to be a good discriminating tool between
quark-gluon plasma and hadron gas \cite{QGP-HRG-fluc}, provided
the signal survives the phase transition and subsequent evolution
of the system. However, in order to properly  assess the
discriminating power of such observables, one should firstly
calculate fluctuations in a hadron gas by including all known
physical effects, such as conservation laws, quantum statistics,
resonance decays, kinematical cuts, finite spatial extension, etc.

Only recently, due to a rapid development of experimental
techniques, first measurements of fluctuations of particle 
multiplicities \cite{fluc-mult} and transverse momenta
\cite{fluc-pT} were performed. And in fact one is tempted to
interpret recent NA49 data on multiplicity fluctuations in most
central Pb-Pb collisions \cite{NA49} as the first observation
\cite{MCEvsData} of the recently predicted canonical suppression
of fluctuations. The most promising region in the phase diagram
for observation of critical phenomena seems to be accessible to
the SPS accelerator \cite{Horn}. A new SPS scan program
\cite{ScanProgram} for different ion sizes as well as
center-of-mass energies has been proposed to study strongly
interacting systems at different energy densities and life times.
This should be as well our main motivation for further
investigation of properties of statistical ensembles. The aim is
the calculation of `base-line' fluctuations on top of which
one hopes to find  unambiguous signals of a phase transition
\cite{PhaseTrans}, a critical point \cite{CriticalPoint}, or
thermal/chemical (local or global) non-equilibrium \cite{non-eq}.

The main subject of the past study has been the mean multiplicity
of produced hadrons. However, there is a qualitative difference in
the properties of the mean multiplicity and the scaled variance of
multiplicity distributions in statistical models. In the case of
the mean multiplicity results obtained in the grand canonical
ensemble (GCE), canonical ensemble (CE), and microcanonical
ensemble (MCE) approach  each other in the large volume limit. One
refers here to as the thermodynamical equivalence of statistical
ensembles. It was recently found \cite{CEfluc_1,MCEfluc_1} that
corresponding results for the scaled variance are different in
different ensembles, and thus the choice of ensemble remains a
crucial one, even for large systems.

In previous publications \cite{SPE, CEfluc_1, MCEfluc_1,
  BGZ, GrandMce, QGas, res, CETurko} the focus was on
asymptotic values of the scaled variance, or calculations were
altogether only performed in the GCE \cite{GCEfluc}. In this
paper we try to present not only approximations to CE and
MCE distributions in the large volume limit, but also to find a reasonable
approximation scheme for finite system size in an analytical, rather than
Monte Carlo \cite{MonteCarlo}, approach. The main subject of this work
will be to identify the GCE partition function with the
characteristic function of a statistical system. This approach is more
effective in terms of mathematical ease as well as in terms of computing time
than previous methods.
Thus we are presenting for the first time analytical
formulas for the final state of a general multi-specie CE or MCE
hadron gas at finite volume.

The paper is organized as follows. In Section \ref{sec-GGPF} the
generalized partition function will be introduced and some basic
assumptions of this paper will be stated. Section \ref{CLTCE}
gives a mathematical approximation to the system partition
function in the form of Laplace expansion  for  the CE, while
Section \ref{CLTMCE} is concerned with  the MCE formulation
of multiplicity distributions in terms of this expansion for
a simple quantum gas of massless particles. A general recipe for
calculation of the scaled variance is presented. Resonance decay
is included into the system partition function in Section
\ref{ResDecay}. In Section \ref{Correspondence} we show a
comprehensive comparison of this method to the previously used
microscopic correlator approach. A method for finite volume
corrections is developed in Section \ref{FiniteVolCorr}. Section
\ref{BoltzmannPionGas} will present an application of this  method
in great detail and aims to give some physical interpretation. A
summary in Section \ref{Conclusion} closes the paper.
\section{Generalized Grand Canonical Partition Function}
\label{sec-GGPF}
In textbooks on statistical mechanics (see e.g.,  Ref.
\cite{Physics}) often  first the MCE is introduced, where exact
conservation laws for energy-momentum and particle number are
imposed. Relaxing the constraints for energy and momentum
constitutes the CE, while allowing additionally particle number to
fluctuate about some mean value introduces the  GCE. In a
relativistic gas of hadrons  quantum numbers (charges)
will be the conserved quantities rather than particle numbers. In
this paper it will prove to be of considerable advantage to start
off with the GCE formulation and imposing exact conservation laws
thereafter. The basic idea is to define the probability of a given
number of  particles of some species, $N_l$, at fixed value of
conserved charge, $Q$, i.e. the CE distribution $P(N_l|Q)$, in
terms of the GCE distributions, $P(N_l,Q)$ and $P(Q)$. Thus, the
GCE partition function will be the basis for all calculations in
this work. Generally the (micro)canonical partition  function is
obtained from the grand canonical one by  multiplication with
Kronecker (or Dirac) delta-functions which  pick out all
microstates consistent with a particular  conservation law. It is
often  more economical to use the Fourier  representations
delta-functions, rather than the delta-function themselves.
A short example will motivate the following general  treatment.
The GCE and the CE partition functions $Z$ and $Z^Q$ are connected
as,
\begin{equation}\label{Z-GCE}
Z(V,T,\mu_Q)~ = ~\sum \limits_{Q=-\infty}^{\infty}
~\exp\left(\frac{Q\mu_Q}{T}\right)~ Z^Q(V,T)~,
\end{equation}
where $V$, $T$, and $\mu_Q$ are respectively  the system
volume, temperature and chemical potential associated with
conserved charge $Q$. The probability of finding the GCE system in
a particular charge state $Q$ equals to the number of all states
with net-charge $Q$ divided by all accessible states:
\begin{equation}\label{PQ}
P(Q)~ = ~\frac{\textrm{all states with charge $Q$}}{\textrm{all
states}}~=~ \frac{\exp\left(\frac{Q\mu_Q}{T}\right)~Z^Q}{Z}~.
\end{equation}
It is important to note that the GCE partition function $Z$ depends on
$\mu_Q/T$, while the usual CE partition function does not. We introduce
the generalized grand canonical partition function (GGPF)
$\mathcal{Z}(\phi_Q)$ by substitution $\mu_Q/T \rightarrow \mu_Q/T + i
\phi_Q$ in $Z$. One then finds:
\begin{equation} \label{PQ-1}
P(Q)~=~\frac{1}{Z}~\int \limits_{-\pi}^{\pi} \frac{d \phi_Q}{2
\pi}~ e^{-i Q  \phi_Q} ~\mathcal{Z}(\phi_Q)~.
\end{equation}
The system with one net-charge, $Q$, could be a quite general one and may
include different  particle species. In the state of chemical equilibrium the
chemical potential of any specie $l$ equals to $\mu_l=q_l\,\mu_Q$,
where $q_l$ is the charge of a particle of specie $l$. Considering only the
distribution of particle species $l$, the joint
probability $P(N_l,Q)$ to find particle number $N_l$ and
net-charge $Q$ in the GCE equals to:
\begin{eqnarray}\label{PNQ-1}
P(N_l,Q) &=& \frac{\textrm{all states with $N_l$ particles and
charge $Q$}}{\textrm{all states}}~\nonumber \\
&=& \frac{1}{Z}~\int \limits_{-\pi}^{\pi} \frac{d \phi_Q}{2 \pi}
\int \limits_{-\pi}^{\pi} \frac{d \phi_{N_l}}{2 \pi}~ e^{-iQ\phi_Q}
~e^{-iN_l\phi_{N_l}}~
  \mathcal{Z}(\phi_Q,\phi_{N_l})~.
\end{eqnarray}
Similar to Eq.~(\ref{PQ-1}) the distribution (\ref{PNQ-1}) is
presented in terms of the GGPF $\mathcal{Z}(\phi_Q,\phi_{N_l})$ through
substitutions $\mu_l/T\rightarrow \mu_l/T+i\phi_{N_l}$, and $\mu_Q/T
\rightarrow \mu_Q/T+i\phi_Q$  in the GCE partition function $Z$. Finally the
canonical (or conditional) multiplicity distribution $P(N_l|Q)$ is given by:
\begin{eqnarray}\label{PNQ-CE}
P(N_l|Q)  &=& \frac{\textrm{all states with $N_l$ particles and
    charge $Q$}}{\textrm{all states with charge $Q$}} 
~=~ \frac{P(N_l,Q)}{P(Q)}~.
\end{eqnarray}
Eq.~(\ref{PNQ-CE}) presents the CE particle number distribution $P(N_l|Q)$ in
terms of two GCE distributions, $P(N_l,Q)$ and $P(Q)$. A detailed account of
this is given in terms of an ideal Boltzmann pion gas in Section
\ref{BoltzmannPionGas}.

It is worth noting that Eqs.(\ref{PQ},\ref{PNQ-1},\ref{PNQ-CE})
are as well the basis for any Monte Carlo approach
\cite{MonteCarlo}. A sampling distribution, usually taken from a
Boltzmann GCE  system, is used to generate a $\lbrace N_l
\rbrace$-tuple of particle multiplicities of all considered
species. All `events' consistent with certain constraints, like a
set of conserved charges, are accepted, while the rest is
rejected. On the basis of this set of all accepted `events` one
constructs an ensemble by using a suitable re-weighting scheme to
account for quantum statistics and proper normalization. One is
now ready to calculate distributions $P(N_l|Q)$, i.e. the
conditional distributions to find particle multiplicity $N_l$ in
the `Monte Carlo ensemble`, while global charge is fixed to $Q$,
and therefore observables like mean multiplicity and multiplicity
fluctuations. The advantage of the method presented here is
certainly that we proceed in a completely  analytical fashion and
therefore much unneeded information, like the exact composition of
$\lbrace N_l \rbrace$, is `integrated away'.

An immediate consequence of
Eqs.(\ref{PQ},\ref{PNQ-1},\ref{PNQ-CE}) is that temperature and
chemical potentials appear in our formulation of (micro)
canonical distributions (as well as in the Monte Carlo approach
\cite{MonteCarlo}). At first sight this seems to be  a
serious problem and an unnecessary  complication of our
initial task to find a reasonable approximation to the CE and MCE
partition functions. However, the main technical challenge when
numerically integrating the original version of the
microcanonical partition function arises from a heavily
oscillating integrant. Auxiliary parameters $T$ and $\mu$ will
produce a very smooth function, for which approximation schemes
can be used (see Section \ref{BoltzmannPionGas}). In fact chemical
potentials and temperature can be factored out, and thus our
partition function is just the original partition function times
some factor (Appendices \ref{CEPartFunc} and \ref{MCEPartFunc}).
Taking the ratio  in Eq.(\ref{PNQ-CE}) artificially
introduced temperature and chemical potential drop out. The
quality of our approximation on the other hand will crucially
depend on their choice. We will show in Section
\ref{FiniteVolCorr} that the requirement of maximizing the GGPF at
some given equilibrium point leads to a unique determination of
thermal parameters and moreover constitutes the optimal choice for
our approximation scheme. This prescription is self-consistent and
can be shown to be not in contradiction to basic thermodynamic
relations known from textbooks \cite{Physics}.

Taking the GCE partition function as a basis, we will throughout this paper
use the language of statistical mathematics for calculation of conditional
multiplicity distributions in any ensemble. So it is worthwhile to give a
general outline. For any probability distribution function (PDF) one can
define an associated characteristic function (CF) by Fourier
back-transformation (Appendix \ref{CharFunc}). In this work we will identify
the GGPF as the CF of any GCE distribution, and find  joint GCE distributions,
e.g. $P(N_l,Q)$, by Fourier analysis of the GGPF.
In the example above the CF associated with  the PDF $P(N_l,Q)$ in
Eq.(\ref{PNQ-1}) is $\mathcal{Z} (\phi_Q, \phi_{N_l})/Z$, while the CF
associated with  the PDF $P(Q)$ in Eq.(\ref{PQ}) is $\mathcal{Z} (\phi_Q)/Z$.
However, only in the simplest cases one can find an analytical solution for
these Fourier integrals. It seems as well to be rather difficult to
obtain the CF directly for the PDF $P(N_l|Q)$, and hence we define $P(N_l|Q) =
P(N_l,Q) / P(Q)$. Provided one has knowledge of all (sufficiently many) moments
of an PDF one can construct the CF and find the exact (approximate) PDF. For
practical  purposes, however, cumulants, which are obtained by Taylor
expansion of the logarithm of the CF, are far better suited.

\section{Canonical Ensemble}
\label{CLTCE}

The method presented here will provide the basis for the following
discussion. We present in detail our calculation of the CE partition
function. In this section we will employ saddle point expansion to
approximate the canonical PDF for an ideal hadron resonance gas and focus on
the asymptotic solution only. We first calculate the 3-dim distribution of
charges in GCE, $P(Q,B,S)$, namely baryon number, strangeness and electric
charge. In fact our calculation is similar to that of Becattini
{\it et.al.} \cite{SPE}, however we start explicitly from a GCE partition
function which includes chemical potentials. We will identify our version of
the CE partition function as the un-normalized probability distribution
function of conserved charges in GCE, and  show that this type of integral
over the GCE partition function generally leads to a Gaussian (multivariate
normal) distribution in the large volume limit. In a next step we use this
result to obtain the 4-dim PDF for GCE, $P(N,Q,B,S)$, hence the probability to
find our GCE hadron-resonance gas in a state $(N,Q,B,S)$. The ratio is the
canonical PDF $P(N|Q,B,S)$, see Eq. (\ref{PNQ-CE}). In particular we will
present a very simple formula for the asymptotic scaled variance of
multiplicity fluctuations.
\subsection{CE Partition Function}
The canonical system partition function of a hadron resonance gas with three
conserved charges, i.e. electric charge $Q$, baryon number $B$, and
strangeness $S$ is given by the triple Fourier integral over GCE partition
function
\cite{CEPF}:
\begin{eqnarray} \label{BSQPartFunc}
\mathcal{Z}^{Q,B,S} &=& \int \limits_{-\pi}^{\pi}
\frac{d\phi_Q}{2\pi} \int \limits_{-\pi}^{\pi}
\frac{d\phi_B}{2\pi} \int \limits_{-\pi}^{\pi}
\frac{d\phi_S}{2\pi} e^{-iQ\phi_Q} e^{-iB\phi_B} e^{-iS\phi_S}
\times ~\exp \left[ \sum_{l} z_l \left( \phi_Q, \phi_B, \phi_S
\right) \right]~.
\end{eqnarray}
Adopting vector notation for the set of conserved charges $(Q,B,S) =\vec Q
=Q^j$ and angles $(\phi_Q, \phi_B,\phi_S) = \vec \phi = \phi_j$ we can write
Eq.(\ref{BSQPartFunc}) in compact form:
\begin{eqnarray}\label{BSQPartFunc_j}
\mathcal{Z}^{Q^j} = \left[ \prod_{j=1}^{3} \int \limits_{-\pi}^{\pi}
  \frac{d\phi_j}{\left( 2\pi \right)} \right]~ e^{-iQ^j\phi_j} ~
 \exp \left[ \sum_{l} z_l \left( \phi_j \right) \right]~,
\end{eqnarray}
where repeated upper and lower indexes $j$ imply summation over $j$. The single
particle partition function of particle specie $l$ is given by:
\begin{eqnarray}\label{QstatsZET}
z_l \left(\phi_j \right) &=&  \frac{g_l V} {\left( 2 \pi\right)^3}
\int d^3p ~\ln \left( 1 \pm  e^{-\left( \varepsilon_l -\mu_l \right) /T}\;
  \; e^{ i q^j_l \phi_j}\right)^{\pm1} 
~\equiv~ V \psi_l \left(\phi_j \right),
\end{eqnarray}
where we have introduced particle $l$'s quantum number
configuration $q^j_l = \vec q_l = (q_l,b_l,s_l)$, it's degeneracy
factor $g_l=(2J_l+1)$, internal angular momentum $J_l$, mass $m_l$, and energy
$\varepsilon_l=\sqrt{p^2+m_l^2}$, the chemical potential vector
${\mu^j} = (\mu_Q,\mu_B,\mu_S)$, and particle $l$'s  chemical
potential $\mu_l = q^j_l \mu_j $. $V$ is the system
volume, and $T$ it's temperature. The summation $\sum_l$ includes
also anti-particles, for which $q^j_l\rightarrow -q^j_l$. The
(Wick rotated) fugacity of particle specie $l$ is thus given by
the substitution $\lambda_l = \exp\left[q_l^j \mu_j/T\right]
\rightarrow \lambda_l = \exp\left[q_l^j\left( \mu_j/T +i \phi_j
\right)\right]$. The upper sign in the Eq.~(\ref{QstatsZET})
denotes Fermi-Dirac statistics (FD), while the lower is used for
Bose-Einstein statistics (BE). The case of Maxwell-Boltzmann
statistics (MB) is analogous and will not be discussed separately.
Together with $\sum_l \psi_l\left(\phi_j \right) = \Psi
\left(\phi_j \right)$ the system partition function can be written
as:
\begin{eqnarray} \label{GeneralPF}
\mathcal{Z}^{{Q^j}} = \left[ \prod_{j=1}^{3} \int \limits_{-\pi}^{\pi}
  \frac{d\phi_j}{\left( 2\pi \right)} \right]~e^{-iQ^j\phi_j}~
 \exp \Big[ V ~ \Psi \left( \phi_j \right) \Big]~.
\end{eqnarray}
The following calculation is based only on the general form of
Eq.(\ref{GeneralPF}), and can easily be generalized to a larger set of
conserved quantities. In the language of statistical mathematics the
function $\Psi \left( \phi_j \right)$ would be called cumulant generating
function (CGF) of the partition function $\mathcal{Z}^{Q^j}$, while $\exp
\left[ V ~ \Psi \left( \phi_j \right) \right]$ is called CF. For large volume,
$V\rightarrow \infty$, the main contribution to the integral
Eq.(\ref{GeneralPF}) comes from a small region around the origin
\cite{SPE}. Thus we proceed by Taylor expansion of the CGF $\Psi \left( \phi_j
\right)$ around $\phi_j = \vec 0$ and introduce cumulant tensors $\kappa$:
\begin{equation}\label{kappatensor}
\kappa_n^{j_1,j_2,\dots,j_n } ~\equiv~ \left(-i\right)^n\frac{\partial^n ~
\Psi \left( \phi_j \right) }{\partial \phi_{j_1}
  \partial \phi_{j_2} \dots \partial \phi_{j_n} } ~\Bigg|_{\phi_j  = \vec 0}~.
\end{equation}
The CGF therefore can be expressed in terms of a Taylor series:
\begin{equation}
\Psi (\phi_j) ~\simeq~ \sum_{n=0}^{\infty} \frac{i^n}{n!} ~
\kappa_n^{j_1,j_2,\dots,j_n } \; \phi_{j_1} \phi_{j_2} \dots \phi_{j_n}~,
\end{equation}
where summation over repeated indices is implied. The cumulant of $0^{th}$
order is just the logarithm of the GCE partition function $Z$ divided by the
volume, $Z \equiv \exp (V \kappa_0)$.  Hence, after extending the limits of
integration to $\pm \infty$, which will introduce a negligible
error, we find:
\begin{eqnarray}\label{CE_PF_1}
\mathcal{Z}^{Q^j} &\simeq& Z ~  \left[ \prod_{j=1}^{3} \int
  \limits_{-\infty}^{\infty} \frac{d\phi_j}{\left( 2\pi \right)} \right] ~\exp
\Big[  -iQ^j\phi_j
~+~V \sum_{n=1}^{\infty} \frac{i^n}{n!} \; \kappa_n^{j_1,j_2,\dots,j_n } \;
\phi_{j_1} \phi_{j_2} \dots \phi_{j_n} \Big] ~.
\end{eqnarray}
It is worth noting that the integrant of Eq.(\ref{CE_PF_1})
is not $2\pi$-periodic anymore, while the one of Eq.(\ref{GeneralPF}) is.
Spelling out the first two terms of the summation yields:
\begin{eqnarray}
\mathcal{Z}^{Q^j} \simeq Z \left[ \prod_{j=1}^{3} \int
  \limits_{-\infty}^{\infty} \frac{d\phi_j}{\left( 2\pi \right)} \right]
~\exp  \Bigg[ -iQ^j\phi_j  + -iV \kappa_1^j \phi_j
- V \frac{\kappa_2^{j_1,j_2}}{2!} \phi_{j_1} \phi_{j_2} \nonumber \\ +V
\sum_{n=3}^{\infty} \frac{i^n}{n!} \; \kappa_n^{j_1,j_2,\dots,j_n } \;
\phi_{j_1} \phi_{j_2} \dots \phi_{j_n} \Bigg].
\end{eqnarray}
Performing now a change of variables will simplify this triple integral.
\begin{equation}\label{theta}
\theta_j ~=~ \sqrt{V} ~\sigma_{j}^{~k} ~\phi_k~,
\end{equation}
where $\sigma_{j}^{~k}$ is the square root of the second rank
tensor $\kappa_2$:
\begin{equation}\label{sigmatensor}
\sigma_{j}^{~k} ~\equiv~ \left( \kappa_2^{~~1/2}\right)_j^{~~k}.
\end{equation}
The element $d \theta_j$ equals to:
\begin{equation}\label{dtheta}
d \theta_j ~=~ \det | \sqrt{V} ~\sigma | ~d \phi_j ~=~ V^{3/2}~ \det |
\sigma | ~d \phi_j~.
\end{equation}
Lastly in terms of this transformation normalized cumulant tensors $\lambda$
are introduced:
\begin{equation}\label{normcum}
\lambda_n^{j_1,j_2,\dots,j_n} ~ \equiv ~ \kappa_n^{k_1,k_2,\dots,k_n}
\left(
  \sigma^{-1}\right)_{k_1}^{\;\;\;j_1} \left(
  \sigma^{-1}\right)_{k_2}^{\;\;\;j_2} \dots \left(
  \sigma^{-1}\right)_{k_n}^{\;\;\;j_n}  ~.
\end{equation}
The new variable $\xi^j$  will be a measure for the distance of the
actual charge vector $Q^j$ to the peak of the distribution of the PDF:
\begin{equation}\label{xi-var}
\xi^j ~=~ \left( Q^k - V \kappa_1^k \right) \left(
  \sigma^{-1}\right)_{k}^{\;\;j}  V^{-1/2}~.
\end{equation}
Including all these steps at once yields:
\begin{eqnarray} \label{CLTfinalZ}
\mathcal{Z}^{Q^j} \simeq \frac{Z}{V^{3/2} \det
|\sigma|} \left[ \prod_{j=1}^{3} \int \limits_{-\infty}^{\infty}
\frac{d \theta^j}{2\pi} \right] ~ \exp \Bigg[ &-& i \xi^j \theta_j -
 \frac{\theta^j \theta_j}{2!} \nonumber \\
&+& \sum \limits_{n=3}^{\infty} i^n
 V^{-\frac{n}{2}+1} \frac{\lambda_n^{j_1,j_2,\dots,j_n}}{n!}
 \theta_{j_1}\theta_{j_2} \dots \theta_{j_n}   \Bigg]~.
\end{eqnarray}
Eq.(\ref{CLTfinalZ}) is the starting point for obtaining an
asymptotic solution in this section as well as for finite volume
corrections in Section \ref{FiniteVolCorr}. Through coordinate
transformation Eq.(\ref{theta}) we have explicitly separated
terms in their power in volume. Thus as system size is increased
influence of higher order normalized cumulants $\lambda_n$
decreases, allowing for truncation of the summation for
sufficiently large volume. A few words on physical units are in
order. The single particle partition function Eq.(\ref{QstatsZET})
$\psi_l[fm^{-3}]$, and therefore all cumulant elements
Eq.(\ref{kappatensor}) $\kappa [fm^{-3}]$ in CE, consequently
entries in Eq.(\ref{sigmatensor}) $\sigma [fm^{-3/2}]$. The
normalization in Eq.(\ref{CLTfinalZ}) $V^{3/2} \det |\sigma|$ for
3-dim $\sigma$, as well as the new variable of integration
Eq.(\ref{theta}) $\theta^j$, are hence dimensionless. The inverse
sigma tensor elements are $\sigma^{-1}[fm^{3/2}]$ and thus the
charge vector Eq.(\ref{xi-var}) $\xi^j$ will be dimensionless.
Finally  normalized cumulants Eq.(\ref{normcum}) are $\lambda_n
[fm^{-3+3n/2}]$, which is canceled by the factor
$V^{-n/2+1}$ in the summation in Eq.(\ref{CLTfinalZ}). Thus all
terms involved in Eq.(\ref{CLTfinalZ}) are dimensionless.

For $V\rightarrow \infty$ one can discard terms of $V^{-1/2}$ and higher:
\begin{equation}\label{CE_Square}
\mathcal{Z}^{Q^j} \simeq \frac{Z}{V^{3/2} \det
|\sigma|} \left[ \prod_{j=1}^{3} \int \limits_{-\infty}^{\infty}
\frac{d \theta^j}{2\pi} \right]~ \exp \left[ - i \xi^j \theta_j -
 \frac{\theta^j \theta_j}{2!}   \right]~.
\end{equation}
Completing the square, the integral (\ref{CE_Square}) can be solved:
\begin{equation}\label{CE_HG_Gauss}
\mathcal{Z}^{Q^j} \simeq Z \frac{\exp \left(-\frac{\xi^j \; \xi_{ j}}{2}
  \right)}{\left(2\pi V \right)^{3/2} \det| \sigma| }~.
\end{equation}
From Section \ref{sec-GGPF} it follows that in GCE one
would find for the 3-dim charge PDF:
\begin{equation}\label{CE_HG_P_Gauss}
P ( Q^j ) ~= ~ \frac{e^{\frac{Q^j\mu_j}{T}}~Z^{Q^j}}{Z} ~=~
\frac{\mathcal{Z}^{Q^j}}{Z} ~\simeq~ \frac{\exp \left(-\frac{\xi^j \;
      \xi_{j}}{2} \right)}{\left(2\pi V  \right)^{3/2} \det| \sigma| }~.
\end{equation}
Please note that Eq.(\ref{CE_HG_P_Gauss}) (albeit in  different
notation) was used as an assumption in the microscopic correlator approach
\cite{ MCEfluc_1,QGas,res,GrandMce,BGZ}.
\subsection{Particle Number Distribution}
Similarly to Eq.(\ref{BSQPartFunc}) the canonical state involving a charge
vector $Q^j$ and a particular number of particles $N$ of some species can be
described by the following 4-dim partition function:
\begin{eqnarray}  \label{BSQNPartFunc}
\mathcal{Z}^{N,Q,B,S}
\;=\; \int \limits_{-\pi}^{\pi} \frac{d\phi_{N}}{2\pi} \int
\limits_{-\pi}^{\pi} \frac{d\phi_Q}{2\pi} \int
\limits_{-\pi}^{\pi} \frac{d\phi_B}{2\pi} \int
\limits_{-\pi}^{\pi} \frac{d\phi_S}{2\pi}\;
 e^{-iN\phi_{N}}
e^{-iQ\phi_Q}  e^{-iB\phi_B} e^{-iS\phi_S} \nonumber \\
\times  \exp \left[ \sum_{l=1} z_l \left(\phi_{N}, \phi_Q,
\phi_B, \phi_S\right) \right] ~,
\end{eqnarray}
where $ (N,Q,B,S) = \left(N,Q^j \right) = \tilde Q^j$. Each particle specie
which is selected receives a Wick rotated fugacity $\exp \left(i \phi_{N}
\right)$ in complete analogy to those associated with charge conservation.
Similar to Eq.(\ref{CE_HG_Gauss}) the canonical partition function
$\mathcal{Z}^{\tilde Q^j}$ can be approximated by:
\begin{equation}
\mathcal{Z}^{\tilde Q^j} ~\simeq~ Z~ \frac{\exp
  \left(-\frac{\tilde \xi^j \; \tilde \xi_{ j}}{2} \right)}{\left(2\pi V
  \right)^{4/2} \det |\tilde \sigma|}\;,
\end{equation}
Therefore in GCE we find the 4-dim PDF of
finding our system in a state $\tilde Q^j$:
\begin{equation}
 P \left( \tilde Q^j \right) ~=~ \frac{\mathcal{Z}^{\tilde Q^j} }{Z} ~\simeq~
 \frac{\exp \left(-\frac{\tilde \xi^j \; \tilde \xi_{ j}}{2}
   \right)}{\left(2\pi V   \right)^{4/2} \det |\tilde \sigma|}  ~.
\end{equation}
 However we are only interested in a 1-dim slice with constant values of
conserved charges $Q^j = (Q,B,S)$, e.g. the conditional PDF $P(N | Q^j) =
\mathcal{Z}^{\tilde  Q^j}/\mathcal{Z}^{ Q^j}$, where the normalization is
Eq.(\ref{CE_HG_Gauss}) from
the previous section. The expansion works best around the peak of the
distribution. If analytical solutions were available this would not matter
much, but here we need to find the peak of the PDF. In the thermodynamic limit
peak  and mean of the distribution coincide (see Section \ref{FiniteVolCorr}),
hence it peaks at $Q^k_{eq} = V \kappa_1^k = (\langle Q \rangle,\langle B
\rangle,\langle S \rangle)$. For the
3-dim distribution Eq.(\ref{CE_HG_Gauss}) from the previous section it follows
that Eq.(\ref{xi-var}) $\xi^j = \left( Q^k_{eq} - V \kappa_1^k \right)
\left(\sigma^{-1}\right)_{k}^{\;\;\;j} V^{-1/2} = 0$. For the 4-dim
distribution of this section we find at $\tilde Q^k_{eq}=(N,Q^k_{eq})$ for
$\tilde \xi^j \tilde \xi_j =  \left(  N - \langle N \rangle \right)^2 V^{-1}
\left( \tilde \sigma^{-1} \right)^{~j}_{1} \left( \tilde \sigma^{-1}
\right)^{~1}_{j}$, and $\langle N \rangle = V \kappa_1^{N}$.  The canonical
PDF can thus be written as:
\begin{eqnarray}
P(N | Q^j_{eq})  &\simeq&  \frac{ \det
|\sigma|}{\left(2\pi V \right)^{1/2} \det |\tilde \sigma|}
\exp \Bigg(- \left( N - \langle N \rangle \right)^2\;
   \frac{ \left(\tilde \sigma^{-1}\right)^{~j}_{1}
  \left( \tilde \sigma^{-1}\right)^{~1}_{j} }{2 V } \Bigg)~.
\end{eqnarray}
Hence one can find the width of the distribution from its
normalization, as well as from its exponential $ \det |  \sigma| /  \det |
\tilde  \sigma| = \sqrt{\left(  \tilde \sigma^{-1}\right)^{~j}_{1} \left(
    \tilde \sigma^{-1}\right)^{~1}_{j}} = (\langle N^2\rangle -\langle
  N\rangle^2 )^{-1/2}~ V^{1/2}$.
The identity
\begin{equation}\label{EqDetIdent}
 \frac{ \det | \kappa_2| }{\det | \tilde \kappa_2| } ~=~
\left(  \tilde \sigma^{-1}\right)^{~j}_{1} \left(  \tilde
  \sigma^{-1}\right)^{~1}_{j} ~.
\end{equation}
is proven in Appendix (\ref{DetIdent}). One gets for the scaled variance of
the particle number distribution:
\begin{equation} \label{SimpleOmega}
\omega ~=~ \frac{\langle N^2 \rangle - \langle N \rangle^2}{\langle N \rangle}
~=~  \frac{V\; \det | \tilde \sigma|^2 }{ V \;
  \kappa_1^{N} \;\det| \sigma |^2 } ~=~ \frac{ \det | \tilde \kappa_2| }{
  \kappa_1^{N} \;\det| \kappa_2|  }~.
\end{equation}
In words, this is the ratio of the product of the eigenvalues of
the $3+1$ dimensional matrix $\tilde \kappa_2$ and the $3$
dimensional matrix $ \kappa_2$ divided by the particle density of
the particle under investigation. The equivalence of
Eq.(\ref{SimpleOmega}) to results of the micro-correlator approach, see
e.g. \cite{ MCEfluc_1,QGas,res,GrandMce,BGZ}, will be shown in Section
\ref{Correspondence}.
\begin{eqnarray}
P(N | Q^j_{eq}) ~ \simeq~  \frac{1}{\left(2\pi \omega \langle N \rangle
  \right)^{1/2}}   \exp  \left(- \frac{ \left( N - \langle N \rangle
    \right)^2}{2 \omega  \langle N \rangle  } \right) ~.
\end{eqnarray}
Thus we obtained the rigorous mathematical proof of the hypothesis
that the multiplicity distributions in a hadron-resonance gas in
thermodynamic limit have Gaussian shape. Diagonalization of
$\kappa_2$ was not at all necessary for arriving at the asymptotic
solution  Eq.(\ref{SimpleOmega}). However for finite volume
corrections one will have to go through the process of  finding
eigenvectors and eigenvalues of $\kappa_2$ (see Section
\ref{FiniteVolCorr}).
\subsection{The Cumulant Tensor}
\label{PrimKappa}
In this section we will calculate the $\kappa$-tensor Eq.(\ref{kappatensor})
only for primordial single particle specie fluctuations. In Section
\ref{ResDecay} this will be extended to fluctuations of a selection of
particles, like `positively charged` and resonance decay will be included.
\begin{equation}
\kappa_1^{j_1} =  \left( -i \frac{\partial}{\partial \phi_{j_1}}
\right) \Psi ~\Big|_{\phi^j = \vec 0} ~,
\end{equation}
and
\begin{equation} \kappa_2^{j_1,j_2} =  \left( -i \frac{\partial}{\partial
      \phi_{j_1}} \right) \left( -i \frac{\partial}{\partial \phi_{j_2}}
  \right) \Psi ~\Big|_{\phi^j = \vec 0} ~.
\end{equation}
Cumulants of order $1$ give GCE expectation values, hence average baryon,
strangeness, electric charge, and particle density. The second cumulant
contains information about GCE fluctuations of some quantity (diagonal
elements), as well as correlations between different quantities (off-diagonal
elements). For the distributions $P( Q^j)$ and
$P(\tilde Q^j)$ the first and second cumulants are:
\eq{
\kappa_1 =
\begin{pmatrix}
\kappa_1^Q, & \kappa_1^B, & \kappa_1^S
\end{pmatrix} ~,&&
\tilde \kappa_1 =
\begin{pmatrix}
\kappa_1^N, & \kappa_1^Q, & \kappa_1^B, & \kappa_1^S
\end{pmatrix}~, \label{CEvector}
}
\eq{ \kappa_2 =
\begin{pmatrix}
\kappa_2^{Q,Q} & \kappa_2^{Q,B} & \kappa_2^{Q,S} \\
\kappa_2^{B,Q} & \kappa_2^{B,B} & \kappa_2^{B,S} \\
\kappa_2^{S,Q} & \kappa_2^{S,B} & \kappa_2^{S,S}
\end{pmatrix}~, &&
\tilde \kappa_2 =
\begin{pmatrix}
\kappa_2^{N,N} & \kappa_2^{N,Q} & \kappa_2^{N,B} & \kappa_2^{N,S} \\
\kappa_2^{Q,N} & \kappa_2^{Q,Q} & \kappa_2^{Q,B} & \kappa_2^{Q,S} \\
\kappa_2^{B,N} & \kappa_2^{B,Q} & \kappa_2^{B,B} & \kappa_2^{B,S} \\
\kappa_2^{S,N} & \kappa_2^{S,Q} & \kappa_2^{S,B} & \kappa_2^{S,S}
\end{pmatrix} ~.\label{CEmatrix}
}
In case two quantities are un-correlated, as for example primordial $\pi^+$
multiplicity and globally conserved strangeness ($\pi^+$ does not carry
strangeness), then the corresponding elements
$\kappa_2^{N,S}=\kappa_2^{S,N}=0$. For clarity some elements are explicitly
given. The primordial mean value of particle number density of particle
species $l$ and the mean charge density are:
\begin{eqnarray}
\kappa_1^{N} &=& \left( -i \frac{\partial}{\partial
\phi_{N}}\right) \Psi ~\Big|_{{\phi_j} = \vec 0}  = \psi^{\prime}_l
\label{k_1_N_p}~, \\
\kappa_1^{Q} &=& \left( -i \frac{\partial}{\partial
\phi_{Q}}\right) \Psi ~\Big|_{{\phi_j} = \vec 0}  = \sum_l q_l~
\psi^{\prime}_l~.  \label{k_1_Q}
\end{eqnarray}
Fluctuation of particle density $\kappa_2^{N,N}$, correlation between particle
number and baryonic charge $\kappa_2^{N,B}$, and correlation between strangeness
and baryonic charge $\kappa_2^{S,B}$ are given by:
\begin{eqnarray}
\kappa_2^{N,N} &=& \left( -i \frac{\partial}{\partial
\phi_{N}}\right)^2 \Psi ~\Big|_{{\phi_j} = \vec 0}  = \psi^{\prime
\prime}_l ~,\label{k_2_NN_p} \\
\kappa_2^{N,B} &=&
 \left( -i \frac{\partial}{\partial \phi_{N}}\right)
 \left( -i \frac{\partial}{\partial \phi_B}\right)
 \Psi ~\Big|_{{\phi_j} = \vec 0}  = b_l ~ \psi^{\prime \prime}_l ~,
 \label{k_2_BN_p}  \\
 \kappa_2^{S,B} &=&
 \left( -i \frac{\partial}{\partial \phi_S}\right)
 \left( -i \frac{\partial}{\partial \phi_B}\right)
\Psi ~\Big|_{{\phi_j} = \vec 0}  = \sum_l s_l ~ b_l ~ \psi^{\prime
\prime}_l  \label{k_2_SB}\;.
\end{eqnarray}
where the first and second derivative of $\psi$ are:
\begin{eqnarray}
\psi^{\prime}_l &=& \frac{g_l}{\left(2\pi \right)^3} \int d^3p~
\frac{e^{-\left(\varepsilon_l -\mu_l \right)/T }}{\left( 1 \pm
  e^{-\left(\varepsilon_l -\mu_l \right)/T } \right)}~, \label{psi_p_g}\\
\psi^{\prime\prime}_l &=& \frac{g_l}{\left(2\pi \right)^3} \int
d^3p~ \frac{e^{-\left(\varepsilon_l -\mu_l \right)/T }}{\left( 1 \pm
  e^{-\left(\varepsilon_l -\mu_l \right)/T } \right)^2} ~.\label{psi_pp_g}
\end{eqnarray}
Please note that the analogs of the matrix $\kappa_2$ and the
vector $\kappa_1$, however in different notation, were used
in both previously published methods for calculation of scaled
variance under the thermodynamic limit, the micro-correlator
approach, see e.g. \cite{ MCEfluc_1,QGas,res,GrandMce,BGZ}, and saddle point
expansion method \cite{SPE}. The advantage of this paper is
certainly the very simple formula (\ref{SimpleOmega}), and the
possibility for calculation of finite volume corrections (see
Section \ref{FiniteVolCorr}),  giving a lower bound for the
validity of the micro-correlator approach and saddle point expansion.
Exact agreement was found with the analytical results for the
scaled variance in the thermodynamic limit \cite{
  MCEfluc_1,QGas,res,GrandMce,BGZ,MCEvsData}. Numerical
calculations give very good agreement  to \cite{SPE},
despite the fact that different particle tables were used in the latter.

\section{Microcanonical Ensemble}
\label{CLTMCE}
In the MCE we additionally enforce kinematic conservation laws.
The simplest example is an ultra-relativistic gas, made up of
one kind of neutral massless particles. The reason for this choice is
that an analytical solution exists, at least for  a
microcanonical ensemble, where energy but not momentum is
conserved \cite{Fermi,GrandMce}, in MB approximation, allowing for a
comparison of analytic results to asymptotic solutions (see Sections \ref{QofA}
and \ref{MCEexpample}). Further we can carry out the integration of the GGPF
in full detail for this example (Appendix \ref{MCEPartFunc}), and highlight
some features of our method, such as the role of temperature in the MCE.

In this section, however, we explicitly include momentum
conservation as well as FD and BE statistics, and restrict ourself
to the asymptotic solution. The basic ideas are essentially the
same as before. The only conserved quantities here then are the
total energy $E$ and the three momentum $\vec P$. In principle we would have
to treat energy and momentum conservation on equal footing and introduce
Lagrange multipliers associated with conserved momenta. However, in the rest
frame of a static thermal source we always find $\langle \vec P
\rangle = \vec 0$. Thus we only need to consider the
$0^{th}$ component, $1/T$, of the `four-temperature`
\cite{MonteCarlo}. The probability to find a microcanonical
system in a state with exactly $N$ particles is defined by:
\begin{equation}\label{P(N|E,p=0)}
P(N|E,\vec P = 0) = \frac{\textrm{number of all states with $N$ particles,
    $E$, and $\vec P = 0$}}{\textrm{number of all states with $E$, and $\vec P
    = 0$}}~.
\end{equation}
Our starting point is again the  GCE partition function, which can be written
as:
\begin{equation}
Z = \exp \left[ V \frac{g}{\left( 2 \pi \right)^3} \int
  d^3p~ f \left(\vec p \right) \right]~,
\end{equation}
where $f(\vec p)$ is the probability of having a particular momentum state
occupied. For massless particles we find $\varepsilon=\mid \vec p \mid =
\sqrt{p_x^2+p_y^2+p_z^2}$, thus $f(\vec p)= \exp \left( - |\vec p|/T
\right)$ in MB approximation, or in the quantum statistical treatment
$f(\vec p)= \ln \left( 1 \pm \exp \left( - |\vec p| /T\right)
\right)^{\pm1}$, where the upper sign denotes FD, while the lower sign stands
for BE statistics. The numerator in
Eq.(\ref{P(N|E,p=0)}) is given by the 5-dim Fourier integral over the GGPF
$\mathcal{Z} \left( \phi_N, \phi_E, \vec \phi_p\right)$:
\begin{equation}
P \left(N, E, \vec P \right) = \frac{1}{ {Z} } ~ \int
\limits_{-\pi}^{\pi} \frac{d \phi_N}{2 \pi}  \int \limits_{-\infty}^{\infty}
\frac{d \phi_E}{2 \pi} \int \limits_{-\infty}^{\infty} \frac{d \vec
  \phi_p}{\left( 2 \pi \right)^3}~ e^{-iN \phi_N}~ e^{-i E \phi_E}~ e^{-i
 \vec P \vec \phi_p}  \exp \left[V \Psi \left(\phi_N,\phi_E, \vec \phi_p
  \right)\right]~.
\end{equation}
Energy and momentum conservation (continuous quantities) require additionally
the use of Dirac $\delta$ functions, rather than the Kronecker $\delta$ for
discrete quantities. We need dimensionless quantities for the delta-functions,
thus $\phi_E$ and $\vec \phi_p$ need to have dimension $[GeV^{-1}]$ (see as
well Section \ref{MCEexpample}, Appendix \ref{MCEPartFunc}, and
ref. \cite{MonteCarlo}). The CGF can be expressed with  a suitable choice for
 $\phi_E$ and $\vec \phi_p$ in MB statistics as:
\begin{equation}
 \Psi \left(\phi_N,\phi_E, \vec \phi_p \right) =
\frac{g}{\left(2\pi \right)^3}
  \int d^3p \; e^{ - |\vec p|/T  } \; e^{i\phi_N} \;
  e^{i |\vec p| \phi_E} \; e^{i \vec p \vec   \phi_p}~,
\end{equation}
or for FD (upper sign) and BE (lower sign) statistics as:
\begin{equation}
\Psi \left(\phi_N,\phi_E, \vec \phi_p \right) =
\frac{g}{\left(2\pi \right)^3}
  \int d^3p \; \ln \left[ 1 \pm e^{ -  |\vec p|/T } \; e^{i\phi_N} \;
  e^{i |p| \phi_E} \; e^{i \vec p \vec   \phi_p} \right]^{\pm1}.
\end{equation}
Using shorthand $\widetilde{Q}^j = \left(N,E,P_x, P_y, P_z \right)$
and $ \tilde \phi_j = \left(\phi_N, \phi_E, \phi_{p_x}, \phi_{p_y},
\phi_{p_z} \right)$ one can simplify the notation, and only use
the asymptotic solution Eq.(\ref{CE_HG_P_Gauss}) for large volumes, derived
in the previous section.
\begin{eqnarray}
P \left(\tilde Q^j \right) &\simeq& \Bigg[ \prod \limits_{j=1}^5 \int
\limits_{-\infty}^{\infty} \frac{d \tilde \phi_j}{\left(2 \pi
 \right)} \Bigg] ~e^{-i \tilde Q^j \tilde \phi_j} ~\exp \left( V
 \sum_{n=1}^{\infty} ~\frac{i^n}{n!}~  \tilde \kappa_n^{j_1 \; \dots \; j_n}
 \phi_{j_1} \; \dots \; \phi_{j_n}\right) \nonumber \\
 &\simeq& \frac{1}{\left( 2 \pi  V \right)^{5/2}} \frac{\exp \left(
 -\frac{\tilde \xi^j \tilde \xi_j}{2}\right)}{\det \mid \tilde \sigma \mid}
~.
\end{eqnarray}
The cumulants needed are now given by the respective derivatives
at the origin. In case FD or BE statistics are used one
will employ the same set of derivative operators, which result in the
usual, yet slightly more difficult, integrals due to the
logarithm. The expectation values are:
\begin{eqnarray}
\kappa_1^N &=& \left( -i \frac{\partial}{\partial \phi_N}\right)
 \Psi ~\Big|_{\phi_j= \vec 0}  ~=~ \frac{g}{\left( 2 \pi \right)^3} \int
d^3p\; f^{\prime} \left( \vec p \right)  ~,
 \\
\kappa_1^E &=& \left( -i \frac{\partial}{\partial \phi_E}\right)
 \Psi ~\Big|_{\phi_j= \vec 0}  ~=~ \frac{g}{\left( 2 \pi \right)^3} \int
d^3p\;  f^{\prime} \left( \vec p \right) ~ |\vec p| ~,
\\
\kappa_1^{p_x} &=&  \left( -i \frac{\partial}{\partial
    \phi_{p_x}}\right)  \Psi ~\Big|_{\phi_j= \vec 0}  ~=~ \frac{g}{\left(
    2 \pi \right)^3} \int d^3p\;  f^{\prime} \left( \vec p \right) ~ p_x =0~,
\end{eqnarray}
while some selected elements from the second rank tensor $\tilde \kappa_2$ are:
\begin{eqnarray}
\kappa_2^{N,N} &=&  \left( -i \frac{\partial}{\partial
\phi_N}\right)^2  \Psi ~\Big|_{\phi_j= \vec 0} ~=~ \frac{g}{\left( 2 \pi
\right)^3} \int d^3p\;  f^{\prime\prime} \left( \vec p \right)~,
\\
\kappa_2^{E,E} &=& \left( -i \frac{\partial}{\partial
\phi_E}\right)^2  \Psi ~\Big|_{\phi_j= \vec 0} ~=~ \frac{g}{\left( 2 \pi
\right)^3} \int d^3p\;  f^{\prime\prime} \left( \vec p \right)~ |\vec p|^2~,
\\
\kappa_2^{p_x,p_x} &=& \left( -i \frac{\partial}{\partial
    \phi_{p_x}}\right)^2  \Psi ~\Big|_{\phi_j= \vec 0} ~=~ \frac{g}{\left(
    2 \pi \right)^3} \int d^3p\;  f^{\prime\prime} \left( \vec p \right)~
p_x^2 ~=~ \frac{1}{3} ~ \kappa_2^{E,E} ~,
\\
\kappa_2^{E,N} &=& \left( -i \frac{\partial}{\partial
    \phi_E}\right) \left( -i \frac{\partial}{\partial \phi_N}\right)
\Psi ~\Big|_{ \phi_j = \vec 0} ~=~ \frac{g}{\left( 2 \pi \right)^3} \int d^3p~
 f^{\prime\prime} \left( \vec p \right)~  |\vec p| ~,
\end{eqnarray}
where similar to Eqs. (\ref{psi_p_g}) and (\ref{psi_pp_g}) we define:
\begin{eqnarray}
f^{\prime}  \left( \vec p \right)~=~ \frac{e^{-  |\vec p|/T}}{1 \pm e^{-
    |\vec p|/T} }~, \qquad  \textrm{and} \qquad f^{\prime \prime} \left(
  \vec p \right) ~=~ \frac{e^{-   |\vec p|/T}}{\left( 1 \pm e^{-
      |\vec p|/T } \right)^2 }~,
\end{eqnarray}
or simply $f\left( \vec p \right)=f^{\prime}  \left( \vec p \right)=f^{\prime
  \prime} \left( \vec p \right)$ in MB approximation. Due to spherical
symmetry of the momentum distribution, we find $\kappa_2^{p_x,p_x} =
\kappa_2^{p_y,p_y} = \kappa_2^{p_z,p_z}$. Further correlation terms are
identical to zero, $\kappa_2^{N,p_x} = \kappa_2^{E,p_x} = \kappa_2^{p_x,p_y} =
0 $. In our previous notation we find the first and second cumulants, $\tilde
\kappa_1$ and $\tilde \kappa_2$, in Boltzmann approximation for the
distribution $P(N,E,\vec P)$:
\eq{ \tilde \kappa_1 ~=~ \frac{g}{\left(2\pi \right)^3}~
\begin{pmatrix}
8 \pi T^3, & 24 \pi T^4, & 0, & 0, & 0
\end{pmatrix}~,\qquad \textrm{and}
\label{MCEvector}
}
\eq{ \tilde \kappa_2 ~=~  \frac{g}{\left(2\pi \right)^3}~
\begin{pmatrix}
8 \pi T^3 & 24 \pi T^4 & 0 & 0 & 0\\
24 \pi T^4 & 96 \pi T^5 & 0 & 0 & 0\\
0 & 0 & 32 \pi T^5 & 0 & 0 \\
0 & 0 & 0 & 32 \pi T^5 & 0 \\
0 & 0 & 0 & 0 & 32 \pi T^5
\end{pmatrix}~.
\label{MCEmatrix}
}
For the normalization we need additionally the second cumulant
$\kappa_2$ of the PDF $P(E,\vec P)$, which is obtained from $\tilde
\kappa_2$ by crossing out the $1^{st}$ column and $1^{st}$ row. The
remaining relevant integrals for FD and BE statistics are
summarized in table \ref{masslessInt}.
\begin{table}[h!]
\begin{center}
\begin{tabular}{c|c|c|c}
~~~~~~~~~~~ & ~~~~MB~~~~ & ~~~~~~FD~~~~~~ & ~~~~~~BE~~~~~~  \\
\hline $ \kappa_1^N $ &  $8 \pi T^3$ & $ 6 \pi \zeta (3) T^3$ & $
8 \pi \zeta(3) T^3$ \\
$ \kappa_1^E $ &  $24 \pi T^4$ & $\frac{7}{30} \pi^5 T^4 $ &
$\frac{4}{15} \pi^5 T^4 $ \\
$ \kappa_2^{N,N}$ & $8 \pi T^3$  & $\frac{2}{3} \pi^3 T^3 $  &
$\frac{4}{3} \pi^3 T^3 $ \\
$ \kappa_2^{E,N}$ & $24 \pi T^4$  & $18 \pi \zeta (3) T^4 $  &
$24 \pi \zeta(3) T^4 $ \\
$ \kappa_2^{E,E}$ & $96 \pi T^5$  & $\frac{14}{15} \pi^5 T^5 $  &
$\frac{16}{15} \pi^5 T^5 $ \\
\end{tabular}
\end{center}
\caption{Selected elements of the first two cumulant tensors for a massless
gas in Fermi-Dirac, Bose-Einstein statistics and Boltzmann approximation. All
entries have to be multiplied by $g / \left( 2 \pi \right)^3$.  The Riemann
Zeta function is $\zeta \left(3 \right) \simeq 1.202$.}
\label{masslessInt}
\end{table}
Following the recipe from above, see Eq.(\ref{SimpleOmega}), the scaled
variance can be expressed as follows:
\begin{equation}
\omega~ =~ \frac{\det|\tilde \kappa_2|}{\kappa_1^N ~\det |\kappa_2|
}~=~\frac{\kappa_2^{E,E} ~ \kappa_2^{N,N} - \left(
\kappa_2^{E,N} \right)^2}{\kappa_1^N~ \kappa_2^{E,E}}~.
\end{equation}
Independent of energy density one finds for the large volume limit the
following asymptotic values for the scaled variance $\omega$, in exact
agreement with \cite{GrandMce}, in table \ref{massless}.
\begin{table}[h!]
\begin{center}
\begin{tabular}{c|c|c|c}
 & MB & FD & BE  \\
\hline
$ \omega_{mce}$ &  $~~~~0.25~~~~$ & $~0.198314~$ & $~0.535463~$ \\
\end{tabular}
\end{center}
\caption{Asymptotic of the scaled variance in a MCE for a neutral
massless gas} \label{massless}
\end{table}
Momentum conservation explicitly drops out in the thermodynamic limit in the
calculation of the ratio of the relevant determinants. This will not
hold true for any finite system size (see Section \ref{FiniteVolCorr}), due to
the appearance of off-diagonal terms symmetric in momentum like
$\kappa_3^{N,p_x,p_x}$. However for large enough volumes, expected to be
created in heavy ion collision experiments, one can probably
safely disregard exact momentum conservation, when considering multiplicity
fluctuations in the full momentum space, i.e. $4 \pi$ yield and
fluctuations. In previous publications on asymptotic multiplicity fluctuations
in the MCE it was only always assumed (without proof) that exact momentum
conservation would not affect the result, and thus for technical reasons not
taken into account.

One last issue should be mentioned. For partition functions for very small
volume one should return to summation over (quantized) momentum states, rather
than integration over (continuous) momentum space, or ideally turn to a
quantum field theoretical frame work of the MCE \cite{QFTMCE}. Usually these
distinctions do not affect the result much, simplify however calculations
considerably. For a massless gas however we  encounter the  problem of a
divergent $\kappa_3^{N,N,N}$ for BE statistics. This problem is  well known
from textbooks and usually overcome by applying a low momentum cut-off
\cite{Physics}. 
\section{Resonance Decay}
\label{ResDecay}
In this section we want to show how to include resonance decay
analytically directly into the system partition function. This will
prove far more efficient than the definition of a generating function
\cite{res} which requires a rather cumbersome calculation of all possible
primordial correlators, see Section \ref{Correspondence_res}. Particle decay
is itself a random process. Nevertheless one can assign a particular volume in
phase space, given by the value of its single particle partition function
$z_l$, Eq.(\ref{QstatsZET}), to one type of resonance $l$. Resonance decay will
now populate this volume in phase space according to some weight factor, the
branching ratio, for each of the possible decay modes. This weight can be
assigned  to the particle type(s) we are set to investigate. Based on the
assumption that detected particles are drawn in the form of a random sample
from all final state particles, e.g. disregarding correlation in momentum
space, this procedure leads to acceptance scaling employed in ref.
\cite{CEfluc_1,res,MCEvsData} (see Section \ref{AppScaling}). Conservation
laws can be imposed on the primordial state (rather than the final state),
since decay channels, which are experimentally measured, do not only obey
charge conservation, but all relevant conservation laws (omitting weak decays).
\subsection{Final State Partition Function}
\label{FinalStatePF}
For calculation of final state distributions, we have to determine {\it
  final} branching ratios of a resonance type into only stable particles. As an
example we consider the decay channel $A \rightarrow B + X$ with branching
ratio $\Gamma_{A\rightarrow B + X} = a$. Resonance  $B$ could  itself be
unstable and subsequently decay via the channel $B \rightarrow Y +Z$ with
branching ratio $\Gamma_{B  \rightarrow Y + Z} = b$. So we define the {\it  final}
branching ratio $\Gamma_{A \rightarrow X+ Y+ Z} = \Gamma_{A \rightarrow B +
X}~\Gamma_{B\rightarrow Y + Z} = ab$. Decay tables in \cite{res,MCEvsData}
have also been generated according to this prescription.

For resonances it seems economical to define further {\it absolute} branching
ratios $\Gamma_l^c$ as the sum over all {\it final} decay channels of
resonance $l$ with a given number $c$ of selected  daughters. Hence
$\Gamma_l^2$ is the sum over all {\it final} decay  channels with two  daughter
particles which are of interest, i.e. two positively charged particles in case
one wants to calculate $\omega^+$. As a consequence of this
definition, branching ratios $\Gamma_l^c$ will depend on which
$\omega$ one is set to calculate.

For the final state one has to take all $C_{l}$ {\it absolute}
decay channels of resonance type $l$ into a number $c$ of selected
stable particles into account. All parent resonances carry 
fictitious particle specific fugacity $\lambda_N$ taken to the
power of the number $c$  of selected daughters in a particular
channel, $\Gamma_l^c ~\left( \lambda_N \right)^c$. For the sake of
a common treatment for all particles and resonances we assign a
`decay` channel to stable particles as well, either $\Gamma_l^1 =
1$ (while $\Gamma_l^{c \not= 1}=0$) if selected, or $\Gamma_l^0 =
1$  (while $\Gamma_l^{c \not= 0}=0$) if not selected. The single
particle partition function reads after substitution $\left(
\lambda_N\right)^c \rightarrow e^{ic\phi_N}$:
\begin{equation}
\psi_l \left(\phi_j,\phi_{N},\Gamma_l^{c} \right) =
\frac{g_l}{\left( 2 \pi \right)^3} \int
    d^3p \ln \left(1 \pm e^{ - \left( \varepsilon_l-\mu_l\right)/T} e^{i
   q_l^j  \phi_j} \left[ \sum_{c=0}^{C_{l}} \; \Gamma_l^{c} \;
e^{i  c  \phi_{N}} \right] \right)^{\pm 1}~,
 \end{equation}
where the form of the vector $q_l^j$ depends on our choice of ensemble. For
instance, we could have for a hadron resonance gas with three conserved
charges $q_l^j = (q_l,b_l,s_l)$ in the CE, or  $q_l^j =
(q_l,b_l,s_l,\varepsilon_l,\vec p_l )$ in the MCE. The sum over all decay
channels which produce {from zero up to} a  number $C_{l}$ particles of the
selected types needs to be one:
\begin{equation}\label{Decay1}
\sum_{c=0}^{C_{l}} \; \Gamma_l^{c} = 1~.
\end{equation}
This is somewhat of a practical challenge, since decay chains of
heavier resonances are not always well established \cite{PDG} and
respective thermal models codes \cite{THERMUS,SHARE} struggle to
implement this. There are several ways to deal with this, the two
extreme ones are 1) rescale all known channels according to
Eq.(\ref{Decay1}), to unity, or 2) assign the missing fraction to
the 'channel' $\Gamma_l^0$, e.g. to the channel without stable
particles of interest.  In case one sets all angles $ (\phi_j,
  \phi_{N}) = \vec 0$, i.e. one returns to $e^{ic\phi_N} \rightarrow \left(
    \lambda_N \right)^c$ , one obtains the GCE partition function:
\begin{equation}
Z = \exp \left[ V \sum_{l} \frac{g_l}{\left( 2 \pi \right)^3} \int
    d^3p \ln \left(1 \pm e^{ - (\varepsilon_l-\mu_l)/T}  \left[
        \sum_{c=0}^{C_{l}} \; \Gamma_l^{c} \; \left( \lambda_N \right)^c
      \right] \right)^{\pm 1} \right]~,
\end{equation}
from which  GCE expectation values can be calculated as $\langle N \rangle =
\lambda_N \frac{\partial }{\partial \lambda_N} \ln Z |_{\lambda_N = 1}$.

\subsection{Limited Acceptance}
\label{FinalDetect}
An imperfect detector will modify our definition of  branching ratio
$\Gamma_l^c$. Based on the assumption of un-correlated particle detection, all
particles have equal probability $q$ of being observed. The corresponding
acceptance distribution of observing $n$ particles when $c$ are produced is
given by a binomial distribution $P_{acc}(n,c) = q^{n} \left(1-q
\right)^{c-n} {c \choose n}$. We define the {\it effective} branching ratios
$ \Gamma_{l,q}^{n}$, which already include the effect of finite acceptance,
as the product of {\it absolute} branching ratio $\Gamma_l^{c}$ and
acceptance distribution:
\begin{equation} \label{Binomial}
 \Gamma_{l,q}^{n} = \sum_{c=n}^{C_{l}}  ~\Gamma_l^{c}~ q^{n} \left(1-q
  \right)^{c-n} {c \choose n} \;.
\end{equation}
For example the {\it effective} branching ratio $\Gamma_{l,q}^{2}$ is the sum
over all {\it absolute} branching ratios $\Gamma_l^{c}$ which produce at least
2 stable selected particle, $\Gamma_{l,q}^{2} = \Gamma_l^2 q^2 + 3 \Gamma_l^3
q^2 (1-q)  + 6 \Gamma_l^4 q^2  (1-q)^2  + \hdots$.  The binomial coefficient
${c \choose n} = \frac{c!}{n!\left(c-n \right)!}$ takes care of the fact that
particles are indistinguishable. In our approximation resonance decay and
particle detection are thus two independent random processes. When both,
particle decay and un-correlated detection, are included, the single particle
partition functions reads:
\begin{equation}
\psi_{l,q} \left(\phi_j,\phi_{N},\Gamma_l^{c},q\right) =
\frac{g_l}{\left( 2 \pi\right)^3} \int
   d^3p \ln \left(1 \pm  e^{ - \left( \varepsilon_l-\mu_l \right) / T} e^{i
   q_l^j \phi_j} \left[ \sum_{n=0}^{C_{l}}  \; \Gamma_{l,q}^{n} \;
e^{i n  \phi_{N}} \right] \right)^{\pm 1}~.
\label{Finalpsidetect}
\end{equation}
The sum over all {\it effective} channels $\Gamma_{l,q}^{n}$ is equal to one,
\begin{equation}
\sum_{n=0}^{C_{l}}~  \Gamma_{l,q}^{n} ~=~
\sum_{c=0}^{C_{l}} ~\Gamma_l^{c} ~\sum_{n=0}^{c}
q^{n} ~\left(1-q \right)^{c-n} {c \choose
  n } ~=~ 1 ~,
\end{equation}
since the second summation ($\sum_{n=0}^{c}$) is equal to unity, while
according to Eq.(\ref{Decay1}) the branching ratios $\Gamma_l^c$ are also
normalized. The final state CGF thus is the sum over all such single particle
partition functions:
\begin{equation}
\Psi ~=~  \sum_{l} ~\psi_{l,q}\left(\phi_j,\phi_{N},\Gamma_{l,q}^{c} \right)
~.
\end{equation}

\subsection{The Cumulant Tensor}
\label{FinalKappa}
The procedure is essentially no different from Sections \ref{CLTCE} and
\ref{CLTMCE}, yet slightly more complicated, due to the various multiplicities
in the decay modes. For clarity some elements are explicitly given. The choice
of ensemble then naturally defines the cumulants needed for
calculations. Both, CE and MCE, are considered. The final mean
values of particle and  energy density are:
\begin{eqnarray}
\kappa_1^N &=& \left( -i \frac{\partial}{\partial \phi_{N}}\right)
\Psi ~\Big|_{\phi_j = \vec 0} ~=~ \sum_l \left[ \sum_{n=0}^{C_{l}}
\Gamma_{l,q}^{n} ~ n \right] ~ \psi^{\left(1,1;0 \right)}_l ~,\label{k_1_N_f}
\\
\kappa_1^{E} &=&  \left( -i \frac{\partial}{\partial
\phi_E}\right) \Psi ~\Big|_{ \phi_j = \vec 0} ~=~ \sum_l
\psi^{\left(1,1;1 \right)}_l ~,\label{k_1_E}
\end{eqnarray}
Fluctuation of particle and energy density, and
correlations between particle and baryon number, particle number and energy,
and between baryon number and energy are given by:

\begin{eqnarray}
\kappa_2^{N,N} &=& \left( -i \frac{\partial}{\partial
\phi_{N}}\right)^2  \Psi ~\Big|_{ \phi_j = \vec 0} ~=~ \sum_l \left[
\sum_{n=0}^{C_{l}} \Gamma_{l,q}^{n} ~ n^2 \right] ~
\psi^{\left(1,1;0 \right)}_l
- \left[ \sum_{n=0}^{C_{l}} \Gamma_{l,q}^{n} ~ n
\right]^2 ~ \psi^{\left(2,2;0 \right)}_l~, \label{k_2_NN_f}\\
\kappa_2^{E,E} &=&  \left( -i \frac{\partial}{\partial
\phi_E}\right)^2 \Psi ~\Big|_{ \phi_j = \vec 0} ~=~  \sum_l
\psi^{\left(1,2;2 \right)}_l~, \label{k_2_EE}\\
\kappa_2^{N,B} &=&\left( -i \frac{\partial}{\partial
\phi_B}\right) \left( -i \frac{\partial}{\partial \phi_{N}}\right)
\Psi ~\Big|_{\phi_j = \vec 0} ~=~  \sum_l ~ b_l ~\left[ \sum_{n=0}^{C_{l}}
  \Gamma_{l,q}^{n} ~ n \right] ~ \psi^{\left(1,2;0 \right)}_l~,
\label{k_2_NB_f} \\
\kappa_2^{N,E} &=&  \left( -i \frac{\partial}{\partial
\phi_E}\right) \left( -i\frac{\partial}{\partial \phi_{N}}\right)
 \Psi ~\Big|_{\phi_j = \vec 0} ~=~ \sum_l \left[ \sum_{n=0}^{C_{l}}
   \Gamma_{l,q}^{n} ~ n \right] ~\psi^{\left(1,2;1\right)}_l~,
 \label{k_2_EN_f}\\
\kappa_2^{B,E} &=&  \left( -i \frac{\partial}{\partial \phi_E}\right) \left(
  -i \frac{\partial}{\partial \phi_B}\right)  \Psi ~\Big|_{\phi_j = \vec 0} ~=~
\sum_l b_l \psi^{\left(1,2;1\right)}_l~,\label{k_2_BE}
\end{eqnarray}
where we use a more general shorthand notation for the derivatives similar to
Eqs.(\ref{psi_p_g}), and (\ref{psi_pp_g}):
\begin{equation}\label{psi_deriv}
\psi^{\left(a,b;c \right)}_l =  \left(\pm 1 \right)^{a+1} ~
\frac{g_l}{\left(2\pi \right)^3} \int d^3p~ ~ \varepsilon_l^c ~ \frac{\left(
e^{-\left( \varepsilon_l -\mu_l\right) /T  } \right)^a} {\left( 1 \pm
  e^{-\left( \varepsilon_l -\mu_l\right) /T} \right)^b}\;.
\end{equation}
It is quite easy to see that higher orders of these elements
become quickly very complicated. For calculation of asymptotic multiplicity
fluctuations, integrated over full momentum space (4$\pi$ fluctuations), it is
not necessary to take  momentum conservation into account (see Section
\ref{CLTMCE}). The implementation of proper kinematical cuts will be the
subject of a forthcoming paper.
\section{Correspondence to Microscopic Correlator Approach}
\label{Correspondence}
One can make a correspondence between different notations in this
paper and in the micro-correlator approach published previously
\cite{res,MCEfluc_1,MCEvsData,QGas,GrandMce,BGZ} for asymptotic fluctuations.
The aim is to show that both methods lead to the same result, as well as handle
resonance decay exactly the same way. This fact should not be taken as
trivial, as both methods appear very differently (not only in terms of
notation). In this work all quantities are derived
from a macroscopic partition function, while in the microscopic correlator
approach one considers average occupation numbers, and their fluctuations, of
individual momentum levels and correlations between different levels.
We restrict the consideration to CE with three conserved charges. Additional
charges or energy (momentum) conservation in the MCE would just lead to
enlargement of the matrixes $\kappa_2$ and $\widetilde{\kappa_2}$.
We start off from Eq.(\ref{SimpleOmega}):
\eq{\label{omega_kappa_A}
  \omega_{c.e.} \;=\;
  \frac{\det|\widetilde{\kappa}_2|}{\kappa_1^{N}\det|\kappa_2|}~,
}
where $\det |\kappa_2|$ and $ \det |\widetilde{\kappa_2}|$ are the
determinants of the matrices Eq.(\ref{CEmatrix}). Expanding the determinants
in terms of their complementary minors, $M_{i,k}$ and $\widetilde{M}_{i,k}$
respectively, yields \cite{Formulas}:
\eq{
  \det |\widetilde{\kappa_2}| \;=\; \sum_{j=1}^4 (-1)^{j+k} ~
  \left(\widetilde{\kappa_2}\right)_{j,k} ~ \widetilde{M}_{j,k}\;,
}
where minors of $\widetilde{\kappa}_2$ can be expressed in terms of the minors
of $\kappa_2$:
\eq{
  \widetilde{M}_{N,N} \;=\; \det|\kappa_2|\;, &&
  \widetilde{M}_{N,Q} \;=\; \kappa_2^{Q,N} M_{Q,Q} \;-\; \kappa_2^{B,N}
  M_{B,Q} \;+\; \kappa_2^{S,N} M_{S,Q}\;, && \textrm{etc.}~.
}
After straightforward calculation Eq.~(\ref{omega_kappa_A}) can be rewritten
as follows:
\eq{
  \omega_{c.e.} & \;=\; \frac{ \kappa_2^{N,N} \widetilde{M}_{N,N} \;-\;
    \kappa_2^{N,Q}
    \widetilde{M}_{N,Q} \;+\; \kappa_2^{N,B} \widetilde{M}_{N,B} \;+\;
    \kappa_2^{N,S} \widetilde{M}_{N,S} }{\kappa_1^N \det|\kappa_2|}
  \nonumber\\
  &\;=\; \frac{\kappa_2^{N,N}}{\kappa_1^N} \;-\; \frac{1}{\kappa_1^N
    \det|\kappa_2|}  \;\Bigg[\; \left(\kappa_2^{N,Q}\right)^2 M_{Q,Q} \;+\;
  \left(\kappa_2^{N,B} \right)^2 M_{B,B}
  \;+\; \left(\kappa_2^{N,S} \right)^2 M_{S,S}  \nonumber \\
  &\qquad\qquad\;\;\,
  \;+\; 2\kappa_2^{N,Q}\kappa_2^{N,S}\,  M_{Q,S}
  \;-\; 2\kappa_2^{N,Q}\kappa_2^{N,B}\,  M_{Q,B}
  \;-\; 2\kappa_2^{N,B}\kappa_2^{N,S}\,  M_{B,S}  \Bigg]~.
  \label{omegai-ce-2}
}
This is the most general case. For a specific calculation we need to specify
the matrix elements.
\subsection{Primordial}
\label{Correspondence_Prim}
In order to change our notation here to that of the microscopic correlator
approach \cite{res,MCEvsData}, we need the derivatives Eq.(\ref{psi_deriv}) of
the single particle partition function $\psi_i$:
\begin{eqnarray}
V\psi_i^{1,1;0} \equiv \sum \limits_p \langle n_{p,i} \rangle~, \qquad
V\psi_i^{1,2;0} \equiv \sum \limits_p \upsilon^2_{p,i}~, \qquad \textrm{and}
\qquad \psi_i^{2,2;0} \equiv \psi_i^{1,1;0} - \psi_i^{1,2;0}~, \nonumber
\end{eqnarray}
where the subscripts $i$ and $p$ denotes particle species and momentum level
respectively. The matrix entries from Section \ref{PrimKappa} are:
\begin{eqnarray}
V \kappa_1^{N} &=& ~\langle N \rangle \equiv ~ \sum_{p,i}
\langle n_{p,i} \rangle\; \qquad V \kappa_2^{N,N}=~\Delta (n^2)\equiv ~\sum_{p,i}
\upsilon_{p,i}^2\; \qquad V \kappa_2^{N,Q}=~\Delta (n q)\equiv ~ \sum_{p,i}
q_i\upsilon_{p,i}^2\; \nonumber \\ V\kappa_2^{Q,Q} &=&  \Delta
(q^2)\equiv ~ \sum_{p,i} q_i^2\upsilon_{p,i}^2\; \qquad  V \kappa_2^{Q,B}
=~\Delta (qb)\equiv
~ \sum_{p,i} q_ib_i\upsilon_{p,i}^2\; \qquad \textrm{etc.~ ,}\nonumber
\end{eqnarray}
where \cite{Physics} $\upsilon_{p,i} = \langle n_{p,i} \rangle (1\pm \langle
n_{p,i} \rangle)$ (upper sign for FD, lower sign for BE). It is easy to see
that our matrix $\kappa_2$ corresponds to the correlation matrix $A \equiv
V\kappa_2$, see e.g. Eq.(12) in \cite{res}. Substituting into
Eq.(\ref{omegai-ce-2}) one finds Eqs.(50,51) from reference \cite{res}:
\begin{equation}
\omega_{c.e.} \;\equiv\;  \frac{1}{\langle N\rangle}\sum_{i,j}\langle \Delta
  N_{i}\Delta N_{j}\rangle_{c.e.}~,
  \label{omegai-ce}
\end{equation}
with the correlator $\langle \Delta N_{i}\Delta N_{j}\rangle_{c.e.}$ being
Eqs.(11,15) from \cite{res}:
\begin{eqnarray}\label{corr}
\langle \Delta N_{i}\Delta N_{j}\rangle_{c.e.} &=&
   \sum \limits_{p,k} \Bigg[ \upsilon_{p,i}^2\,\delta_{ij}\,\delta_{pk}
 -\;  \frac{\upsilon_{p,i}^2v_{k,j}^2}{|A|}
 \Big[ q_iq_j M_{Q,Q} + b_ib_j M_{B,B} + s_is_j M_{S,S} \\
&+&  \left(q_is_j + q_js_i\right) M_{Q,S}
 - \left(q_ib_j + q_jb_i\right) M_{Q,B} - \left(b_is_j +
b_js_i\right) M_{B,S}
 \Big] \Bigg]\;. \nonumber
\end{eqnarray}
Eq.~(\ref{omegai-ce}) can be simplified for the case of single-specie
fluctuations:
\eq{
  \omega^{j}_{c.e.}~=~\omega^j_{g.c.e.}~\Bigg[ ~1 ~-~ \frac{\sum_k
    ~v_{k,j}^2}{|A|}~& \Bigg( q_j^2 M_{Q,Q} + b_j^2 M_{B,B} + s_j^2 M_{S,S}
  \nonumber\\
 & +  2q_js_j M_{Q,S} - 2q_jb_j M_{Q,B} - 2b_js_j  M_{B,S} \Bigg)~\Bigg]~,
}
which coincides with Eq.(16) from \cite{res}, where $\omega^j_{g.c.e.} \equiv
\sum_p ~v_{p,j}^2 / \sum_p \langle n_{p,j} \rangle$~.
\subsection{Resonance Decay}
\label{Correspondence_res}
In order to account for resonance decay we need the matrix elements stated in
Section \ref{FinalKappa} together with:
\begin{eqnarray}
\sum \limits_{n=0}^{C_{l}} \Gamma_l^{n} n^2 = \sum \limits_i \sum
\limits_j \langle n_i n_j \rangle_l~,  \qquad \textrm{and } \qquad \left(\sum
  \limits_{n=0}^{C_{l}} \Gamma_l^{n} n \right)^2 = \sum \limits_i
\sum \limits_j \langle n_i \rangle_l \langle n_j \rangle_l ~,
\end{eqnarray}
where the summations are to be taken over all stable hadrons $i$ and $j$ which
one is set to consider, and $n=\sum_i n_i$ is the total number of selected
particles in this channel.  The matrix element $V \kappa_2^{N,N}=\Delta \left(
  n^2 \right)$  then splits into summations over stable hadron $i,j$ and
resonances $R$~,
\begin{eqnarray}
\Delta \left( n^2 \right) &=& \sum_{i,p} v^2_{i,p} ~+~ \sum_{R,p} v^2_{R,p}
\sum_{i,j} \langle n_i \rangle_R \langle n_j \rangle_R ~+~  \sum_{R,p} n_{R,p}
\sum_{i,j} \langle \Delta n_i \Delta n_j \rangle_R~,
\end{eqnarray}
while the correlation terms $V \kappa_2^{N,Q}=\Delta \left( nq \right)$ are of
the form
\begin{eqnarray}
\Delta \left( nq \right) &=& \sum_{i,p} q_i~v^2_{i,p} ~+~ \sum_{R,p} q_R ~
v^2_{R,p} \sum_{i} \langle n_i \rangle_R~.
\end{eqnarray}
Substituting into Eq.(\ref{omegai-ce-2}) gives Eq.(47) in
reference \cite{res}:
\eq{\label{res-old}
 & \langle \Delta N_i\,\Delta N_j\rangle_{c.e.}
 \;=\; \langle\Delta N_i^* \Delta N_j^*\rangle_{c.e.}
  \;+\; \sum_R \langle N_R\rangle\; \langle \Delta n_{i}\; \Delta
  n_{j}\rangle_R
 \;+\; \sum_R \langle\Delta N_i^*\; \Delta N_R\rangle_{c.e.}\; \langle
 n_{j}\rangle_R
  \; \nonumber \\
 &+\; \sum_R  \langle\Delta N_j^*\;\Delta N_R\rangle_{c.e.}\; \langle
 n_{i}\rangle_R \;+\; \sum_{R, R'} \langle\Delta N_R\;\Delta
 N_{R'}\rangle_{c.e.} \; \langle n_{i}\rangle_R\; \langle
 n_{j}\rangle_{R^{'}}\;.
 }
Correlation terms $\langle\Delta N_i^* \Delta N_R\rangle_{c.e.}$, and
$\langle\Delta N_R \Delta N_{R'}\rangle_{c.e.}$ appear in the CE (and the MCE)
due to products of $\Delta \left( nq \right) \Delta \left(ns \right)$, etc.,
in Eq.(\ref{omegai-ce-2}), and are absent in the GCE.
\subsection{Acceptance Scaling}
\label{AppScaling}
Starting off again from our approximation of un-correlated particle detection,
from Section \ref{FinalDetect}, we find for the first two moments
\cite{Formulas} of the binomial distribution of detected particles produced by
decay of resonance type $l$:
\begin{eqnarray}
\langle n \rangle_l &=& \sum \limits_{c=0}^{C_l} ~ \Gamma^c_l~\sum
\limits_{n=0}^{c}~n~q\left(1-q \right)^{c-n}
{c \choose n}~=~ \sum \limits_{c=0}^{C_l} ~\Gamma^c_l~ qc
~=~q\langle c\rangle_l~,\\
\langle n^2 \rangle_l &=& \sum \limits_{c=0}^{C_l}~\Gamma^c_l ~\sum
\limits_{n=0}^{c}~n^2~q\left(1-q \right)^{c-n}
{c \choose n}~=~ \sum \limits_{c=0}^{C_l} ~ \Gamma^c_l~ \Big[
  q \left(1-q \right)c +q^2c^2\Big]\nonumber \\
&=&q\left(1-q \right)\langle c\rangle_l ~+~ q^2\langle c^2\rangle_l~.
\end{eqnarray}
Comparing with Eqs.(\ref{k_1_N_f}), (\ref{k_2_NB_f}), and (\ref{k_2_NN_f}) we
find for the cumulant tensor elements,
\begin{eqnarray}
 \kappa_1^{N} &=& \sum_l ~q~ \langle c_l \rangle~  \psi_l^{\left(
    1,1;0\right)} ~=~q ~(\kappa_{1}^N)_{4\pi}\;,  \qquad \textrm{and}
    \\
 \kappa_2^{B,N} &=&
\sum_l ~q ~b_l~ \langle c_l \rangle~  \psi_l^{\left(1,2;0\right)}
 ~=~ q~(\kappa_{1}^{B,N})_{4\pi}\;, \qquad \textrm{etc.~,}
 \\
 \kappa_2^{N,N} &=& \sum_l \Bigg[q\left(1-q \right) \langle c_l \rangle
   + q^2\langle c^2_l \rangle \Bigg] \psi_l^{\left( 1,1;0\right)}
  ~-~q^2 \langle c^2_l \rangle \psi_l^{\left( 2,2;0\right)} \nonumber
  \\
 &=& q (\kappa_{1}^N)_{4\pi} - q^2 (\kappa_{1}^N)_{4\pi}
  + q^2 (\kappa_{2}^{N,N})_{4\pi} ~.
\end{eqnarray}
The index `$4\pi$` denotes quantities that would be measured by
the ideal detector with full $4\pi$ acceptance. Substituting into
Eq.~(\ref{omegai-ce-2}), we obtain the acceptance scaling formula
form Ref. \cite{CEfluc_1}:
\begin{equation}
\omega^{acc} ~=~1-q+q\omega^{4\pi}~.
\end{equation}

\section{Finite Volume Corrections}
\label{FiniteVolCorr}
Considering finite system size effects on distributions, we leave the region
were the thermodynamic limit approximation is valid. Chemical potentials
$\mu_j$ and temperature $T$ do not correspond to the physical ones, which
would be found in the GCE, anymore, but have to be thought of as Lagrange
multipliers, used to maximize the partition function for a given
(micro)canonical state. First we derive some volume dependent
corrections terms, and then find a condition that defines the correct values of
$\mu_j$ and $T$. The correct choice allows to write down the thermodynamical
potentials, the Helmholtz free energy $F$ for CE, and the entropy $S$ for the
MCE, in terms of the generalized partition function. Some general criterion for
the validity of the expansion is given. We will compare CE and MCE
results with scenarios which are accessible to analytical methods in Section
\ref{QofA}.
\subsection{Gram-Charlier Expansion}
In Section \ref{CLTCE} we have shown that in the thermodynamic limit any
equilibrium multiplicity distribution can be approximated by a Gaussian
Eq.(\ref{CE_HG_Gauss}). Further parameters, describing the shape of the
distribution, skewness ($\kappa_3$), or excess ($\kappa_4$), tend to zero as
volume is increased. We return to a generalized version of
Eq.(\ref{CLTfinalZ}) with a number $J$ of conserved quantities and use
Gram-Charlier expansion \cite{MATH}. Here we need to explicitly find the
inverse square root $\sigma^{-1}$ of  $\kappa_2$ for the calculation of the
normalized cumulants $\lambda_n$ Eq.(\ref{normcum}), see Appendix
\ref{DetIdent}. For every  considered charge one will pick up an  additional
factor of $\sqrt{V}$ from the volume element $d\theta_j$, Eq.(\ref{dtheta}),
\begin{eqnarray}
\mathcal{Z}^{ Q^j} \simeq \frac{ Z} { V^{J/2} \det|
\sigma | }
 \left[ \prod \limits_{j=1}^{J}
  \int \limits_{-\infty}^{\infty} \frac{d \theta_j}{ 2\pi } \right] ~ \exp
\Bigg[ &-& i  \xi^j \theta_j - \frac{\theta^j  \theta_j}{2!} \nonumber \\
  + &&\sum \limits_{n=3}^{\infty} ~i^n~ V^{-\frac{n}{2}+1}
  ~\frac{\lambda_n^{j_1,j_2,\dots,j_n}}{n!} ~\theta_{j_1}\theta_{j_2} \dots
  \theta_{j_n}   \Bigg]~.
\end{eqnarray}
Expanding the exponential in terms of powers in volume, we find:
\begin{eqnarray}\label{Vcorrection}
\mathcal{Z}^{Q^j} &\simeq& \frac{ Z}{ V^{J/2} \det|
\sigma |} \left[ \prod \limits_{j=1}^{J}\int
\limits_{-\infty}^{\infty}
    \frac{d \theta_j}{2\pi} \right]~ \exp \left[ - i \xi^j \theta_j -
    \frac{\theta^j \theta_j}{2!} \right]
 \times \Bigg[ 1 + \frac{\lambda_3^{j_1,j_2,j_3}}{3!}
  ~\frac{ i^3 \theta_{j_1} \theta_{j_2} \theta_{j_3} }{ V^{1/2}} \nonumber
\\
 &+& \frac{\lambda_4^{j_1,j_2,j_3,j_4}}{4!}
  ~\frac{ i^4 \theta_{j_1} \theta_{j_2} \theta_{j_3} \theta_{j_4} }{ V} +
  \frac{1}{2!}
  \frac{\lambda_3^{j_1,j_2,j_3}}{3!} ~\frac{\lambda_3^{j_4,j_5,j_6}}{3!}~
  \frac{i^6 \theta_{j_1} \dots \theta_{j_6} }{V} +
  \mathcal{O} \left( V^{-3/2}\right) \Bigg]~.
\end{eqnarray}
Correction terms in Eq.~(\ref{Vcorrection})
can be obtained by differentiation of $\exp \left[ - i \xi^j \theta_j \right]
$ with respect to $\xi^j$. One  can thus reverse the order by first
integrating and then again differentiating. Using generalized Hermite
polynomials,
\begin{equation} \label{Hermite}
\left( H_n \left( \xi \right)\right)_{j_1,j_2,\dots,j_n} = \left(
-1 \right)^n ~ \exp \left[ ~\frac{\xi^j \xi_j}{2}~ \right]~
\frac{d^n}{d\xi_{j_1} \; d\xi_{j_2} \dots d\xi_{j_n} }~
\exp \left[ -\frac{\xi^j\xi_j}{2} \right]~,
\end{equation}
with the adjusted shorthand notation,
\begin{eqnarray}
h_3 \left( \xi \right) &=&
\frac{\lambda_3^{j_1,j_2,j_3}}{3!} ~ \left(  H_3
  \left(\xi \right) \right)_{j_1,j_2,j_3}~,  \\
h_4 \left( \xi \right) &=&
\frac{\lambda_4^{j_1,j_2,j_3,j_4}}{4!} ~ \left( H_4
  \left(\xi \right)\right)_{j_1,j_2,j_3,j_4} ~+~ \frac{1}{2!}~
\frac{\lambda_3^{j_1,j_2,j_3}}{3!} ~
\frac{\lambda_3^{j_4,j_5,j_6}}{3!} ~\left(
  H_6 \left(\xi \right) \right)_{j_1,\dots j_6}~,  \\
h_5 \left( \xi \right) &=& \frac{\lambda_5^{j_1,\dots,j_5}}{5!}
~\left( H_5 \left(\xi\right) \right)_{j_1,\dots, j_5} ~+~
\frac{\lambda_3^{j_1,j_2,j_3}}{3!} ~
\frac{\lambda_4^{j_1,j_2,j_3,j_4}}{4!}~
\left( H_7 \left(\xi \right) \right)_{j_1,\dots,j_7}  \\
& & \quad +~ \frac{1}{3!}~ \frac{\lambda_3^{j_1,j_2,j_3}}{3!}
~\frac{\lambda_3^{j_4,j_5,j_6}}{3!}
~\frac{\lambda_3^{j_7,j_8,j_9}}{3!}~ \left(
  H_9 \left(\xi \right) \right)_{j_1,\dots,j_9}~,\nonumber
\end{eqnarray}
the partition function for finite volume can be approximated by:
\begin{equation} \label{GramCharlier}
\mathcal{Z}^{Q^j}~ \simeq~ Z ~\frac{ e^{- \frac{\xi^j
\xi_j}{2}}}{\left(2 \pi V  \right)^{J/2}\det| \sigma |  }  \left[1~ +~
\frac{h_3 \left( \xi \right)}{\sqrt{V}} ~+~ \frac{h_4 \left(\xi \right)}{V} +
\frac{h_5 \left( \xi \right)}{V^{3/2}} ~+~ \mathcal{O} \left( V^{-2}\right)
\right]~.
\end{equation}
Considering the simplest case of only one conserved charge, it is evident from
Eq.(\ref{Hermite}), that the first order correction term in
Eq.(\ref{GramCharlier}) is a polynomial of order 3 in $\xi$, while the second
order correction term is a polynomial of order 4, etc. Hence for large values
of $\xi$, e.g. a multiplicity state far from the peak of the distribution will
lead to a bad approximation, and even to negative values for $P(N)$. The
validity of this approximation is thus restricted to the central region of the
distribution. We will compare CE and MCE
results with scenarios which are accessible to analytical methods in Section
\ref{QofA}. In order to distinguish approximations which include corrections
up to different orders in volume in Eq.(\ref{GramCharlier}), we denote the 
asymptotic solution as CLT (central limit theorem), including terms up to
$\mathcal{O} \left( V^{-1/2}\right)$ as GC3 (Gram-Charlier 3), including terms
up to $\mathcal{O} \left( V^{-1}\right)$ as GC4, and including terms up to
$\mathcal{O} \left( V^{-3/2}\right)$ as GC5.

\subsection{Chemical and Thermal Equilibrium}
Here we address the question of how to choose the optimal values
for $T$ and $\mu_j$. If exact solutions were available, distributions, however
{\it not} thermodynamic potentials, would be independent of this choice (see
as well Section \ref{BoltzmannPionGas}). Our
postulate is that our (micro)canonical equilibrium state should be as well
the most likely state in the GCE. For an isotropic momentum distribution, the
macroscopic state $\vec P= \vec 0$ is always the most probable state, since
all odd cumulants involving only momenta vanish $\kappa_1^{p_x} =
\kappa_3^{p_x,p_x,p_x}=\hdots=0$. On the other hand, we know
that the expansion works best around the peak of the distribution. So we
choose $T$ and  $\mu_j$ such that we maximize the partition function at
some point  equilibrium $Q_j^{eq}$. Taking  terms up to $\mathcal O
(V^{-1/2})$ into account, the first derivative of the partition function
Eq.(\ref{CE_PF_1}) reads:
\begin{equation}\label{DiffPF}
  \frac{\partial \mathcal{Z}^{Q^j}}{\partial Q^j} = \frac{e^{-\frac{\xi^j
\xi_j}{2}}} {\left(2 \pi \right)^{J/2} V^{\left(J+1\right)/2} \det| \sigma|} \left[
\xi_k \left(\sigma^{-1} \right)^{ \; k}_{\;\; j} +
 \frac{\lambda_3^{k_1,k_2,k_3}}{3!\sqrt{V}}  \left(\sigma^{-1}
\right)^{ \; k_4}_{\;\; j} H_4\left( \xi \right)_{k_1,k_2,k_3,k_4}
+ \mathcal{O} \left( V^{-1}\right) \right].
\end{equation}
The chemical potentials should be chosen such that the first
derivative Eq.(\ref{DiffPF}) of $\mathcal{Z}^{Q^j}$ with respect
to the conserved quantities $Q^j$ vanishes, hence we 
maximize Eq.(\ref{GramCharlier}) at the point $Q_j^{eq.}$:
\begin{equation}\label{gChemPot}
\frac{\partial \mathcal{Z}^{Q^j}}{\partial Q^j}~\Bigg|_{Q_j^{eq.}}= \vec 0~.
\end{equation}
Using only the asymptotic solution, valid in the thermodynamic limit, this
condition leads to:
\begin{equation}\label{CLTmu}
\xi_k = \left( Q_j - V \kappa_{1,j} \right) \left(
  \sigma^{-1}\right)^{j}_{\;\;k}  = \vec 0~.
\end{equation}
Hence the partition function is maximal at the point $Q^{eq}_j=V
\kappa_{1,j}$. Charge and energy density correspond thus to the GCE values, and
$\mu^j \rightarrow \mu^j_{gce}$ and $T \rightarrow T_{gce}$. While when taking
the first finite volume correction term in Eq.(\ref{DiffPF}) into account we
obtain:
\begin{equation}\label{GCmu}
\xi_k \left(\sigma^{-1} \right)^{ \; k}_{\;\; j} +
 \frac{\lambda_3^{k_1,k_2,k_3}}{3!\sqrt{V}}  \left(\sigma^{-1}
\right)^{  k_4}_{\;\; j} H_4\left( \xi \right)_{k_1,k_2,k_3,k_4} = \vec 0~,
\end{equation}
rather than Eq.(\ref{CLTmu}), and $\mu^j \not= \mu^j_{gce}$, and  $T \not=
T_{gce}$. The recipe for calculation of distributions for finite volume system
thus goes as follows. One should find chemical potentials that satisfy
condition (\ref{gChemPot}). Then keep in mind that they are chemical
potentials only in the thermodynamic limit, while for finite volume they are
simply Lagrange multipliers. Then one should calculate the normalization
$\mathcal{Z}^{Q^{j,eq}}$ and the distribution $\mathcal{Z}^{\tilde
  Q^{j,eq}}$ using chemical potentials and temperature obtained from
Eq.(\ref{gChemPot}). Their ratio gives the distribution $P(N|Q^{j,eq})$ of
particles of the selected species. A technical comment is in order. From
Eq.(\ref{DiffPF}) it is evident that the first order correction term to the
derivative of the partition function is a polynomial of order 4 in
$\xi$, while the second one is of order 5, etc. It is therefore crucial to
find in numerical calculations the correct maximum.

\subsection{Quality of Approximation}
\label{QofA}
\begin{figure}[ht!]
\epsfig{file=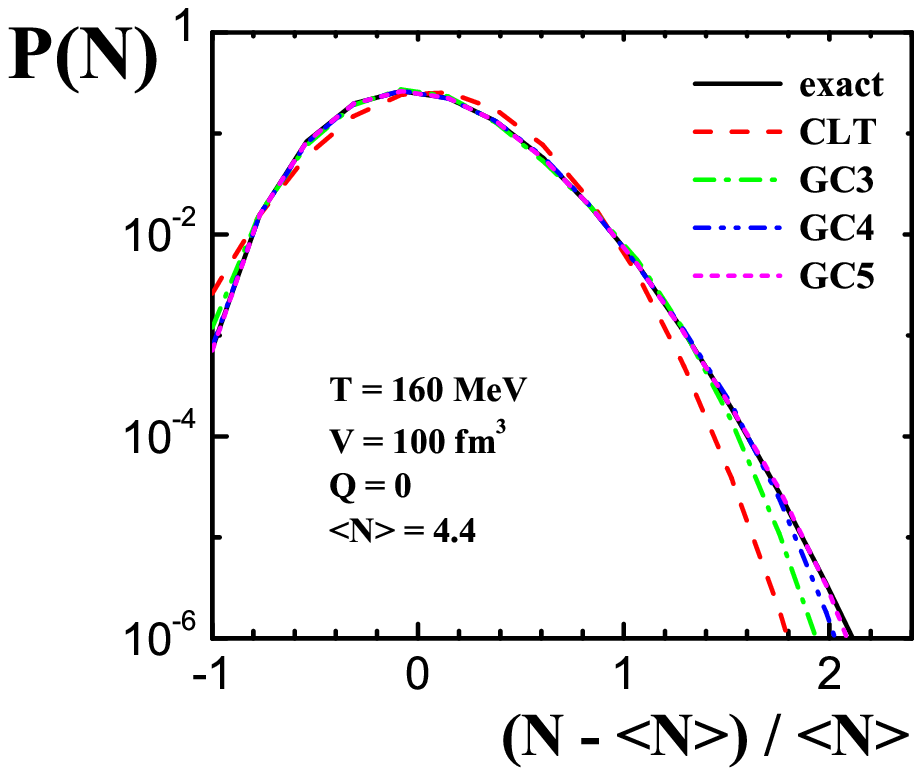,width=7.5cm}
\epsfig{file=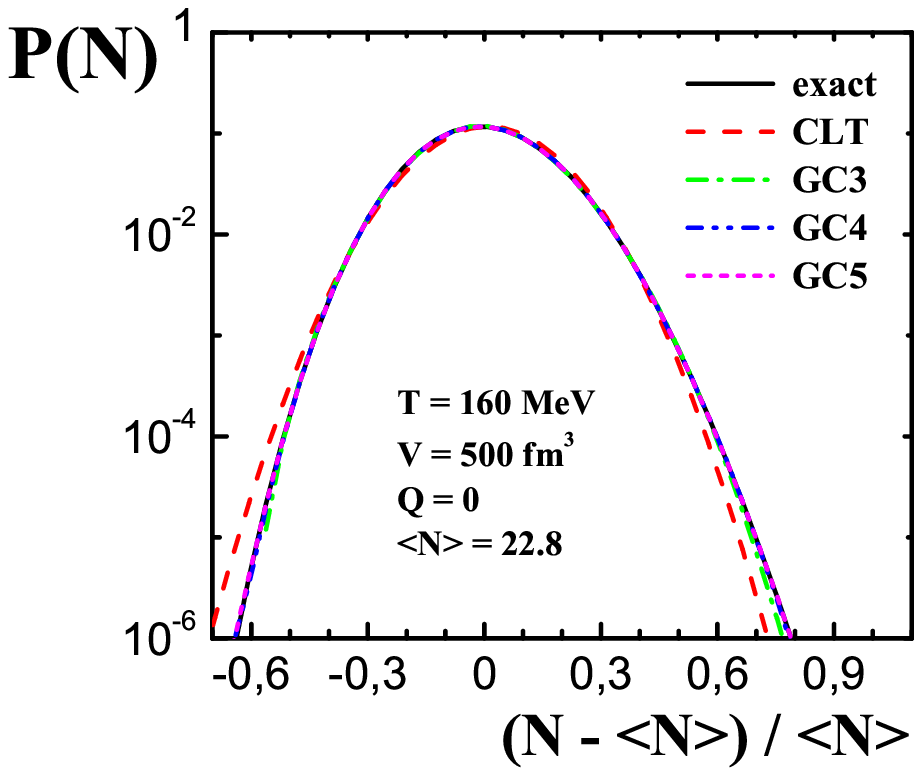,width=7.5cm}
\epsfig{file=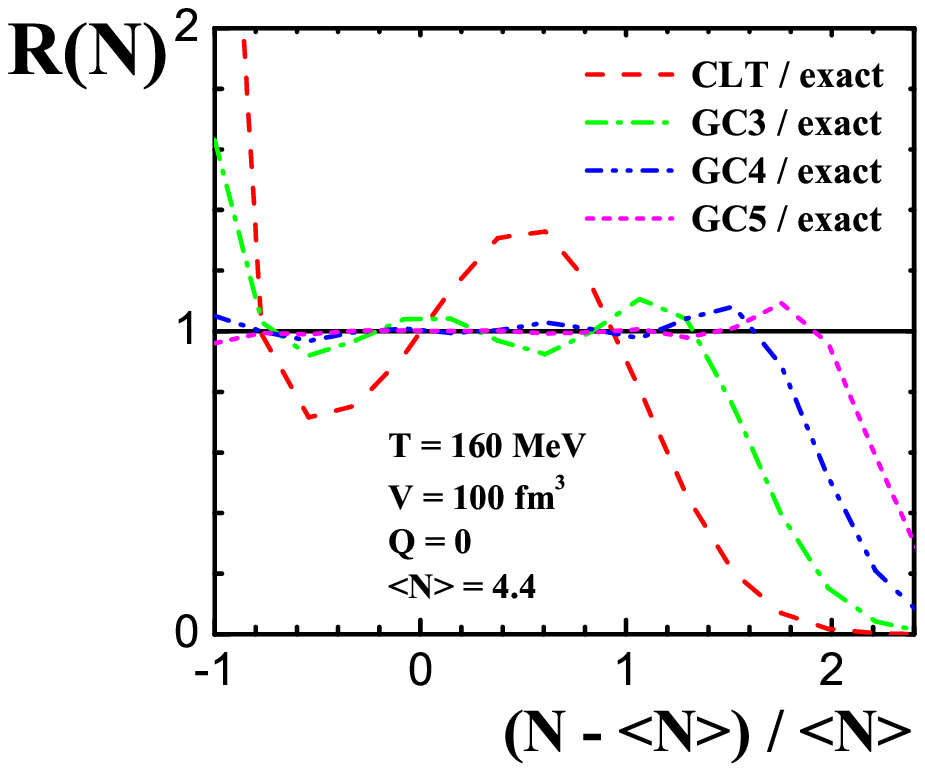,width=7.5cm}
\epsfig{file=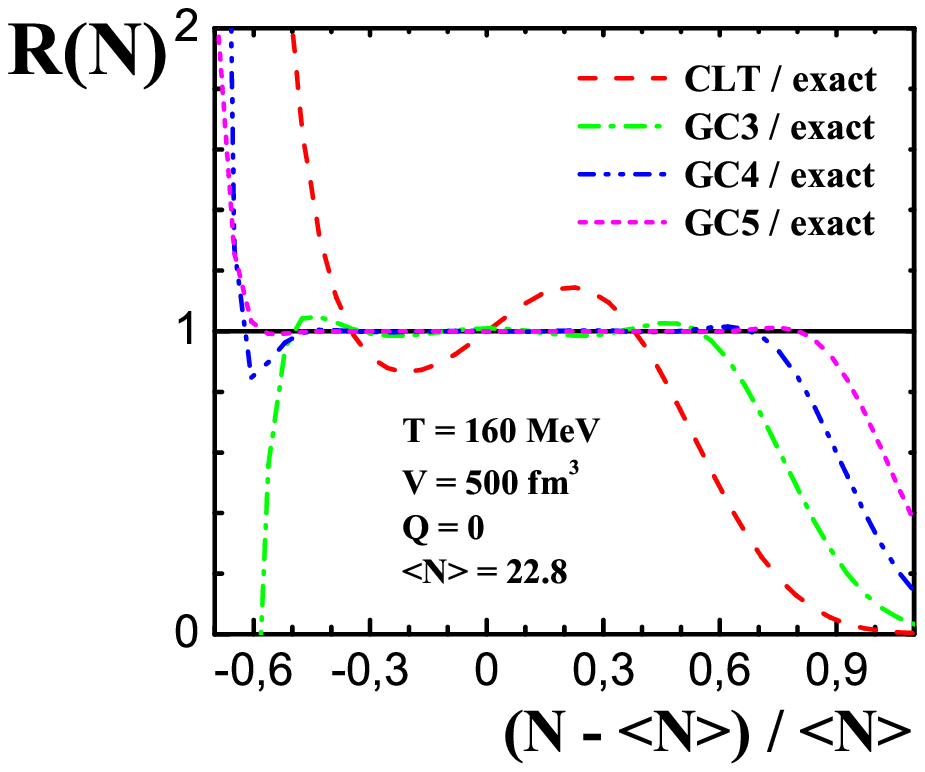,width=7.5cm}
\caption{{\it top}: Boltzmann CE $\pi^{+}$ multiplicity distribution for
  $T=160MeV$, $Q=0$ for  $V=100fm^3$ ({\it left}), and   $V=500fm^3$ ({\it
    right}). Exact solution (solid), and in CLT (dash), GC3   (dash-dot),
  GC4 (dash-dot-dot), and GC5 (dot) approximations. {\it  bottom}: same as
  top, but ratio of exact solution to approximation.}
\label{PN1}
\end{figure}

\begin{figure}[ht!]
\epsfig{file=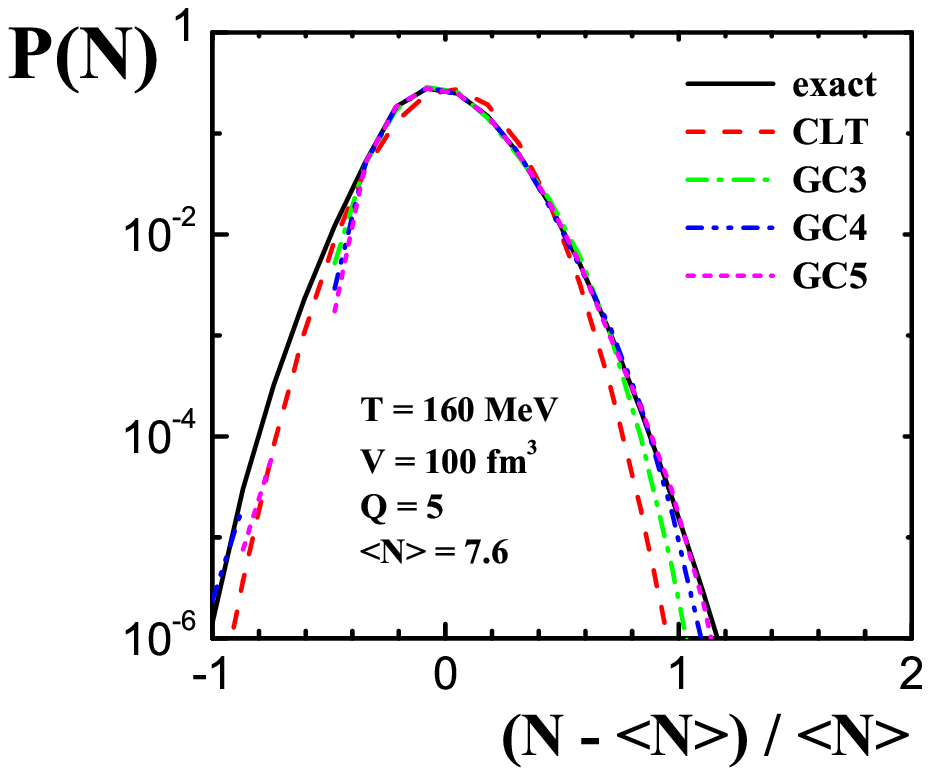,width=7.5cm}
\epsfig{file=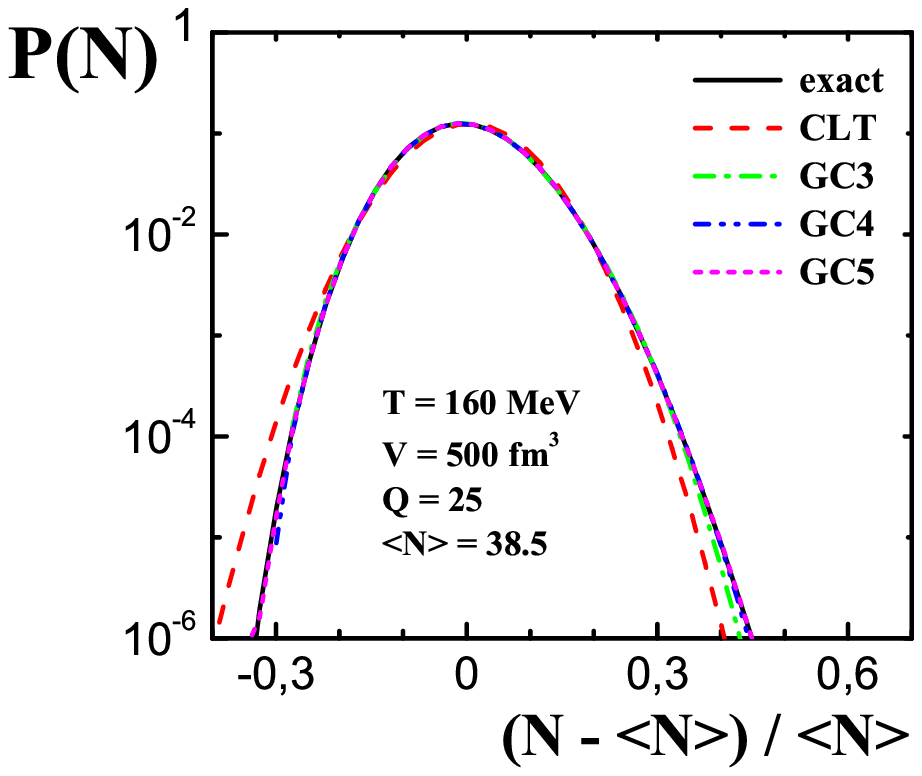,width=7.5cm}
\epsfig{file=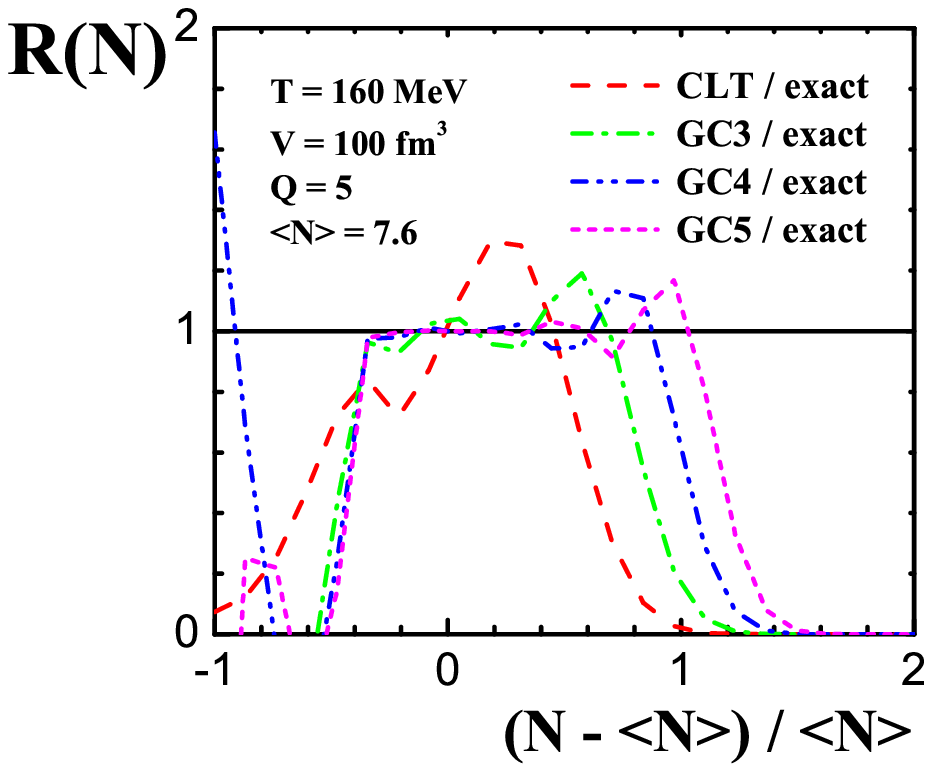,width=7.5cm}
\epsfig{file=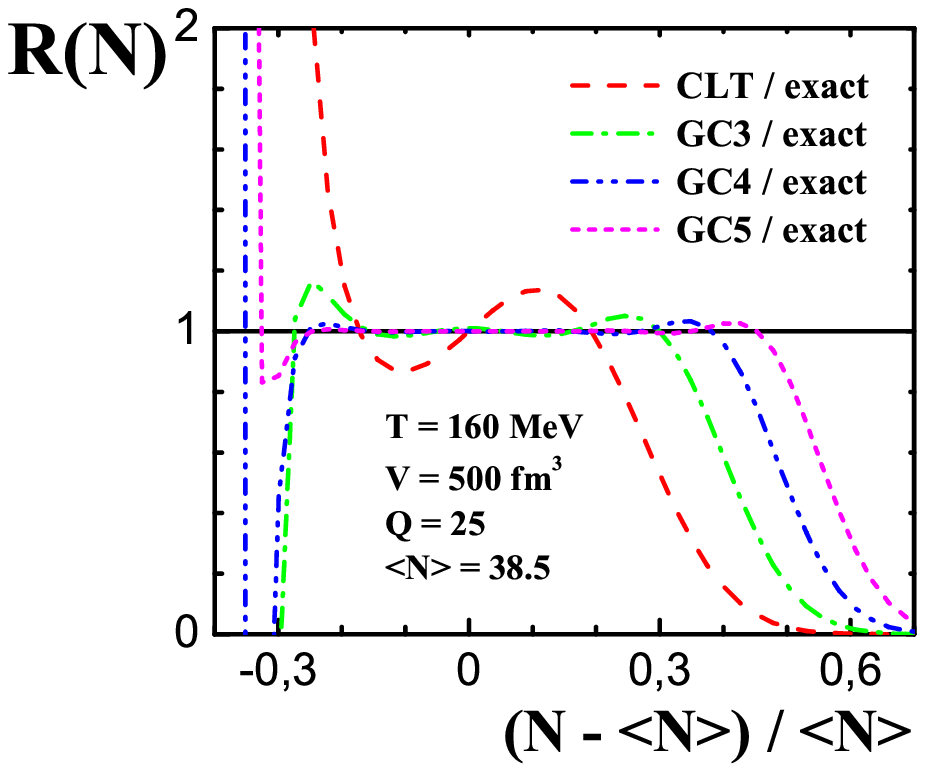,width=7.5cm}
\caption{same as \ref{PN1}, but with $Q=5$ ({\it left}), and $Q=25$ ({\it
    right})}
\label{PN2}
\end{figure}

\begin{figure}[ht!]
\epsfig{file=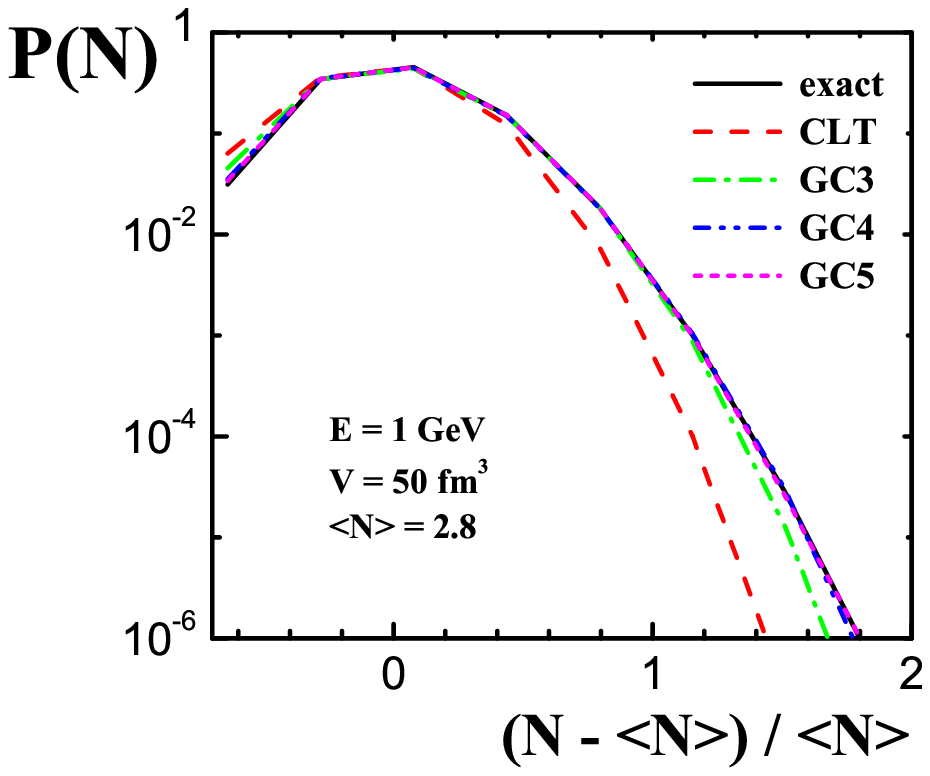,width=7.5cm}
\epsfig{file=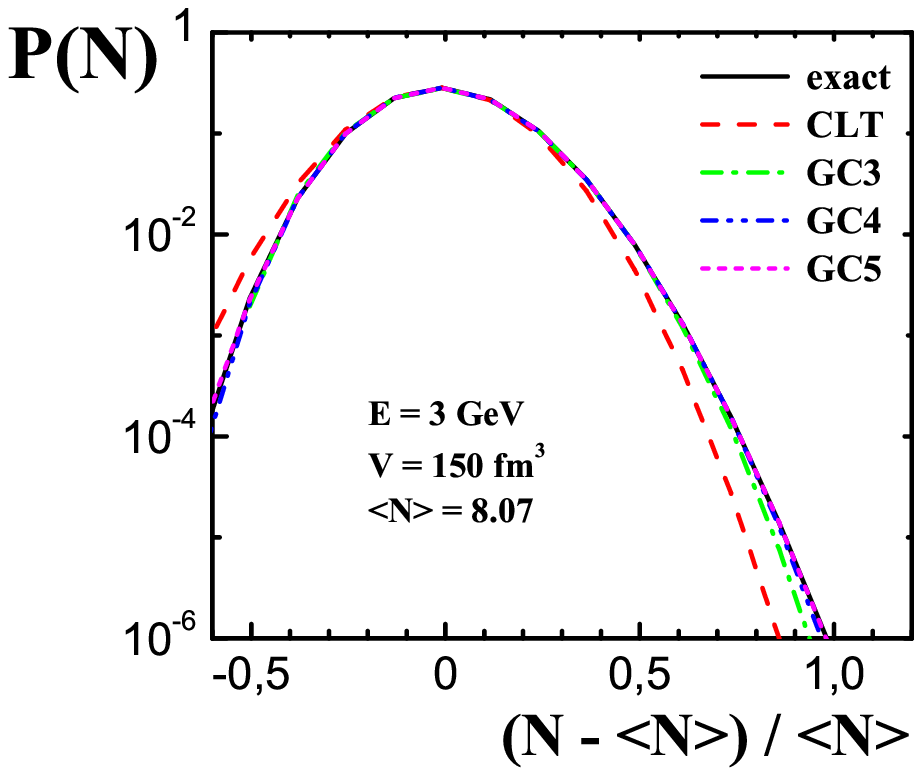,width=7.5cm}
\epsfig{file=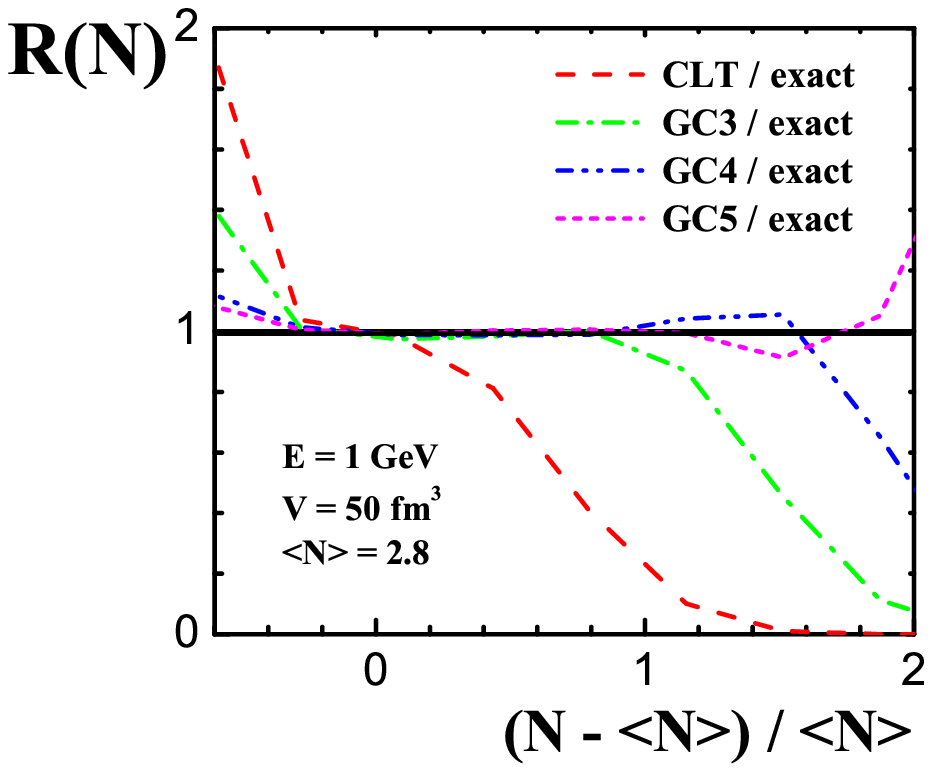,width=7.5cm}
\epsfig{file=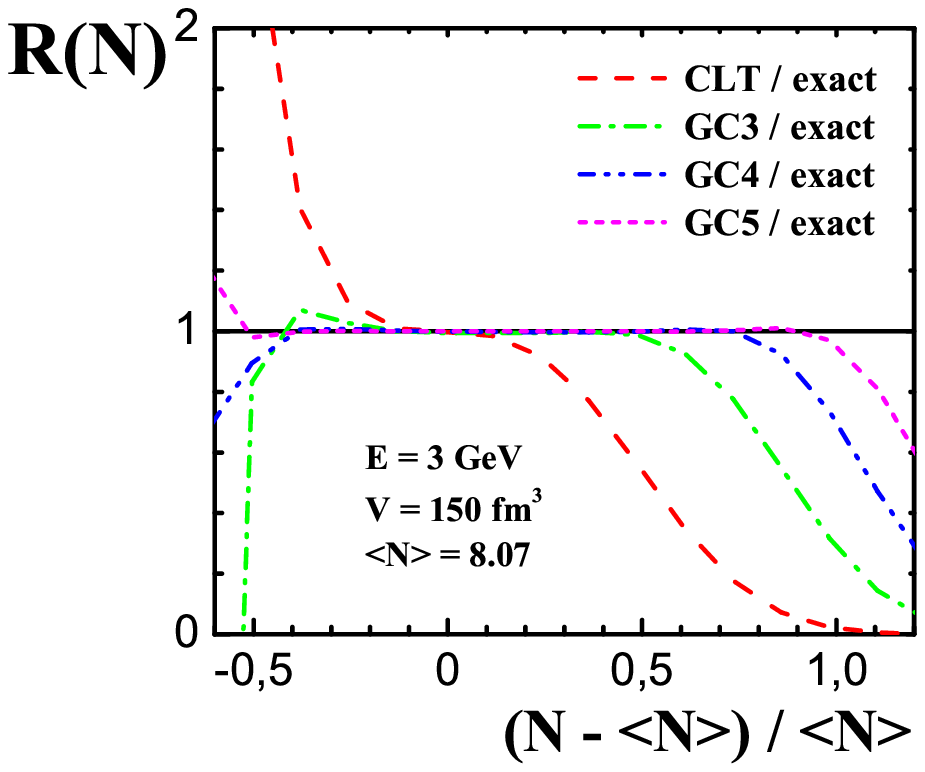,width=7.5cm}
\caption{{\it top}: Particle number distribution in the MCE without momentum
  conservation for a ultra-relativistic Boltzmann gas. $E=1GeV$, $V=50fm^3$
  ({\it left}), and  $E=3GeV$, $V=150fm^3$
  ({\it right}). Exact solution (dash), and in CLT (dash-dot), GC3
  (dash-dot), GC4 (dash-dot-dot), and GC5 (dot) approximations. {\it bottom}:
  same as top, but ratio of exact solution to approximation.}
\label{PN3}
\end{figure}

To test the quality of our approximation for multiplicity distributions at
finite volume, Eq.(\ref{GramCharlier}), for (very) small systems, we compare
to analytical solutions for a CE classical particle-anti-particle gas, and a
classical MCE (without momentum conservation) ultra-relativistic gas.  The
exact solutions are given by Eq.(\ref{PNpos-CE}) and Eq.(\ref{P(N|E)})
respectively. Figure \ref{PN1} shows on the top row the
multiplicity distribution of positively charges particles for various system
sizes in the exact form Eq.(\ref{PNpos-CE}) and in different orders of
approximation Eq.(\ref{GramCharlier}). On the bottom row the ratio of
approximation to exact solution is taken. In figure \ref{PN2} the same
physical system is shown for a (relatively large) positive net-charge. Due
to a one-to-one correspondence between the distributions of negatively
(suppressed) and positively (enhanced) particles we find the distribution
$P(N_+)$ generally more narrow than in the case of a neutral system. In
particular towards the edge of the body of the distribution the approximation
is worse.  For  the MCE  massless gas we compare again approximations to
$P(N)$ and ratios to the exact solution on top and bottom row of figures
\ref{PN3} respectively for different system sizes.
A few general comments attempt to summarize these figures. Our
first observation is that indeed as system size is increased we find a better
description of the central region in terms of the asymptotic
solution CLT (Gaussian with width given by Eq.(\ref{SimpleOmega})).  Our
second observation is that even for systems with a very small number
(in the order of $5$) of produced particles we find a good
approximation in terms of Gram-Charlier expansion. In particular GC5 provides
a very accurate description of the central region with deviations in the order
of a few percent. This is quite remarkable given the fact that multiplicity
distributions for such small systems are not smooth and continuous functions
of multiplicity, while our approximation Eq.(\ref{GramCharlier})
is. Furthermore we have implicitly introduced the concepts
of chemical potential and temperature for systems with small particle number,
which may be in contradiction to the common believe that
these parameters can only be meaningful when the number of involved particles
becomes very large, i.e. in the thermodynamic limit. Our last observation is
that indeed (see bottom rows  of figures \ref{PN1}-\ref{PN3}) finite volume
corrections given in terms of polynomials lead only to good results for the
central region of the distribution.

To give an estimate for a region in which the approximation is reliable, we
note that our finite volume approximation scheme begins to break down when the
first expansion term in Eq. (\ref{GramCharlier}) becomes unity. In the
one-dimensional case this would be:
\begin{equation}
\frac{h_3 \left( \xi_{max} \right)}{\sqrt{V}} \sim \mathcal{O}
\left(1 \right)~. \label{unityCond}
\end{equation}
Approximating the Hermite polynomial $H_3 \sim \xi^3$, one can get
an estimate for $\xi_{max}$.
\begin{equation}
\xi_{max} \simeq \left(\frac{3!}{\lambda_3} \right)^{1/3} V^{1/6}~.
\end{equation}
While, when switching back to the definition of $\xi= \frac{Q - V
\kappa_1 }{\sigma \sqrt{V}} $, Eq.(\ref{xi-var}), the width of the central
region can be estimated by:
\begin{equation}
\frac{|Q - V \kappa_1|_{max} }{\sigma} \simeq
\left(\frac{3!}{\lambda_3} \right)^{1/3} V^{2/3}~.
\end{equation}
Hence the width of the central region scales as $V^{2/3}$, while
the width of the curve should scale as $V^{1/2}$ and our approximation should
be quite good. Even though  larger volumes work better, they will still be
sufficiently small  enough to allow for calculation of distributions relevant
for heavy ion collisions. However we want to stress that there is no simple
criterion for what is a `small` or a `large` volume for a particular physical
system. Formally the existence and finiteness of the (at least) first three
cumulants $\kappa$ is sufficient for application of the asymptotic
solution \cite{MATH}. Considering the simple case of a multiplicity
distribution of BE particles in the GCE we find from Eq.(\ref{QstatsZET})
$\kappa_1^N < \kappa_2^{N,N} < \kappa_3^{N,N,N}<\cdots$. Hence we find, in
particular when finite chemical potentials are involved, cumulants growing
with order, implying that apart from mean and variance further parameters like
skewness and excess \cite{Formulas} of the distribution will remain important
quantities.
\subsection{Temperature and Chemical Potentials in MCE and CE}
The introduction of chemical potentials in the CE and 
temperature in the MCE was first and foremost a mathematical trick
which allowed to conveniently integrate partition functions for
which otherwise no analytical solution could be obtained. However our
generalized partition function is self-consistent and not in contradiction
to the common definition of temperature and chemical potential. We will show
in the following that our definition of $T$ and $\mu$ through
Eq.(\ref{gChemPot}) coincides with expressions well known from textbooks
\cite{Physics}.

\subsubsection{Canonical Ensemble}
The canonical partition function known from textbooks and our generalized
version are connected as follows (see Appendix \ref{CEPartFunc}):
\begin{equation}
{Z}^{CE} \left(Q,V,T\right)~\equiv~ \mathcal{Z}^{Q}\left(Q,V,T, \mu_Q
\right) ~ e^{-Q\frac{\mu_Q}{T}}~.
\end{equation}
The Helmholtz free energy $F$ is the thermodynamic potential relevant for CE,
\begin{equation}
F ~\equiv~ - T \ln {Z}^{CE}~.
\end{equation}
Using the first law of thermodynamics $ dE = TdS - PdV + \mu_Q dQ$, and $F=
E-TS$, where $P$ is the pressure, and $E$ and $S$ are total energy and
entropy, respectively, we can write for the differential $dF = -S dT - P dV +
\mu_Q dQ$. The effective chemical potential $\mu_Q$ associated with conserved
(electric) charge $Q$ is thus defined by:
\begin{equation} \label{CEchemPot}
\left( \frac{\partial F}{\partial Q} \right)_{V,T}~=~ -T~ \frac{ \frac{\partial
    \mathcal{Z}^{Q}}{\partial Q} ~  e^{-Q\frac{\mu_Q}{T}}~ -~ \frac{\mu_Q}{T}
  ~\mathcal{Z}^{Q} ~ e^{-Q\frac{\mu_Q}{T}} } {\mathcal{Z}^{Q} ~
  e^{-Q\frac{\mu_Q}{T}}} ~=~ \mu_Q~,
\end{equation}
where we have used condition (\ref{gChemPot}). Thus the correct choice for
the effective chemical potential is $\frac{\partial \mathcal{Z}^{Q}}{\partial Q}
= 0$, which coincides with $\mu_Q = \left( \frac{\partial F}{\partial Q}
\right)_{V,T}$. In the thermodynamic limit this is
equivalent to, Eq.(\ref{CLTmu}), $\mu_Q \rightarrow \mu_{Q,gce}$. The
subscript in
Eq.(\ref{CEchemPot}) is used to indicate that the derivative with respect to
the conserved charge has to be taken at fixed values of $V$ and $T$. In the
non-relativistic case, where particle number $N$, rather than quantum numbers,
is conserved, the corresponding relation to Eq.(\ref{CEchemPot}) would be
$\left( \frac{\partial F}{\partial N} \right)_{V,T} =\mu_N$
\cite{Physics}. For a detailed presentation of CE calculations please see
Section \ref{BoltzmannPionGas}.
\subsubsection{Microcanonical Ensemble}
The common MCE partition function can be obtain by integration
of the GGPF $\mathcal{Z} \left( \phi_E \right)$ and multiplication by the
inverse Boltzmann factor (see Appendix \ref{MCEPartFunc} for details),
\begin{equation}
{Z}^{MCE} \left(E,V \right) ~\equiv~ \mathcal{Z}^{E}
\left(E,V,T\right) ~ e^{\frac{E}{T}}~.
\end{equation}
The relevant thermodynamic potential in the MCE is the entropy $S$,
\begin{equation}
S ~\equiv~ \ln {Z}^{MCE}~.
\end{equation}
 The effective temperature is \cite{Physics}:
\begin{equation} \label{MCETemp}
\left( \frac{\partial S}{\partial E} \right)_V~=~ \frac{ \frac{\partial
    \mathcal{Z}^{E}}{\partial E} ~  e^{\frac{E}{T}} ~+~ \frac{1}{T}
  ~\mathcal{Z}^{E} ~ e^{\frac{E}{T}} } {\mathcal{Z}^{E} ~
  e^{\frac{E}{T}}} ~=~ \frac{1}{T}~,
\end{equation}
where we used condition (\ref{gChemPot}), $ \frac{\partial
  \mathcal{Z}^{E}}{\partial E} = 0$. The subscript in Eq.(\ref{MCETemp}) is
used to indicate that the derivative with respect to $E$ has to be taken at
fixed volume $V$. Thus Eq.(\ref{MCETemp}) resembles the optimal choice of an
effective temperature for our approximation scheme. In the thermodynamic
limit, $V\rightarrow \infty $, we find $T\rightarrow T_{gce}$, due to
Eq.(\ref{CLTmu}).
\subsubsection{Grand Canonical Ensemble}
Conventionally, e.g. in textbooks, first the MCE is introduced. Summation over
energy, with temperature being a Lagrange multiplier, used to maximize the
entropy, introduces the CE. Additionally dropping the constraint of exact
charge conservation leads to the GCE. Here the chemical potential $\mu_Q$ is the
Lagrange multiplier. For the MCE (without momentum conservation) and one
conserved charge $Q$ the GCE and CE partition functions are defined by:
\begin{equation}
Z^{GCE} (V,T,\mu_Q) ~=~ \sum_Q ~e^{Q\frac{\mu_Q}{T}}~ Z^{CE}(V,T,Q) ~=~  \sum_{Q,E}
~e^{Q\frac{\mu_Q}{T}} ~e^{-\frac{E}{T}}~ Z^{MCE} (V,E,Q)~.
\end{equation}
While in our notation this line would read:
\begin{equation}
\mathcal{Z} (V,T,\mu_Q) ~=~ \sum_Q  \mathcal{Z}^{Q}(V,T,\mu_Q) ~=~  \sum_{Q,E}
~ \mathcal{Z}^{E,Q} (V,T,\mu_Q)~.
\end{equation}
The thermodynamic potential for the GCE is the grand
potential $\Omega$:
\begin{equation}
\Omega ~\equiv~ -T \ln Z^{GCE} ~=~ -T \ln \mathcal{Z} \mid_{\phi
=0}~.
\end{equation}
Again, if exact solutions of the canonical or microcanonical partition
functions were available this reversal would not have been necessary. However
this redefinition of the GCE partition function is entirely consistent and
simplifies calculations considerably. Whenever an exact solution to our
generalized partition function is possible, all the above relations would
hold exactly. We believe this interpretation of the GCE
partition function as the generating (or characteristic) function of a
statistical system to be quite useful, even in more general cases than the one
presented here.

\section{The Simplest Example}
\label{BoltzmannPionGas}

\subsection{An Exact Solution}
To clarify the procedure described in the previous sections, let
us consider a simple example of an ideal Boltzmann
particle-anti-particle gas ($q_+=1,~q_-=-1$). In this simple case
the calculations discussed above can be done explicitly. The GCE
partition function  reads:
\begin{equation}
Z ~=~\sum_{n_+,n_-=0}^{\infty}\frac{z_+^{n_+}}{n_+!}
\frac{z_-^{n_-}}{n_-!}
 \;\equiv\; \sum_{n_+,n_-=0}^{\infty} Z(n_+,n_-)
 \;=\; \exp\left[\; z_++z_-\;\right]
  ~\equiv~\exp \left[2 z~ \cosh \left(
\frac{\mu_Q}{T} \right) \right]~,
\end{equation}
where $z_{\pm}=z ~\exp(\pm\mu_Q/T)$, and $z$ is a single particle
partition function in MB approximation:
\begin{equation} \label{BoltzZET}
z= \frac{gV}{2 \pi^2} \int \limits_{0}^{\infty} p^2 dp
~\exp\left(-~\frac{\sqrt{p^2+m^2}}{T}\right)~ =~ \frac{gV}{2
\pi^2}~ m^2 T~ K_{2}\left(\frac{m}{T} \right)~,
\end{equation}
where $g$ and $m$ are respectively the degeneracy factor and
particle mass, and $K_2$ is the modified Hankel function. The
conserved charge $Q$ is just the difference of $N_+$ and $N_-$. To
be definite we discuss the distribution $P(N_+)$ of positively
charged particles. In the GCE it can be easily found:
\eq{\label{PN-GCE}
 P(N_+)
 &~=~  \frac{\textrm{all states with $N_+$ particles}}{\textrm{all states }}
 \;=\; \frac{\sum_{n_+,n_-=0}^{\infty} Z(n_+,n_-)\;\delta\left(N_+-n_+\right)}
       {\sum_{n_+,n_-=0}^{\infty} Z(n_+,n_-)}\nonumber
 \\
 &~=~ \frac{z_+^{N_+}}{N_+!}~\exp(-z_+)~.
 }
As can be seen it has the form of the Poisson distribution with
the following first two moments:
 \eq{\label{N-GCE} \langle N_+
\rangle_{g.c.e.}~ =~ \sum_{N_+}^{\infty} N_+~
P(N_+)~=~z_+~,~~~~~\langle N_+^2 \rangle_{g.c.e.}~ =~
\sum_{N_+}^{\infty} N_+^2 ~ P(N_+)~=~z_+^2~+~z_+~.
}
This gives the GCE scaled variance:
\eq{ \label{omega-GCE}
\omega_{g.c.e.}^+~\equiv~\frac{\langle N_+^2
\rangle_{g.c.e.}~-~\langle N_+ \rangle_{g.c.e.}^2} {\langle N_+
\rangle_{g.c.e.}}~=~1~.
}
In the thermodynamic limit $z\rightarrow \infty$ the Poisson
distribution (\ref{PN-GCE}) can be transformed into the Gauss one.
Using Stirling's formula, $N_+! \simeq \sqrt{2\pi N_+} ~ \exp
\left(N_+\ln N_+~-~N_+ \right)$, for $N_+\gg 1$ one finds at
$|N_+-z|\ll z$:
\eq{
P(N_+)~=~\frac{z_+^{N_+}}{N_+!}~\exp(-~z_+)~\simeq~\frac{1}{\sqrt{2\pi~z_+}}~\exp
\left[-~\frac{\left(N_+~-~z_+\right)^2}{2z_+}\right]~.
 }
The distribution of net charge $P(Q)$ and the joint distribution
$P(N_+,Q)$ can be found by the straightforward calculations:
\eq{\label{PQ-2}
 P(Q) &~=~\frac{1}{Z} \sum_{n_+,n_-=0}^{\infty}
 \delta \left(Q-[n_+-n_-]\right)\;
 \frac{z_+^{n_+}}{n_+!} \frac{z_-^{n_-}}{n_-!}
 ~=~ \frac{e^{Q\,\mu_Q/T}}{Z}~I_{Q} \left(2z \right) \;,
 \\
 P(N_+,Q) &\;=\;\frac{1}{Z} \sum_{n_+,n_-=0}^{\infty}
 \delta \left(N_+-n_+\right)\; \delta \left(Q-[n_+-n_-]\right)\;
 \frac{z_+^{n_+}}{n_+!} \frac{z_-^{n_-}}{n_-!}
 \;=\;
 \frac{e^{Q\,\mu_Q/T}}{Z}~\frac{z^{2N_+-Q}}{N_+!(N_+-Q)!}\; ,
 \label{PNQ-2}
 }
where $I_Q(2z)$ is the modified Bessel function. The
Eqs.~(\ref{PQ-2}, \ref{PNQ-2}) are the simplest appearance of the
general Eqs.~(\ref{PQ}, \ref{PNQ-1}). One can also notice that for
this example the CE partition function from the Eq.~(\ref{PQ})
equals the modified Bessel function $Z^Q=I_Q(2z)$.

From the Eqs.~(\ref{PQ-2}, \ref{PNQ-2}) one finds the CE particle
number distribution (see the Eq.~(\ref{PNQ-CE})):
\begin{equation}\label{PNpos-CE}
P(N_+|Q)~ =~ \frac{P(N_+,Q)}{P(Q)} = \frac{z^{N_+}}{N_+!} ~
\frac{z^{N_+-Q}}{(N_+-Q)!} ~\left[I_Q \left(2z
\right)\right]^{-1}~.
\end{equation}
The first and second moments of the CE multiplicity distribution
can be easily found \cite{BGZ}:
\eq{ \langle N_+ \rangle_{c.e.}~& =~ \sum_{N_+=Q}^{\infty} N_+~
P(N_+|Q)~=~ z~\frac{I_{Q-1}(2z)}{I_Q(2z)}~,
\\
\langle N_+^2 \rangle_{c.e.}~& =~ \sum_{N_+=Q}^{\infty} N_+^2 ~
P(N_+|Q)~=~
z~\frac{I_{Q-1}(2z)}{I_Q(2z)}~+~z^2~\frac{I_{Q-2}(2z)}{I_Q(2z)}~.
}
This leads to the CE scaled variance:
\eq{
\omega_{c.e.}^+~\equiv~\frac{\langle N_+^2
\rangle_{c.e.}~-~\langle N_+ \rangle_{c.e.}^2} {\langle N_+
\rangle_{c.e.}}~=~1~-~z~\left[\frac{I_{Q-1}(2z)}{I_Q(2z)}~-~\frac{I_{Q-2}(2z)}{I_{Q-1}(2z)}\right]~.
} In the thermodynamic limit when $z\rightarrow \infty$ and
$Q/2z=y$ one finds \cite{BGZ}:
\eq{
\langle N_+\rangle_{c.e.}~=~z~(y~+~\sqrt{1+y^2})~, ~~~~
\omega_{c.e.}^+~=~\frac{1}{2}~-~\frac{y}{2\sqrt{1+y^2}}~\label{omega-CE}.
}
Let us compare the results for the average particle number
$\langle N_+\rangle$ and for the  scaled variance $\omega^+$
calculated in the GCE and CE. We fix the $V$ and $T$ parameters to
be the same in the GCE and CE. The relation between the chemical
potential $\mu_Q$ of the GCE and the charge $Q$ of the CE
formulation is obtained from the following requirement:
\eq{\label{Q-GCE}
 Q~=~ \langle Q\rangle~=~\langle
N_+\rangle_{g.c.e.}~-~\langle
N_-\rangle_{g.c.e.}~=~z_+~-z_-~=~2z~\sinh\left(\frac{\mu_Q}{T}\right)~.
}
This gives,
$\exp\left(\mu_Q/T\right)=y+\sqrt{1+y^2}$,
and leads to
$\langle N_+\rangle_{c.e.} \simeq \langle N_+\rangle_{g.c.e.}$,
which means the thermodynamic equivalence of the CE and GCE.
Comparing Eqs.~(\ref{omega-GCE}) and (\ref{omega-CE}) one finds
$\omega^+_{c.e.}\ne \omega^{+}_{g.c.e.}$, thus, the scaled
variances are not equivalent even in the thermodynamic limit.

We have used the GCE distributions $P(Q)$ and $P(N_+,Q)$ to
calculate the CE distribution $P(N_+|Q)$. The Eq.~(\ref{PNpos-CE})
demonstrates that the $\mu_Q$-dependence has completely
disappeared in $P(N_+|Q)$ if exact analytical results in the GCE
for $P(Q)$ (\ref{PQ-2}) and $P(N_+,Q)$ (\ref{PNQ-2}) are used.
This means that the GCE with arbitrary value of $\mu_Q$ (and,
thus, the corresponding value of $\langle Q\rangle$) can be used
for the {\it exact} CE calculations of $P(N_+|Q)$.

\subsection{The Saddle Point Expansion}

We will discuss now the {\it approximate} calculations of $P(Q)$
(\ref{PQ-2}) and $P(N_+,Q)$ (\ref{PNQ-2}) in the thermodynamic
limit. We replace the corresponding delta-functions which fix the
value of net charge, $Q$, and positively charged particle number,
$N_+$ by their Fourier representations as the $\phi_Q$ and
$\phi_+$ integrations. This method can be used for more
complicated cases when exact analytical results can not be
obtained. In these asymptotic calculations the role of the
chemical potential will be demonstrated. The Eq.~(\ref{PQ-2}) can
be rewritten as follows:
\eq{\label{P(Q)-example}
 P(Q) &~=~\frac{1}{Z}~\int \limits_{-\pi}^{\pi}
 \frac{d \phi_Q}{2 \pi}~\exp(-i Q
  \phi_Q) ~\mathcal{Z}(\phi_Q)
 \\
 &\;=\; \frac{1}{Z}~\int \limits_{-\pi}^{\pi} \frac{d \phi_Q}{2\pi}~ \exp(-i Q
  \phi_Q)~ \exp\left[~z~ \exp\left(\frac{\mu_Q}{T}~+~i \phi_Q\right)~
  +~ z~ \exp\left(-\frac{\mu_Q}{T}~ -~i \phi_Q\right)\right]~.\nonumber
  }
In thermodynamic limit, $z\rightarrow\infty$, one can expand
$\exp\left(\pm\left[ \mu_Q/T + i \phi_Q\right]\right)$ in the
Taylor series and leave only the terms up to $\phi_Q^2$, because
for $z\rightarrow\infty$ the main contribution comes from the
$\phi_Q=0$ region. Then the distribution $P(Q)$ becomes a
Gaussian:
 \eq{\label{PQ-3}
 P(Q)&~=~\int \limits_{-\infty}^{\infty}
         \frac{d \phi_Q}{2 \pi}~ \exp\left[-i
   \left(Q - 2z\, \sinh(\mu_Q/T)\right)\phi_Q \;-~z\,\cosh(\mu_Q/T)~\phi_Q^2
   \;+\;\ldots\right]
\nonumber \\
& \;\simeq~P_G(Q) \;=~ \left[4\pi z\cosh(\mu_Q/T)\right]^{-1/2}
~\exp\left[-~\frac{(Q~-~2z~\sinh(\mu_Q/T))^2}{4
z\cosh(\mu_Q/T)}\right]~. }
For $P(N_+,Q)$ at $z\rightarrow\infty$ one similarly finds:
 \eq{\label{PNQ-3}
 & P(N_+,Q)~ =~\frac{1}{Z}~\int \limits_{-\pi}^{\pi}
  \frac{d \phi_Q}{2 \pi}~e^{-i Q
  \phi_Q} ~\int \limits_{-\pi}^{\pi} \frac{d \phi_+}{2 \pi}~e^{-i N_+
  \phi_+}~\mathcal{Z}(\phi_Q,\phi_+)\nonumber \\
  &=~
  \frac{1}{Z}  \int \limits_{-\pi}^{\pi}  \frac{d \phi_Q}{2 \pi}
 \int \limits_{-\pi}^{\pi} \frac{d \phi_+}{2 \pi}\;
 \exp\left[-i Q \phi_Q  -  i N_+ \phi_+
 + z \exp\left(\frac{\mu_Q}{T} + i \phi_Q + i \phi_+\right)
 + z \exp\left(-\frac{\mu_Q}{T} - i \phi_Q\right) \right]
 \nonumber
 \\
 ~ &\simeq~P_{G}(N_+,Q)~\equiv~ \frac{1}{Z}~\int
\limits_{-\infty}^{\infty} \frac{d \phi_Q}{2 \pi}~
 \int \limits_{-\infty}^{\infty} \frac{d \phi_+}{2 \pi}~
 \exp\Bigg[i\phi_Q\left(z_+-z_- - Q\right)  \;+\; i\phi_+\left(z_+ -
N_+\right)
 \nonumber
 \\
 &\qquad\qquad\qquad\qquad\qquad\qquad\qquad\qquad\qquad\quad
  -~(z_++z_-)\;\frac{\phi_Q^2}{2}
 \;-\; z_+\;\frac{\phi_+^2}{2} \;-\; z_+\;\phi_Q\phi_+\Bigg]\;.
 }

The integration over $\phi_Q$ and $\phi_+$ in Eq.~(\ref{PNQ-3})
gives:

\eq{ \label{PNQ-4}
  P_{G}(N_+,Q) 
 \;=\; \frac{1}{2\pi z}\,
 \exp \Bigg[ &-\frac{1}{2z}\exp(\mu_Q/T)\left(Q-\langle Q\rangle\right)^2
 \;+\; \frac{1}{z}\exp(\mu_Q/T) \left(Q-\langle Q\rangle\right)
       \left(N_+-\langle N_+\rangle\right)
 \nonumber
 \\
 &- \frac{1}{z}\cosh(\mu_Q/T)~\left(N_+-\langle N_+\rangle\right)^2
 \Bigg]~,
 }
where $\langle Q\rangle$ and $\langle N_+ \rangle$ in
Eq.~(\ref{PNQ-4}) correspond to the GCE values,
\eq{\label{QN-GCE}
\langle Q\rangle~=~2z~\sinh(\mu_Q/T)~,~~~~\langle N_+ \rangle
~=~z~\exp(\mu_Q/T)~.
 }
The Eq.~(\ref{PNQ-4}) has the form of Bivariate Normal
Distribution i.e. Gauss (Normal) distribution in two dimensions
\cite{Formulas}. The CE distribution (\ref{PNpos-CE}) is then
approximated as,
\eq{\label{PNQ-5}
 &P\left(N_+|Q \right)~ = ~\frac{P\left(N_+,Q \right)}{P\left(Q
\right)}~ \simeq ~ P_{G}(N_+|Q)~\equiv~\frac{P_{G}\left(N_+,Q
\right)}{P_G\left(Q \right)}~\nonumber \\
&=~ \frac{1}{\sqrt{\pi
    z~\cosh^{-1}\left[\mu_Q/T\right]}} ~\exp \Bigg[-~
\frac{\cosh \left( \mu_Q/T \right)}{z} ~ \left(N_+- \langle N_+
\rangle\right)^2
\nonumber \\
& + ~\frac{\exp \left(  \mu_Q/T \right)}{z} ~ \left(N_+- \langle
N_+ \rangle\right)
 \left(Q- \langle Q\rangle \right)~
 ~ - ~ \frac{\exp \left(2
 \mu_Q/T \right)}{2z~\cosh\left(\mu_Q/T \right)} ~ \left(Q- \langle
Q\rangle\right)^2
 \Bigg]~.
}

Comparing the distributions (\ref{PQ-3}) and (\ref{PNQ-4}) with
the exact expressions for $P(Q)$ (\ref{PQ-2}), and $P(N_+,Q)$
(\ref{PNQ-2}) one can notice that the Eqs.~(\ref{PQ-2}), and
(\ref{PNQ-2}) contain the dependance on the chemical potential
$\mu_Q$ just as the factor $e^{\mu_Q/T}$. Thus if one succeeds in
the exact calculations the resulting CE distribution $P(N_+|Q)$
(\ref{PNpos-CE}) does not include $\mu_Q$ dependence in contrast
to (\ref{PNQ-5}). It means that the choice of the chemical
potential is irrelevant for the exact CE calculations (usually it
is chosen equal to zero), while the value of $\mu_Q$ is crucial
for the approximate calculations. The saddle point expansion works
if $\mu_Q$ chosen to fix $\langle Q\rangle=Q$ in thermodynamic
limit, i.e. for $\mu_Q/T=\textrm{arc}\sinh(Q/2z)$. Thus the GCE
should be thermodynamically equivalent to the CE with fixed $Q$
net charge.

\begin{figure}[ht!]
 \epsfig{file=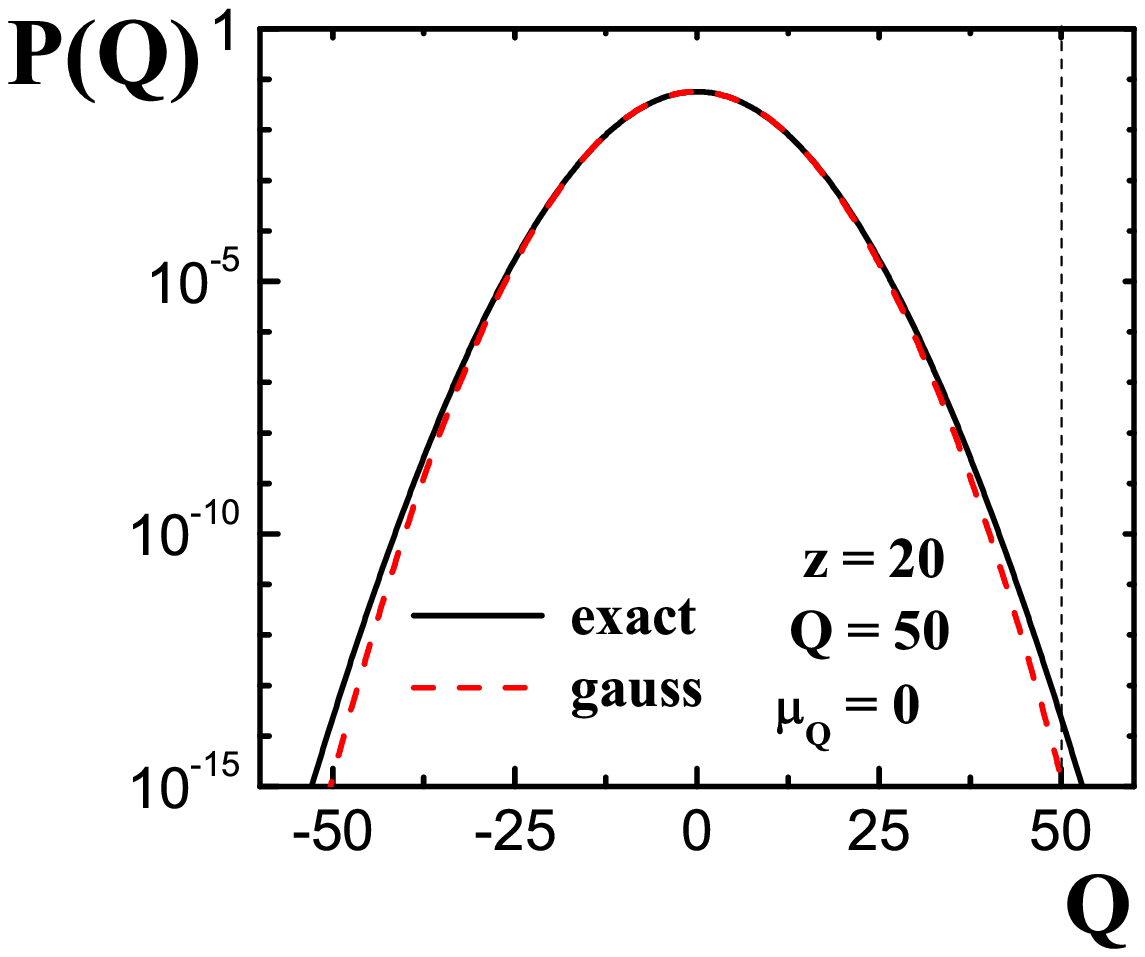,width=5.5cm}
 \epsfig{file=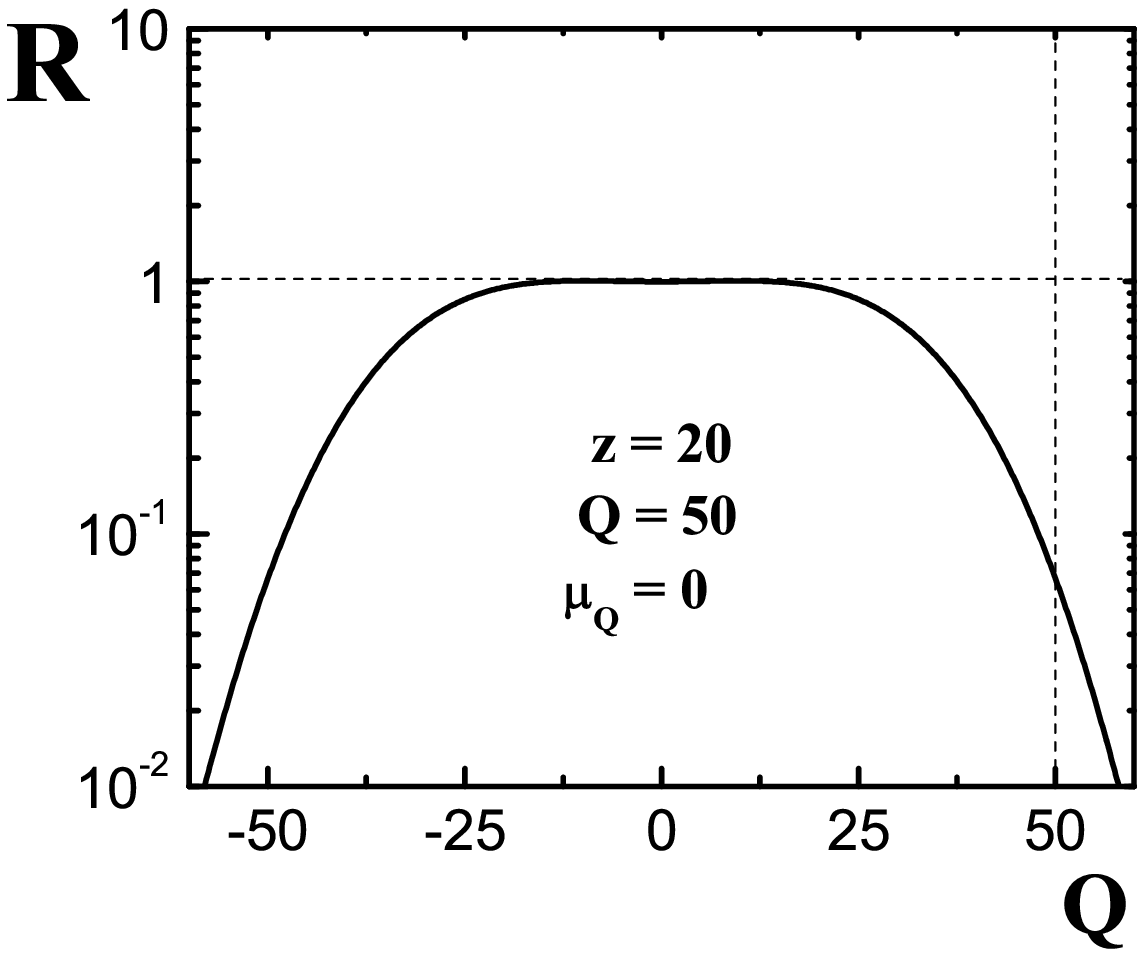,width=5.5cm}
 \epsfig{file=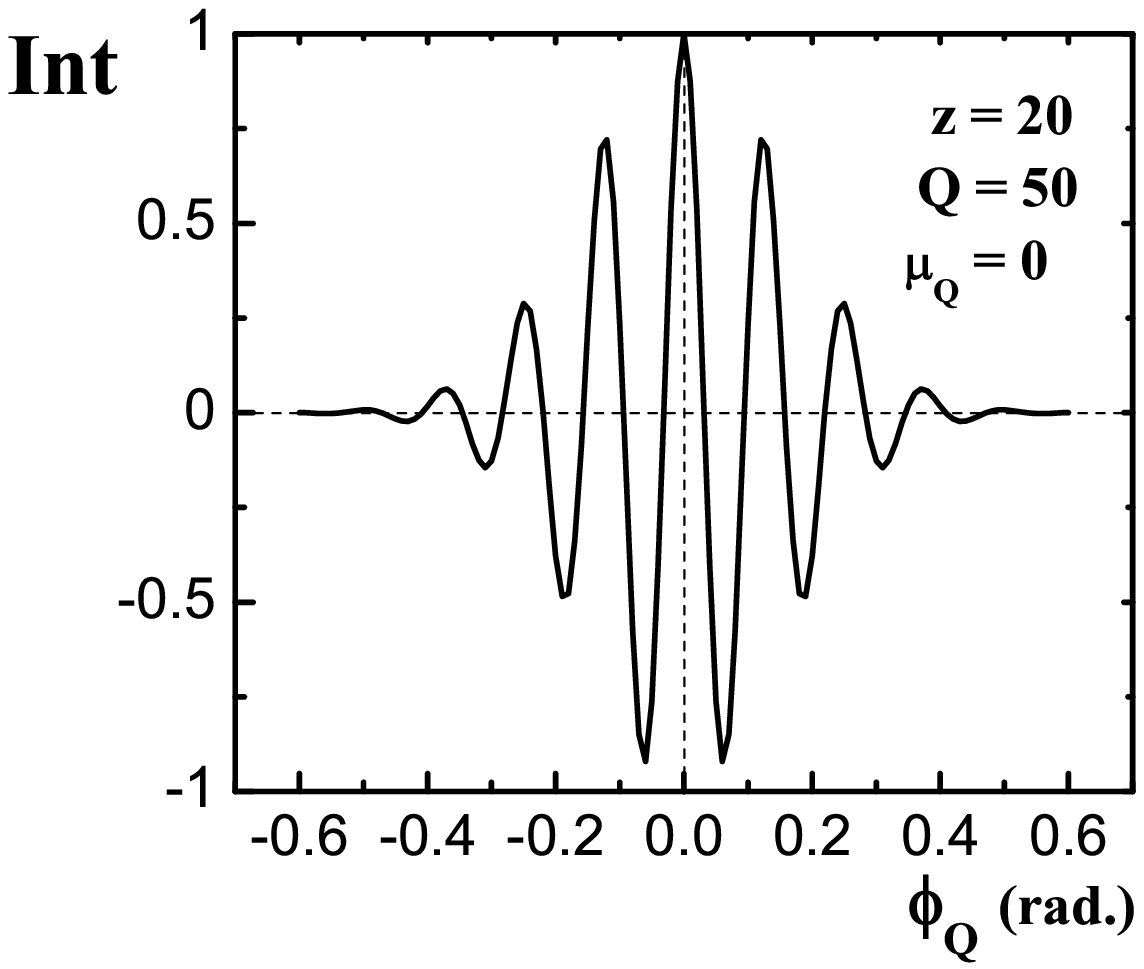,width=5.5cm}
 \vspace{-0.5cm}
 \caption{The exact and approximate distribution $P(Q)$ (left),
 their ratio (center), and the integrant (Int) from the Eq.~(\ref{PQ-3}) (right) for the 'wrong'
chemical potential $\mu_Q/T\neq\textrm{arc}\sinh(Q/2z)$.}
\label{PRI-Q-mu0}
\end{figure}

\begin{figure}[ht!]
 \epsfig{file=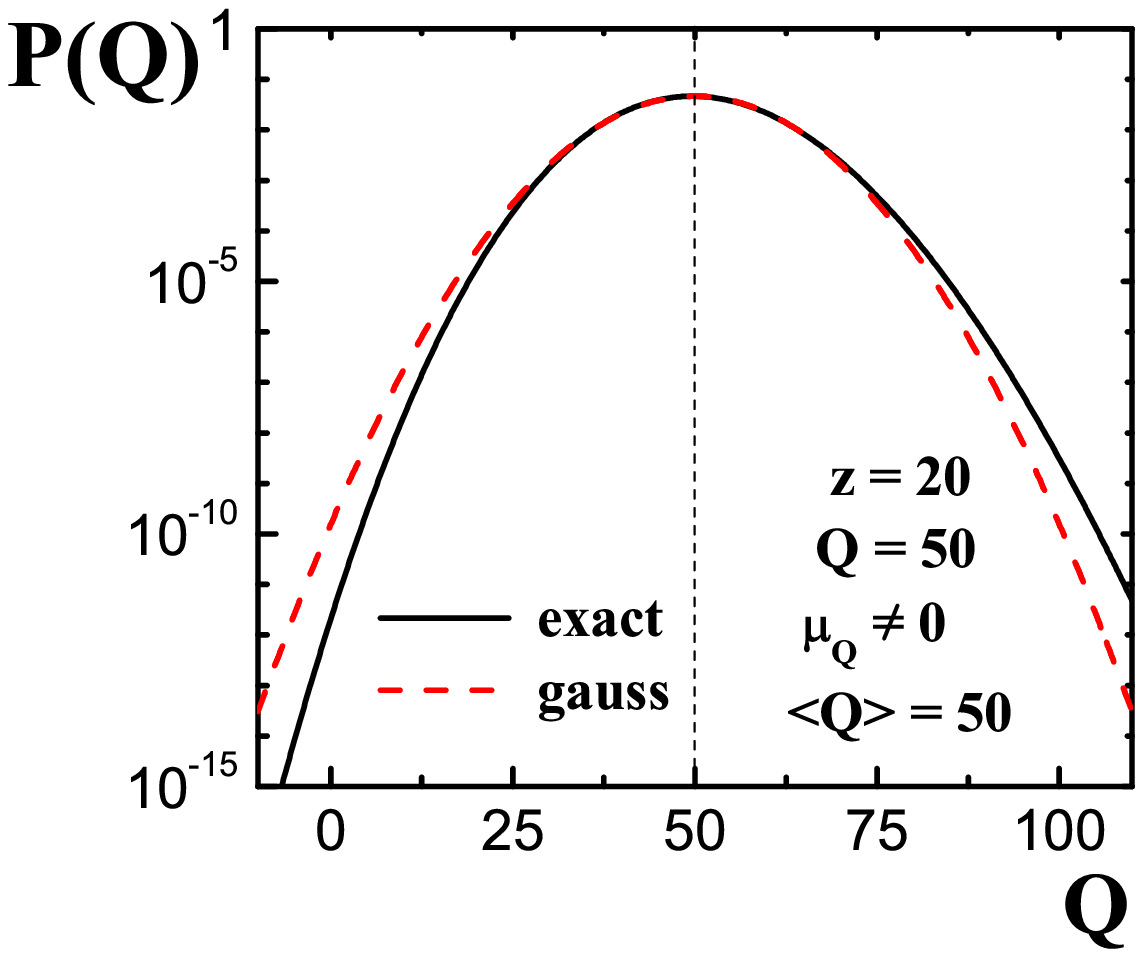,width=5.5cm}
 \epsfig{file=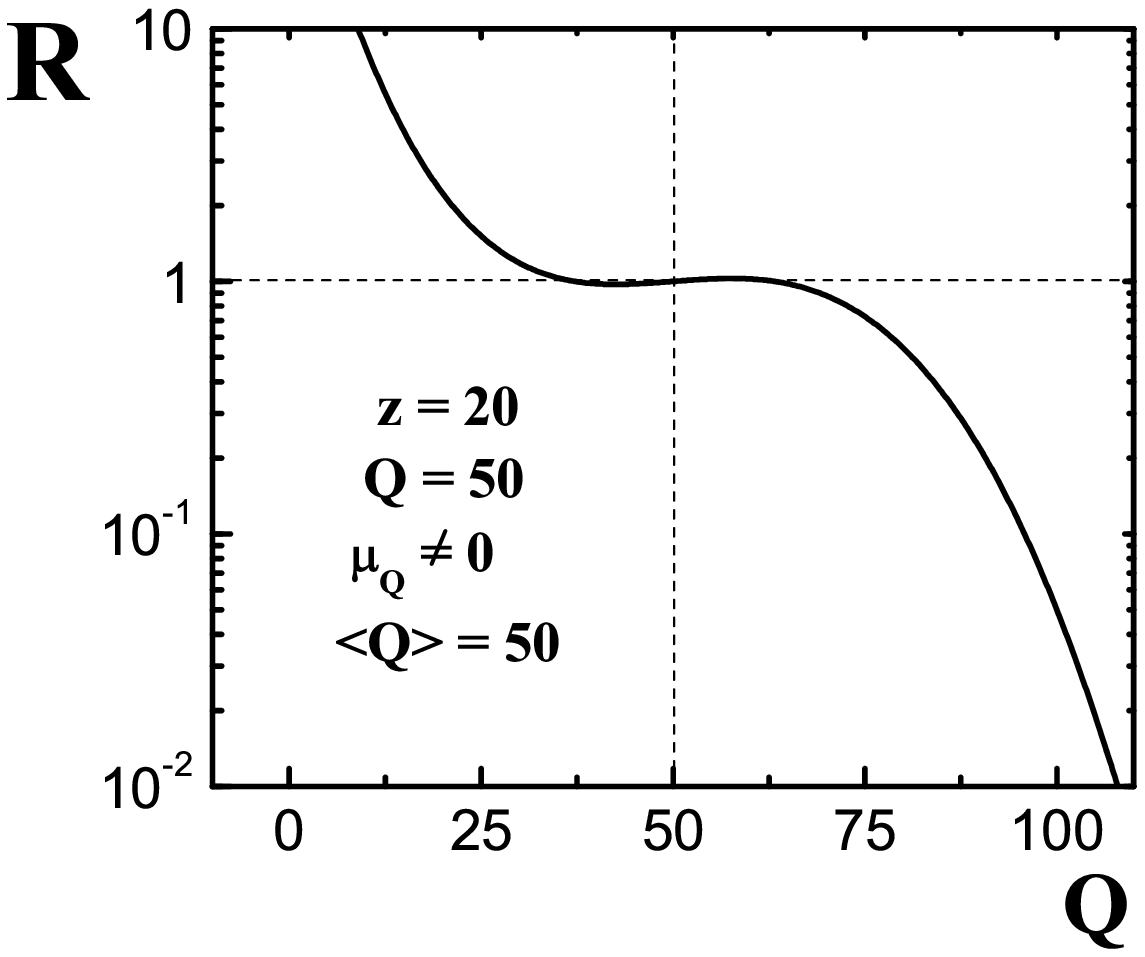,width=5.5cm}
 \epsfig{file=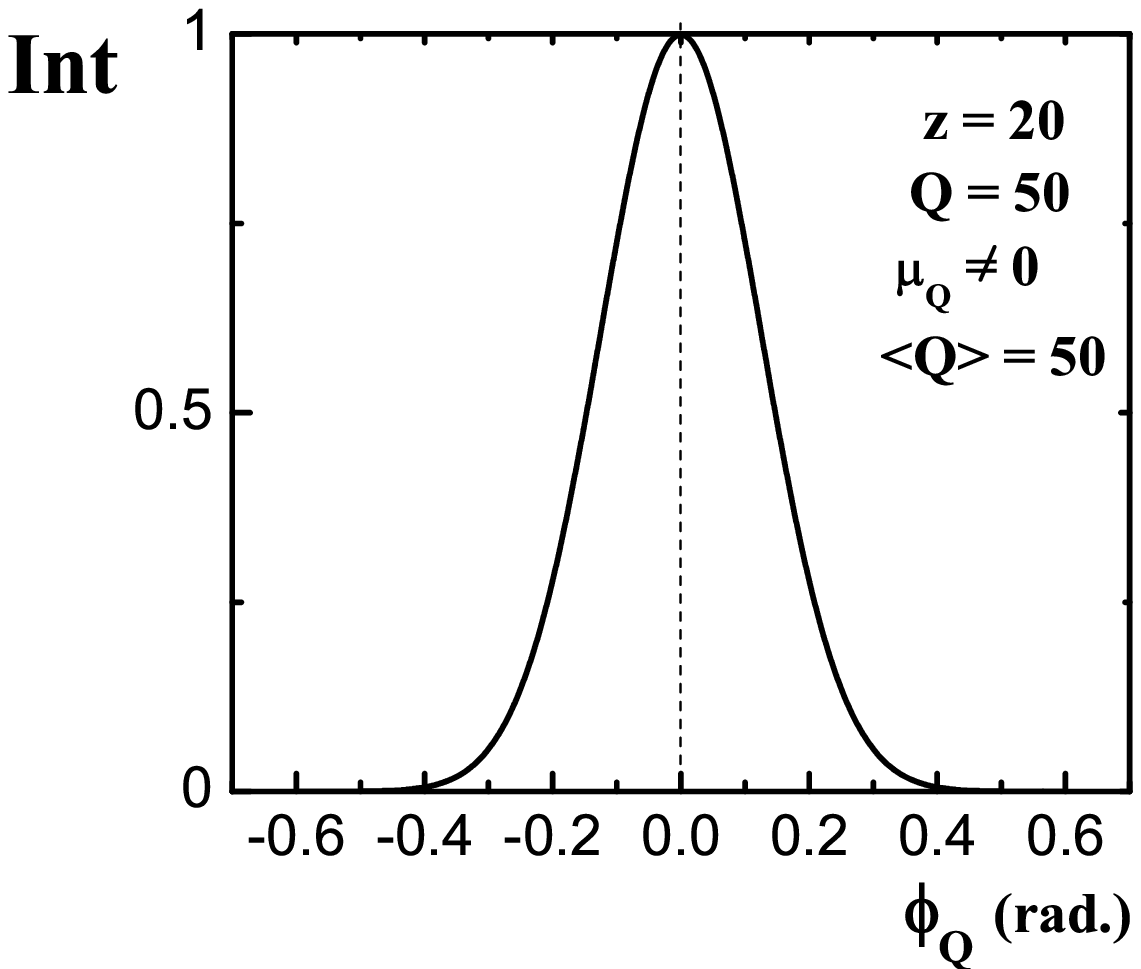,width=5.5cm}
 \vspace{-0.5cm}
 \caption{The same as in Fig.~\ref{PRI-Q-mu0} but for correct
$\mu_Q=\textrm{arc}\sinh(Q/2z)$.} \label{PRI-Q-mu}
\end{figure}

Let us illustrate these statements. We plotted the exact
(\ref{PQ-2}) and approximate (Gauss) distributions $P(Q)$
(\ref{PQ-3}) for the arbitrarily chosen values $z=20$ and $Q=50$
with zero chemical potential $\mu_Q=0$ (Fig.~\ref{PRI-Q-mu0},
left) and for the chemical potential $\mu_Q\neq 0$ that
corresponds to the condition $\langle Q\rangle=Q$
(Fig.~\ref{PRI-Q-mu}, left).

One can see that if one chooses the wrong chemical potential
$\mu_Q=0$ than the approximate distribution (\ref{PQ-3}) is
approximately 30 times smaller than the exact one (\ref{PQ-2}) at
the point $Q=50$ (see the ratios $P_G(Q)/P(Q)$
Fig.~\ref{PRI-Q-mu0}, \ref{PRI-Q-mu}, center). For non zero
chemical potential $\mu_Q\neq 0$ that satisfy the condition
$\langle Q\rangle=Q$, i.e. for the chemical potential
$\mu_Q/T=\textrm{arc}\sinh(Q/2z)$, the exact and the approximate
distribution almost coincide in a wide range near $Q=50$ (see
Fig.~\ref{PRI-Q-mu}, left and center).

The mathematical meaning of this physical requirement, $\langle
Q\rangle=Q$, is the following. The approximate formulas always
have an oscillating part
$\int_{-\infty}^{\infty}\exp\left[iA\phi-B\phi^2\right]d\phi
=\int_{-\infty}^{\infty}\cos\left[A\phi\right]\exp\left[-B\phi^2\right]d\phi$
(the imaginary part includes $\sin[A\phi]$ that gives zero after
integration because of symmetric integration bounds). This
oscillating part is nonzero $A\neq 0$ for $\langle Q\rangle \neq
Q$ thus we need to take the terms higher than $\phi^2$ in order to
make a good approximation. The integrant (Int)
from the Eq.~(\ref{PQ-3}), $Int=\exp\left[-i \left(Q - 2z\,
\sinh(\mu_Q/T)\right)\phi_Q
\;-~z\,\cosh(\mu_Q/T)~\phi_Q^2+\ldots\right]$, is shown in the
Figs.~\ref{PRI-Q-mu0}, \ref{PRI-Q-mu} (right). One can see that
for $\langle Q\rangle \neq Q$ (Figs.~\ref{PRI-Q-mu0},
\ref{PRI-Q-mu}, center) the Int oscillates around zero fast, thus
even small mistake in numerical integration, that is unavoidable
in real calculations, gives the wrong result for $P(Q)$. However
if we take the correct $\mu_Q$ that satisfy the condition $\langle
Q\rangle = Q$, then
$Int=\exp\left[-~z\,\cosh(\mu_Q/T)~\phi_Q^2+\ldots\right]$ becomes
smooth and the term $\phi_Q^2$ is enough (see Fig.\ref{PRI-Q-mu},
right).

At the point $\langle Q\rangle=Q$ from the Eq.~(\ref{PNQ-4}) one
finds:
\eq{\label{PNQ-6}
 P_G(N_+|Q) ~&=~ \frac{1}{\sqrt{\pi
    z~\cosh^{-1}\left(\mu_Q/T\right)}}
    \exp \left[-~\frac{\left(N_+~-~\langle N_+ \rangle
\right)^2}{z~\cosh^{-1}
    \left( \mu_Q/T \right)} \right]
   \nonumber \\
   & =~\frac{1}{\sqrt{2\pi
   ~\omega^+_{c.e.}\cdot \langle N_+
\rangle}} \exp \left[-~\frac{\left(N_+~-~\langle N_+ \rangle
\right)^2}{ 2 ~\omega^+_{c.e.}\cdot \langle N_+ \rangle }
\right]~.
}
In Eq.~(\ref{PNQ-6}) the asymptotic values for multiplicity (valid
in the GCE and CE), $\langle N_+ \rangle = z \exp\left[ \mu_Q/T
\right]$, and for the scaled variance in the CE, $\omega_{c.e.}^+
= \left[2 \exp \left( \mu_Q/T \right) \cosh\left( \mu_Q/T \right)
\right]^{-1}$ (see Eq.~(\ref{omega-CE})), have been used.

The above procedure can be generalized for the case of $n$
integrations using the following formula:
 \eq{\label{N_Gauss}
 \left[\prod_{i=1}^n\int_{-\infty}^{\infty}\frac{d\phi_i}{2\pi}\right]\;
 \exp\left[i\,\vec{A}^{\;T}\vec{\phi}
 \;-\; \frac{1}{2}\;\vec{\phi}^{\;T}B\;\vec{\phi} \;\right]
 \;=\;
 \frac{\exp\left[ - \frac{1}{2}\;\vec{A}^{\;T}B^{-1}\vec{A} \;\right]}
 {(2\pi)^{n/2}\sqrt{\det|B|}}\;,
 }
where $T$ means transposed vector and $B^{-1}$ is the inverse of a
nonsingular matrix $B$. For example, the distribution $P(N_+,Q)$
has the following parameters:
 \eq{\label{matrixes}
 \vec{A}=
 \begin{pmatrix}
 z_+-z_--Q\\
 z_+-N_+
 \end{pmatrix},
 &&
 \vec{\phi}=
 \begin{pmatrix}
 \phi_Q\\
 \phi_+
 \end{pmatrix},
 &&
 B=\begin{pmatrix}
 z_++z_- & \;z_+ \\
 z_+ & z_+
 \end{pmatrix},
 && \det|B|=z_+z_-=z^2\;.
 }
One can easily check that the substitution of the
Eqs.~(\ref{matrixes}) into Eq.~(\ref{N_Gauss}) gives $P(N_+,Q)$
from the Eq.~(\ref{PNQ-4}).

%
\subsection{Microcanonical Ensemble}
\label{MCEexpample}
%

\begin{figure}[ht!]
 \epsfig{file=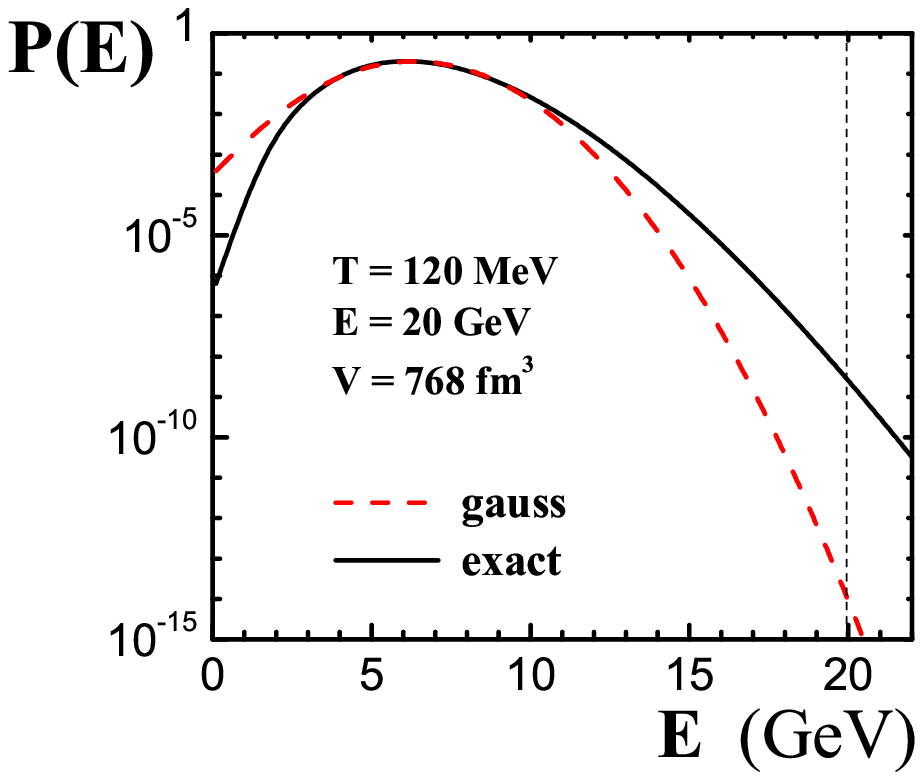,width=5.5cm}
 \epsfig{file=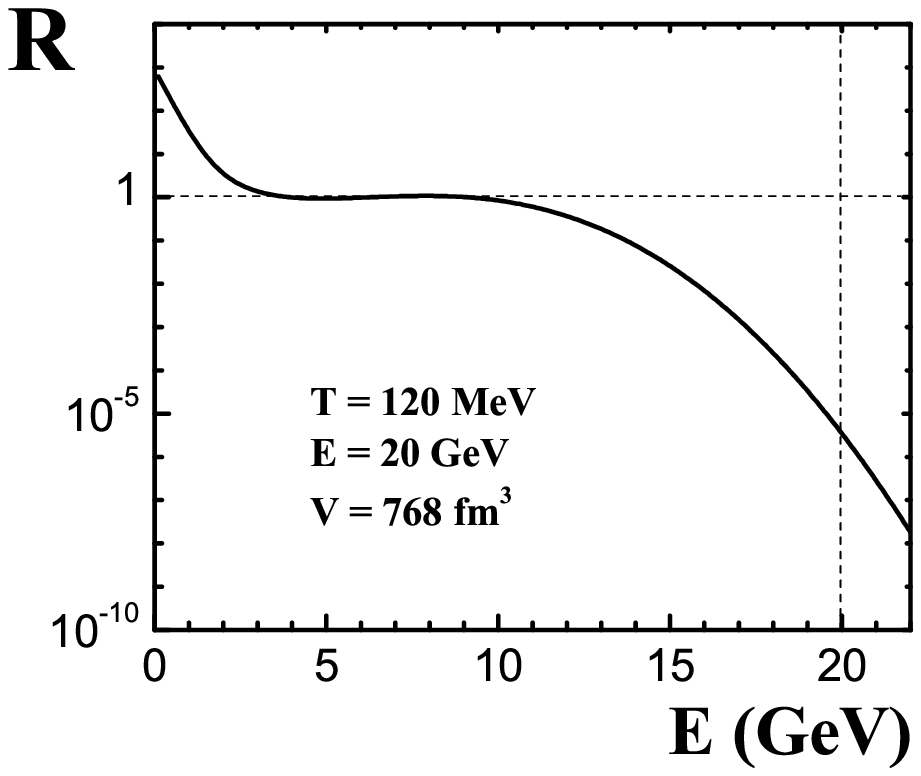,width=5.5cm}
 \epsfig{file=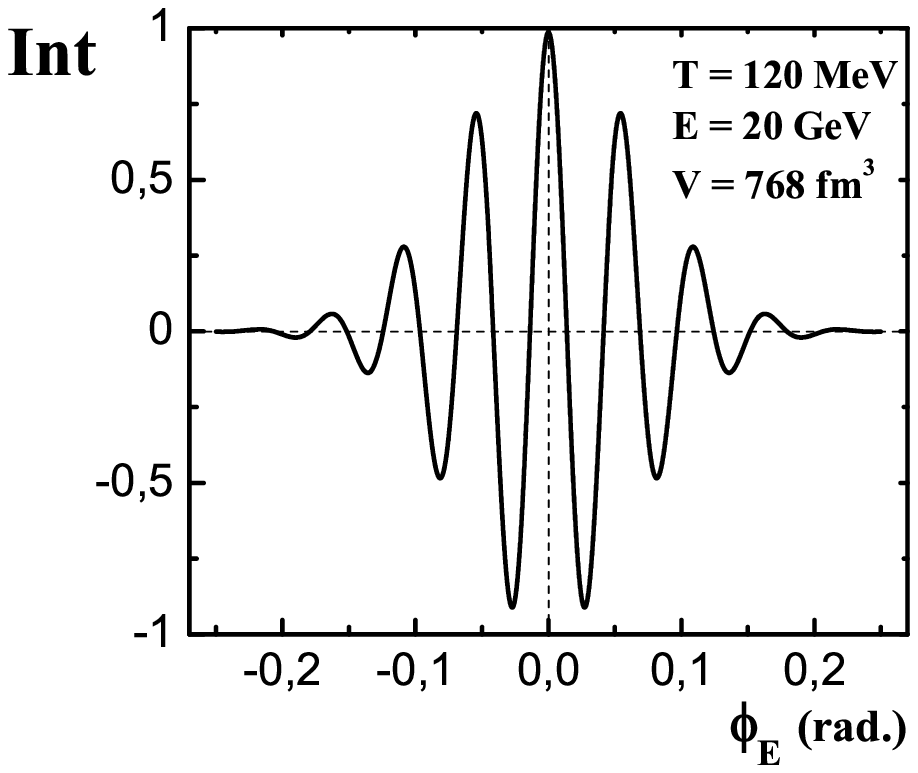,width=5.5cm}
 \vspace{-0.5cm}
 \caption{The comparison of the exact distribution $P(E)$ and its
Gauss approximation $P_G(E)$ (left), their ratio $R=P_G(E)/P(E)$
(center) and the integrant $Int=Re[e^{\left(3\langle
N\rangle-E/T\right)i\varphi_E}]$ for the parameters $\langle
E\rangle=20GeV$, $T=120$~MeV} \label{PRI-E-T50}
\end{figure}

\begin{figure}[ht!]
 \epsfig{file=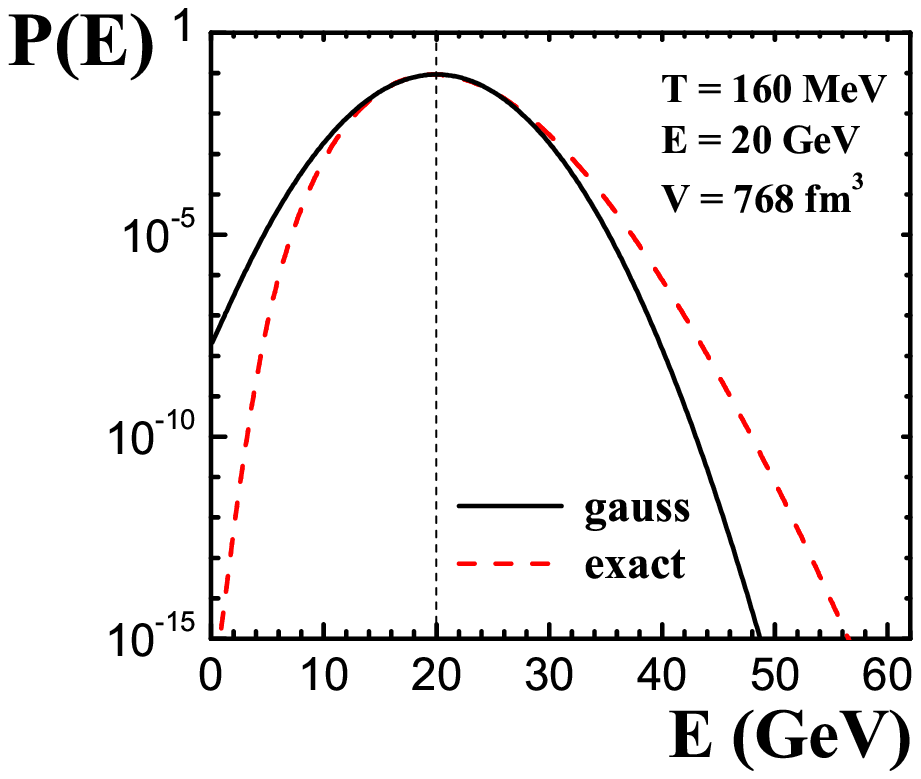,width=5.5cm}
 \epsfig{file=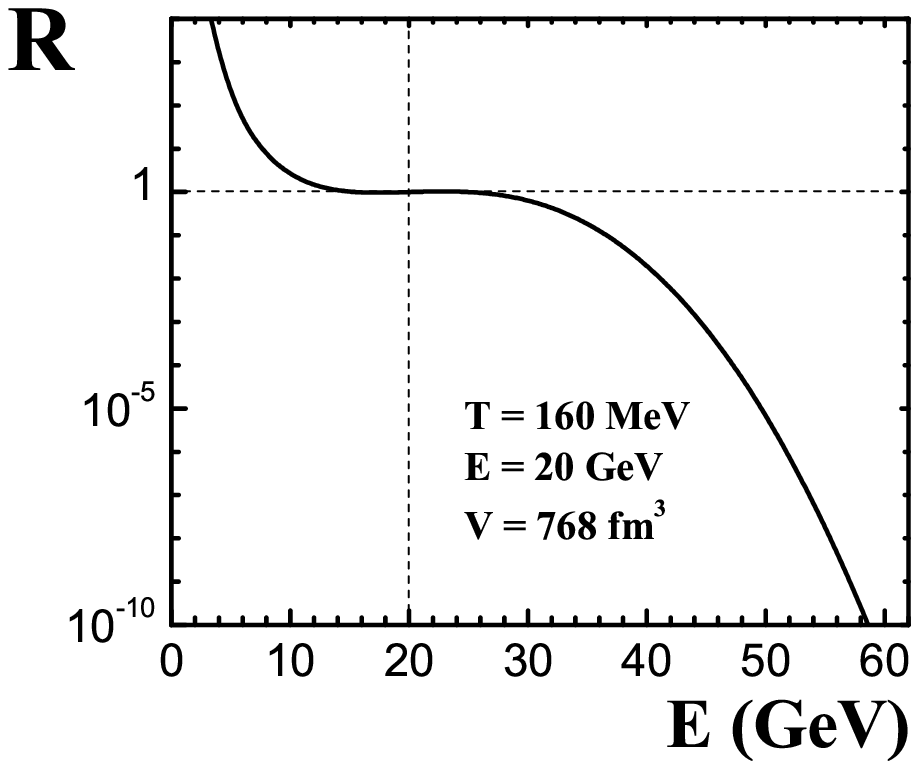,width=5.5cm}
 \epsfig{file=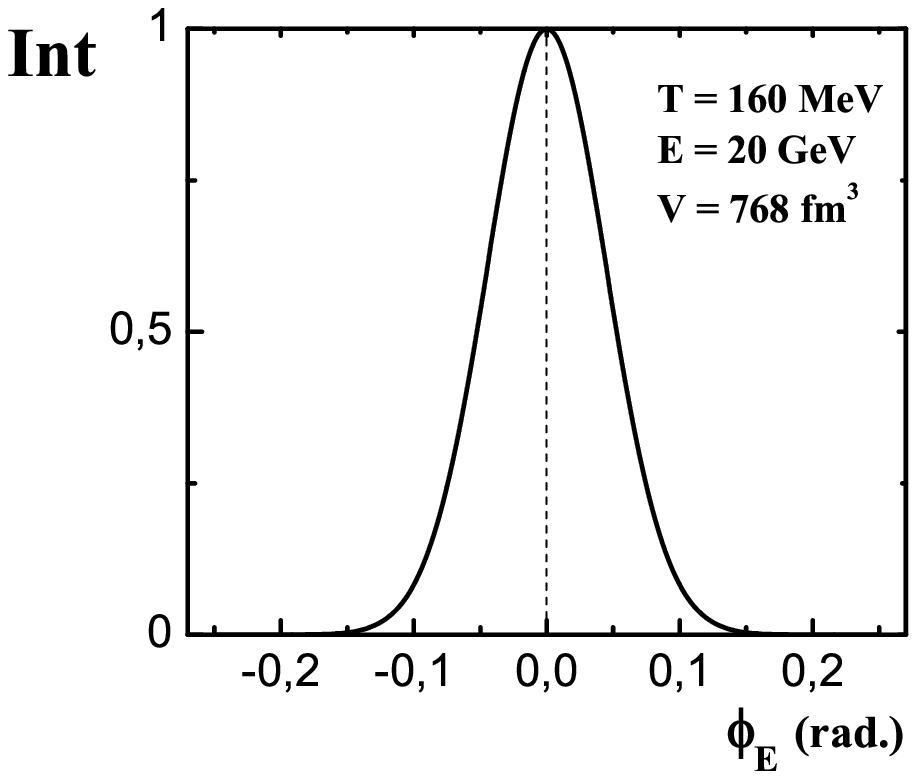,width=5.5cm}
 \vspace{-0.5cm}
 \caption{The same as Fig.~\ref{PRI-E-T50} but for $T=160$~MeV.}
 \label{PRI-E-T150}
\end{figure}

The microcanonical partition function can be calculated in the
same way. In the simplest case of massless Boltzmann particles we
know the exact answer \cite{GrandMce}:
 \eq{\label{P(N|E)}
 P(N|E)\;=\;\frac{P(N,E)}{P(E)}
  \;=\;\frac{1}{Z^{MCE}}\frac{x^N}{E\,(3N-1)!N!}\;,
 }
where $Z^{MCE}=\frac{x}{2E}\;_0F_3
\left(;\,\frac{4}{3},\frac{5}{3},2;\,\frac{x}{27}\right)$ is the
MCE partition function, $_0F_3$ is a generalized hypergeometric
function, and $x \equiv gVE^3/\pi^2$. Let us consider $P(E)$ in
Eq.~(\ref{P(N|E)}). One can obtain it by
straightforward iterative calculations similar to the
Ref.~\cite{GrandMce}:
 \eq{\label{P(E)}
 P(E)
 &\;=\; Z^{-1}\sum_{N=1}^{\infty}\frac{1}{N!}\,
 \frac{gV}{(2\pi )^3}\int\!\! d^3p_1\ldots
       \frac{gV}{(2\pi )^3}\int\!\!
       d^3p_N\;\exp[-\sum\limits_{k=1}^N|\vec{p}_k|\big/T]\;
       \delta(E-\sum_{k=1}^N|\vec{p}_k|)
 \nonumber \\
 &\;=\; Z^{-1}\sum_{N=1}^{\infty}\frac{1}{N!}\,
 \left(\frac{gV}{2\pi^2}\right)^N \int_0^{\infty}\!\! p_1^2 dp_1 \ldots
 \int_0^{\infty}\!\! p_{N-1}^2 dp_{N-1} \;e^{-E\big/T}
       (E-\sum_{k=1}^{N-1}p_k)^2\;.
 \nonumber \\
 &\;=\;
   \frac{e^{-E\big/T}Z^{MCE}}{Z}
 }
The saddle point expansion for $P(E)$ gives:
 \eq{
 P(E)
 \;&=\; Z^{-1}\int_{-\infty}^{\infty} \frac{d\phi_E}{2\pi}\,
        \sum_{N=1}^{\infty}\frac{1}{N!}\,\frac{gV}{(2\pi )^3}\int\!\! d^3p_1\ldots
         \frac{gV}{(2\pi )^3}\int\!\! d^3p_N\;
         \exp[-\sum\limits_{k=1}^N|\vec{p}_k|\big/T]\;
        e^{-i\phi_E (E-\sum_{k=1}^N|\vec{p}_k|)}
  \nonumber \\
  \;&=\;
     Z^{-1}\int_{-\infty}^{\infty} \frac{d\phi_E}{2\pi}\,
     e^{-iE\phi_E \;+\; \frac{\langle N\rangle}{ \left(1-iT\phi_E\right)^3}}
 \;=\; Z^{-1}\int_{-\infty}^{\infty} \frac{d\varphi_E}{2\pi T}\;
       e^{-iE\phi_E \;+\; \langle
         N\rangle\left(1\;+\;3iT\varphi_E-6T^2\varphi_E^2\;+\;\ldots\right)} 
 \nonumber\\
 \;&\simeq\;
 P_G(E)
 \;=\; \frac{1}{\sqrt{8\pi T\langle E\rangle}}\;
 e^{-\frac{(E-\langle E\rangle)^2}{8T\langle E\rangle}}\;,
 }
where $\langle N\rangle=gVT^3/\pi^2$, $\langle E\rangle=3\langle
N\rangle T$ and $Z=e^{\langle N\rangle}$. The exact distribution
$P(E)$ and its Gauss approximation $P_G(E)$, their ratio
$R=P_G(E)/P(E)$, and the integrant $Int=Re[e^{\left(3\langle
N\rangle-E/T\right)i\varphi_E}]$ (the imaginary part vanish
because of symmetric integral bounds) are shown in the
Figs.~\ref{PRI-E-T50}, \ref{PRI-E-T150} for the parameters
$E=20$~GeV, $V\simeq 768 fm^3$ and for two different temperatures:
the 'equilibrium' temperature $T=160$~MeV that gives $\langle
E\rangle=E$ and for 'non-equilibrium' temperature $T=120$~MeV that
gives $\langle E\rangle=6.32$~GeV$\neq E$.

Comparing the Figs.~\ref{PRI-Q-mu0}, \ref{PRI-Q-mu} and
Figs.~\ref{PRI-E-T50}, \ref{PRI-E-T150} one can see that if one
choose the wrong value of temperature than the integrant $Int$ in
$P(E)$ oscillate fast and Gauss approximation gives wrong result
for the MCE system see Fig.~\ref{PRI-E-T50}. It means that if we
calculate the MCE distribution $P(N|E)$ by means of the auxiliary
GCE distributions $P(N,E)$ and $P(E)$ than the temperature play
the role of the 'chemical potential' similarly to the CE.


\subsection{Resonance Decay}\label{ResDecay-Example}
In this subsection we would like to provide a connection between
probability distribution $P$, cumulant generation function $\Psi$,
and the generating function $G$ \cite{res} that was firstly used
to take into account resonance decays analytically. Let us
consider again a simple example of an ideal Boltzmann gas of
positively and negatively charged particles, i.e.
particle-antiparticle gas, and, additionally, 'allow' the decays
of positively charged particles as well as their antiparticles.
The formula for $P(Q)$, (\ref{P(Q)-example}), remains the same,
while $P(N_+,Q)$, (\ref{PNQ-3}), changes:
 \eq{\label{P(N,Q)-G}
 &P(N_+,Q)
 ~=~
 \frac{1}{Z}  \int \limits_{-\pi}^{\pi}  \frac{d \phi_Q}{2 \pi}
 \int \limits_{-\pi}^{\pi} \frac{d \phi_+}{2 \pi}\;
 \exp\bigg[-i Q \phi_Q  -  i N_+ \phi_+
  \\
 &\;+\; z \exp\left(\frac{\mu_Q}{T} + i \phi_Q \right)
 \sum_c \Gamma_+^c \exp\left(i\,c \phi_+\right)
 \;+\; z \exp\left(-\frac{\mu_Q}{T} - i \phi_Q\right)
 \sum_c \Gamma_-^c \exp\left(i\,c \phi_+\right) \bigg]\;,
 \nonumber
 }
where $c$ is a multiplicity of a particular decay channel and
$\Gamma_{\pm}^c$ is a branching ratio that corresponds to the
decay of positively/negatively charged resonance.

There is no any real system that corresponds to the partition
function (\ref{P(N,Q)-G}), but it is the simplest example where we
can show how these formulas work.

We can expand the exponent in the Eq.~(\ref{P(N,Q)-G}):
 \eq{\label{P(N+,Q)-res}
 P(N_+,Q) \;=\;\frac{1}{Z}
 &\int \limits_{-\pi}^{\pi}  \frac{d \phi_Q}{2 \pi}
  \int \limits_{-\pi}^{\pi} \frac{d \phi_+}{2 \pi}\;
  \exp\big[-i (Q-[n_+-n_-])\, \phi_Q  \;-\;  i N_+ \phi_+\big]
 \nonumber \\
 \times
 & \sum_{n_+,n_-=0}^{\infty}
 \frac{z_+^{n_+}}{n_+!} \frac{z_-^{n_-}}{n_-!}\;
 \left(\sum_c \Gamma_+^c \exp\left(i\,c \phi_+\right)
 \right)^{n_+}
 \left(\sum_c \Gamma_-^c \exp\left(i\,c \phi_+\right)
 \right)^{n_-}
 }
and immediately recognize the generating function that was used in
our previous work \cite{res}:
 \eq{
 G(\phi_+) \;\equiv\; \left(\sum_c \Gamma_+^c \exp\left(i\,c \phi_+\right)
\right)^{n_+}\left(\sum_c \Gamma_-^c \exp\left(i\,c \phi_+\right)
 \right)^{n_-}\;.
 }
The difference from the generation function used in \cite{res} is
that here we study the case with only two types of resonances,
$n_{\pm}$, and we also use the different definition for a decay
channel.

From the Eq.~(\ref{P(N+,Q)-res}) one can also see that the
cumulant generating function $\Psi$ is connected with the
generating function $G$ in a very simple way:
 \eq{\label{Psi-G}
 \Psi(\phi_+) \;=\; \ln\left[\sum_{n_+,n_-=0}^{\infty}
 \frac{z_+^{n_+}}{n_+!} \frac{z_-^{n_-}}{n_-!}\;
 G(\phi_+)\right]\;.
 }
In order to make a comparison with the Ref. \cite{res} one should
also remember some technical moment that in the paper \cite{res}
we used the auxiliary parameters $\lambda$ that are the same as
$\exp(i\phi_+)$ in this paper, however
 \eq{
 -i\,\frac{\partial}{\partial\phi_+}
  \;=\; \lambda_+\frac{\partial}{\partial\lambda_+}\;,
 \quad \text{and} \quad
 \left(-i\frac{\partial}{\partial\phi_+}\right)^2
  \;=\; \lambda_+\frac{\partial}{\partial\lambda_+}
    \left(\lambda_+\frac{\partial}{\partial\lambda_+}\right)\;.
 }
From the Eq.~(\ref{Psi-G}) it also can be easily seen that we can
multiply the generation function by any other normalized
distribution for example by the finite acceptance probability
distribution and it will enter the partition function in the same
way as the generating function $G$ (see Section
\ref{FinalDetect}).
%
\section{Summary and Conclusion}
\label{Conclusion}

We have presented an analytical expansion method for calculation
of distributions at finite volume for the canonical as well as the
microcanonical ensembles of the ideal relativistic hadron resonance gas. The
introduction of temperature into the microcanonical partition function and
chemical potentials into the canonical partition function have lead to the
identification of the grand canonical partition function with the
characteristic function of associated joint probability distributions. The
microcanonical and canonical multiplicity distributions could
thus be defined through the conditional probability distributions
of finding a certain multiplicity while other parameters (global
charge or energy) were taken to be fixed. 

We have shown that in the thermodynamic limit the central
region of the canonical and microcanonical partition functions
can be approximated by multivariate normal distributions. Multiplicity
distributions tend thus to Gaussians as system size is increased. In
particular, we could find a simple formula for the scaled 
variance, hence the width of this Gaussian, in this limit. It was
further possible to show the equivalence of our results to those
obtained from the microscopic correlator approach.

Apart from providing a simplified formula for asymptotic multiplicity 
fluctuations, the approach presented in this work owns a few
conceptual advantages. In considering finite volume corrections to
the system partition function, thus relaxing the assumption of
thermodynamic equivalence of different statistical ensembles, one
is lead to demanding that the partition function should be
maximized for a particular set of conserved charges. It turned out
that this requirement is entirely equivalent to the well known
textbook definitions of chemical potential in the canonical
ensemble as the derivative of Helmholtz free energy with respect
to conserved charge and temperature in the microcanonical
ensemble through differentiation of entropy with respect to
conserved energy.

Our method is based on Fourier analysis of the grand canonical partition
function. Conventionally one would not introduce chemical potentials and
temperature into these calculations. However one then faces the problem of a
heavily oscillating (or even irregular) integrant, making numerical integration
unpractical. Artificially introduced temperature and chemical potentials,
correctly chosen, produce a very smooth integrant allowing for expansion of
the integrant in powers of volume. Analytical solutions to asymptotic
multiplicity distributions could thus be found in terms of Laplace's
expansion, while finite volume corrections could be obtained from
Gram-Charlier expansion. A first comparison with available analytical
solutions to simple statistical systems suggests that good results can be
expected even for rather small volume. One drawback is that the results can
only be applied to the central region of the distribution, owning
to the fact that finite volume correction terms appear in the form
of Hermite polynomial of low order.

Additionally we have included resonance decay directly into the
system partition function of the hadron resonance gas model. This
treatment has proven to be more economical than the previous
handling through generating functions or Monte Carlo techniques.
Neglecting correlation in momentum space a first order
approximation to the effect of finite acceptance can be made.

This paper provides a connection between the two previously published methods
for calculation of multiplicity distributions, the microscopic correlator
approach and saddle point expansion, and should be of great practical use.

\begin{acknowledgments}
We would like to thank F.~Becattini, M.~Bleicher, E.L. Bratkovskaya,
A.I.~Bugrij, M.~Ga\'zdzicki, W.~Greiner,  A.P.~Kostyuk,
M.~Martinez, G.~Torrieri, G.~Tupper, and S.~Wheaton for
discussions. V.V. Begun would like to thank for the support of The
International Association for the Promotion of Cooperation with
Scientists from the New Independent States of the Former Soviet
Union (INTAS), Ref. Nr 06-1000014-6454.
\end{acknowledgments}
\appendix
\section{Characteristic Function}
\label{CharFunc}
Let us consider a probability distribution of some observable
$\chi$ which can take values $f_{\chi}(x)$ on the real $x$ axis
with the following normalization:
\begin{equation}
\int \limits_{-\infty}^{\infty} dx~ f_{\chi}(x) =1.
\end{equation}
Hence  $f_{\chi}(x)$ defines the probability of finding a value
$x$ for the observable $\chi$. One can define the characteristic
function $\Phi_{\chi}(i \theta)$ as the Fourier back
transformation of the probability distribution $f_{\chi}(x)$:
\begin{equation}
\Phi_{\chi} (i \theta) ~\equiv~ \hat f_{\chi}(x) ~=~ \int
\limits_{-\infty}^{\infty}dx~ e^{i x \theta }~ f_{\chi}(x)~ .
\end{equation}
The moments $\mathcal{E}^n (\chi)$ can directly be calculated from
the characteristic function from the respective derivatives at the
origin:
\begin{eqnarray}
i^n \mathcal{E}^n (\chi) ~\equiv~ \frac{d^n \Phi_{\chi}(i \theta)}{d
\theta^n} \bigg |_{\theta=0}~ =~ i^n \int \limits_{-\infty}^{\infty}
dx~ x^n~ f_{\chi}(x)~.
\end{eqnarray}
Once the characteristic function $\Phi_{\chi}(i \theta)$ is known, the
probability distribution $f_{\chi}(x)$ can be found by  Fourier
transformation:
\begin{equation}
f_{\chi}(x) ~=~ \int \limits_{-\pi}^{\pi} \frac{d \theta}{2 \pi}
~ e^{-i x \theta} ~ \Phi_{\chi}(i \theta) ~.
\end{equation}

\section{CE Partition Function}
\label{CEPartFunc}
The (generalized) system partition function of a system with conserved charge
vector $Q^j = \left(B,S,Q \right)$, Eq.(\ref{BSQPartFunc_j}), reads:
\begin{equation}\label{AppendixCEPF}
\mathcal{Z}^{Q^j} = \left[ \prod_j \int \limits_{-\pi}^{\pi} \frac{d
    \phi_j}{2\pi}\right] e^{- i Q^j \phi_j} \exp \left[ \sum_l z_l
  \left(\phi_j,\mu_j \right) \right]~.
\end{equation}
The single particle partition function Eq.(\ref{QstatsZET})
includes chemical potentials as well as Wick rotated fugacities. Expanding the
logarithm in Eq.(\ref{QstatsZET}) yields:
\begin{eqnarray}
z_l \left(\phi_j \right) &=&  \frac{g_l V} {\left( 2 \pi\right)^3}
\int d^3p ~\ln \left( 1 \pm  e^{-\frac{\varepsilon_l}{T}}\; e^{\frac{\mu_l}{T}}
  \; e^{ i q^j_l \phi_j}\right)^{\pm1} \\
&=& \sum_{n_l=1}^{\infty} \left( \mp 1\right)^{n_l+1} \frac{g_l V} {\left( 2
    \pi\right)^3} \int d^3p  \frac{1}{n_l}~e^{-\frac{n_l \varepsilon_l}{T}}~
e^{n_l ~q_l^j ~\left( \frac{\mu_j}{T} +  i \phi_j \right) }~,
\end{eqnarray}
where the effective chemical potential of particle species $l$ is given by
$\mu_l = q_l^j \mu_j$, and the single particle partition function of a `lump`
of mass $n_l~m_l$ of species $l$ is given by \cite{MonteCarlo}  $z_{n_l}=
\frac{g_l V} {\left( 2 \pi\right)^3} \int d^3p ~ n_l^{-1} ~e^{-\frac{n_l
    \varepsilon_l}{T}}$. Eq.(\ref{AppendixCEPF}) can be thus written as:
\begin{eqnarray}
\mathcal{Z}^{Q^j} &=& \left[ \prod_j \int \limits_{-\pi}^{\pi} \frac{d
    \phi_j}{2\pi}\right] e^{- i Q^j \phi_j} \exp \left[ \sum_l
 \sum_{n_l=1}^{\infty} \left( \mp 1\right)^{n_l+1}~ z_{n_l} ~e^{n_l q_l^j \left(
     \frac{\mu_j}{T} +
      i \phi_j \right) } \right] \\
 &=& \left[ \prod_j \int \limits_{-\pi}^{\pi} \frac{d
    \phi_j}{2\pi}\right] e^{- i Q^j \phi_j}  \prod_l \prod_{n_l=1}^{\infty}
\sum_{k_{n_l}=0}^{\infty} \frac{ \left(\left( \mp 1\right)^{n_l+1} z_{n_l}
  \right)^{k_{n_l}}}{k_{n_l}!}~ e^{k_{n_l} n_l q_l^j \left( \frac{ \mu_j}{T}
     + i \phi_j \right) }
\end{eqnarray}
Only sets of numbers $\lbrace k_{n_l} \rbrace$ that meet the requirement:
\begin{equation}
\sum_l \sum_{n_l=1}^{\infty} k_{n_l}~ n_l ~q_l^j - Q^j ~=~ 0
\end{equation}
have a non-vanishing contribution to the integral. For any such set
one finds:
\begin{equation} \label{ChemFactor}
\sum_l \sum_{n_l=1}^{\infty} k_{n_l} ~n_l ~q_l^j ~\frac{ \mu_j}{T} ~=~ Q^j
\frac{ \mu_j}{T}~.
\end{equation}
Therefore one can pull the factor Eq.(\ref{ChemFactor}) in front of the
integral (\ref{AppendixCEPF}):
\begin{equation}
\mathcal{Z}^{Q^j} = e^{Q^j \frac{ \mu_j}{T}} \left[ \prod_j \int
  \limits_{-\pi}^{\pi} \frac{d \phi_j}{2\pi}\right] e^{- i Q^j \phi_j}
\prod_l \prod_{n_l=1}^{\infty} \sum_{k_{n_l}=0}^{\infty} \frac{ \left( z_{n_l}
  \right)^{k_{n_l}}}{k_{n_l}!} e^{ i k_{n_l} n_l q_l^j \phi_j  }~.
\end{equation}
However this is just the standard definition of the system partition function
times the factor Eq.(\ref{ChemFactor}), thus:
\begin{equation}
\mathcal{Z}^{Q^j}~ =~ e^{Q^j \frac{ \mu_j}{T}} Z^{Q^j}~, \qquad \qquad
\textrm{and}
\qquad Z^{GCE} ~=~ \Bigg[ \prod_{j=1}^J~\sum_{Q^j= -\infty}^{\infty} \Bigg]~
\mathcal{Z}^{Q^j}.
\end{equation}
A similar calculation can be shown to hold for the MCE.
\section{MCE Partition Function}
\label{MCEPartFunc}
Having introduced temperature in the MCE partition is probably a somewhat
unusual approach. Here we consider this in detail in the example of a
massless gas in Boltzmann approximation without momentum conservation.
We will first state and solve our version of the MCE partition function and
compare to the analytical solution known from textbooks (e.g. \cite{Physics}).
We adopt the notation of \cite{GrandMce}, $\mathcal{Z}^{E,N} = \mathcal{W}_N
\left(E,V \right)$ and $\mathcal{Z}^{E} = \mathcal{W} \left(E,V \right)$.
The number of states consistent with the constraints of fixed energy $E$ and
particle number $N$ in a GCE system is given by the double Fourier integral
over its GGPF $\mathcal{Z} \left(\phi_N, \phi_E \right)$:
\begin{equation}\label{funnyW}
\mathcal{W}_N \left(E,V,T \right) ~\equiv~ \int \limits_{-\pi}^{\pi} \frac{d
    \phi_N}{2 \pi} ~ \int \limits_{-\infty}^{\infty} \frac{d \phi_E}{2 \pi} ~
  e^{-iN \phi_N} ~e^{-iE \phi_E} ~\exp \left[ \frac{gV}{2 \pi^2} \int
    \limits_{0}^{\infty} p^2 dp ~e^{-\beta p}~ e^{i p \phi_E}~ e^{i N \phi_N}
  \right].
\end{equation}
Expanding the exponential and solving the integral over $\phi_N$ yields:
\begin{equation}
\mathcal{W}_N \left(E,V,T \right) ~=~ \frac{1}{N!} \left(\frac{gV}{2 \pi^2}
\right)^N \int \limits_{-\infty}^{\infty} \frac{d \phi_E}{2 \pi} ~e^{-iE
  \phi_E} ~ \left[ \int \limits_{0}^{\infty} p^2 dp ~e^{-\beta p}~ e^{i p
    \phi_E}~ \right]^N.
\end{equation}
Further now solving the integral over momentum gives:
\begin{equation}\label{Pole}
\mathcal{W}_N \left(E,V,T \right) ~=~ \frac{1}{N!} \left(\frac{gV}{2 \pi^2}
\right)^N \int \limits_{-\infty}^{\infty} \frac{d \phi_E}{2 \pi} ~e^{-iE
  \phi_E} ~ \left[ \frac{2}{ \left(\beta-i\phi_E \right)^3}\right]^N.
\end{equation}
Eq. (\ref{Pole}) has obviously a pole of order $3N$ at $\phi_E = - i \beta$. So
we close the integration over the lower hemisphere and use the residue
theorem.
\begin{eqnarray}
\mathcal{W}_N \left(E,V,T \right) &=& \frac{-i}{N!} \left(\frac{gV}{\pi^2}
\right)^N  ~ \textrm{Res} ~\left[ \frac{e^{-iE \phi_E} }{\left( \beta-i\phi_E
    \right)^{3N}}~;~ \phi_E = - i  \beta \right] \nonumber \\
&=& \frac{-i}{N!} \left(\frac{gV}{\pi^2}
\right)^N  \frac{1}{\left(3N-1 \right)!} ~ \lim_{\phi_E \to - i
  \beta} ~\left[ \frac{d^{3N-1}}{d\phi_E^{3N-1}} \left(\phi_E + i \beta
  \right)^{3N}  \frac{e^{-iE \phi_E} }{\left( \beta-i\phi_E \right)^{3N}}
\right] \nonumber \\
&=& \left(\frac{gV}{\pi^2} \right)^N   ~ \frac{E^{3N-1}}{N!\left(3N-1
  \right)!} ~~~ e^{-\beta E}.
\end{eqnarray}
Multiplication with the inverse Boltzmann factor yields the result known from
the literature \cite{Fermi,GrandMce,Physics}, $W_N \left(E,V\right) \equiv
\mathcal{W}_N \left(E,V,T \right)~ e^{\beta E}$ :
\begin{equation} \label{MCEPFN}
W_N \left(E,V\right) ~=~\left(\frac{gV}{\pi^2} \right)^N   ~
\frac{E^{3N-1}}{N!\left(3N-1 \right)!}.
\end{equation}
Using condition Eq.(\ref{gChemPot}) it further follows that the temperature is
given by the average energy per particle $T \equiv E/(3N-1)$. In complete
analogy to what was presented before:
\begin{eqnarray}  \label{MCEPF}
W \left(E,V\right) ~=~ \sum \limits_{N=1}^{\infty} W_N \left(E,V\right), \qquad
\textrm{and} \qquad \mathcal{W} \left(E,V,T\right) ~=~ \sum
\limits_{N=1}^{\infty} \mathcal{W}_N \left(E,V,T \right).
\end{eqnarray}
However in both case we define the MCE multiplicity distribution by:
\begin{equation}\label{gMCEPN}
P \left( N | E \right) ~=~ W_N \left(E,V\right)~ W^{-1} \left(E,V\right)  ~=~
\mathcal{W}_N \left(E,V,T\right) ~ \mathcal{W  }^{-1} \left(E,V,T\right).
\end{equation}
The introduction of temperature in Eq.(\ref{funnyW}) does not necessarily
simplify this calculation (it even drops out entirely from the r.h.s of
Eq.(\ref{gMCEPN})), however makes approximations possible.
The GCE partition function finally is given by:
\begin{eqnarray}
Z^{GCE} \left(V,T \right) ~=~ 1 + \int \limits_{0}^{\infty}dE~ W
\left(E,V\right) e^{-\beta E} ~=~ 1+ \int \limits_{0}^{\infty}dE~ \mathcal{W}
\left(E,V,T\right) ~=~ \exp \left[ \frac{gVT^3}{\pi^2}\right].
\end{eqnarray}
The additional term `$+1$` arises form the fact that the MCE partition function
only includes states with at least one particle, while the state of zero
energy and zero particle number contributes to the GCE partition
function.
\section{Width and Normalization}
\label{DetIdent}
The inverse of a nonsingular $J \times J$ matrix $\tilde \kappa_2$ can be
obtained from its adjoint and its corresponding minors \cite{Formulas}:
\begin{equation}
\left( \tilde \kappa_2^{-1} \right)_{i,j} ~=~ \frac{\textrm{adj}~
\tilde
  \kappa_2}{\det | \tilde \kappa_2|} ~=~ \frac{ \small[ ~\small(\tilde M
  \small)_{i,j}~ \small]^{T}}{\det | \tilde \kappa_2|}~.
\end{equation}
In our case only the minor $\small( \tilde M \small)_{1,1} \equiv \det
|\kappa_2|$ is of interest, hence we find for the element in the
upper left corner of the inverse of $\tilde \kappa_2$:
\begin{equation}
\left( \tilde \kappa_2^{-1} \right)_{1,1} ~=~  \frac{\det
|\kappa_2|}{\det | \tilde \kappa_2|}.
\end{equation}
Generally we can find the inverse of a matrix by multiplication of
a diagonal matrix, with the inverses eigenvalues on its diagonal
$\tau^{-1}_{a,b}$, with orthonormal transformation matrices, which
can be formed from the eigenvectors $\vec v_{a}$ of $ \tilde
\kappa_2$,
\begin{equation}
\left( \tilde \kappa_2^{-1} \right)_{i,j} ~=~ \sum \limits_{a=1}^J
~ \sum \limits_{b=1}^J ~ v_{i,a} ~ \left( \tau^{-1} \right)_{a,b} ~ v^T_{b,j},
\end{equation}
where $v_{i,a}$ is the $a^{th}$ component of the $i^{th}$ eigenvector,
$v^T_{b,j}$ is the transpose of  $v_{i,a}$, hence a matrix with eigenvectors
in its columns, and $\left(  \tau^{-1} \right)_{a,b}$ is a matrix with the
corresponding inverse of the $a^{th}$ eigenvalues $t_a^{-1}$ on its
diagonal. We find for the upper left most entry:
\begin{equation} \label{detdivdet}
\left( \tilde \kappa_2^{-1} \right)_{1,1} ~=~  \sum
\limits_{a=1}^J ~ v^2_{1,a} ~t_a^{-1} ~=~  \frac{\det
|\kappa_2|}{\det | \tilde \kappa_2|}.
\end{equation}
Likewise, we can express the inverse sigma tensor in term of
eigenvalues and eigenvectors of $ \tilde \kappa_2$ (see Section
\ref{FiniteVolCorr}):
\begin{eqnarray}
\left( \tilde \sigma^{-1}\right)_{i,j} ~=~ \sum \limits_{a=1}^J ~
\sum \limits_{b=1}^J ~ v_{i,a} ~ \left( \tau^{-1/2}
\right)_{a,b} ~ v^T_{b,j} ~=~ \sum \limits_{a=1}^J ~  v_{i,a}
~v_{j,a}~ t^{-1/2}_{a}.
\end{eqnarray}
We just need the sum of squares of the entries of the left most column:
\begin{eqnarray}
\sum \limits_{i=1}^J~ \left( \tilde \sigma^{-1}\right)^2_{i,1} ~=~
\sum \limits_{i=1}^J~ \sum \limits_{a=1}^J~ \sum \limits_{b=1}^J ~  v_{i,a}
~v_{1,a}~ t^{-1/2}_{a} ~  v_{i,b}
~v_{1,b}~ t^{-1/2}_{b} ~=~ \sum \limits_{a=1}^J ~ v^2_{1,a}~ t^{-1}_{a},
\end{eqnarray}
since $ \sum_{i=1}^j~  v_{i,a}  v_{i,b} = \delta_{a,b}$. This coincides with Eq.
(\ref{detdivdet}) and proofs Eq.(\ref{EqDetIdent}),
\begin{equation}
\sum \limits_{i=1}^J~ \left( \tilde \sigma^{-1}\right)^2_{i,1} ~=~
\frac{\det |\kappa_2|}{\det |\tilde \kappa_2|},
\end{equation}
in the most general case of a $J$ dimensional $2^{nd}$ rank tensor
$\tilde \kappa_2$.



\begin{thebibliography}{100}
\bibitem{Fermi}
E.Fermi, Progr.Theor. Phys. {\bf 5} (1950) 570.
\bibitem{hagedorn}
R. Hagedorn, Nucl. Phys. B {\bf 24}, 93 (1970).
\bibitem{GSIfits}
J. Cleymans, D. Elliott, A. Keranen, E. Suhonen, Phys.Rev. C {\bf 57} (1998)
3319;
J. Cleymans, H. Oeschler, K. Redlich, Phys.Rev. C {\bf 59} (1999) 1663;
R. Averbeck, R. Holzmann, V. Metag, R.S. Simon,
Phys.Rev. C {\bf 67} (2003) 024903.
\bibitem{AGSfits}
P.~Braun-Munzinger, J.~Stachel, J.~P.~Wessels and N.~Xu,   Phys.\ Lett.\  B
{\bf 344}, 43 (1995).
\bibitem{SPSfits}
P.~Braun-Munzinger, J.~Stachel, J.~P.~Wessels and N.~Xu, Phys.\ Lett.\  B {\bf
  365} (1996) 1;
P.~Braun-Munzinger, I.~Heppe and J.~Stachel,   Phys.\ Lett.\  B {\bf 465}, 15
(1999);
F. Becattini, M. Gazdzicki, A. Keranen, J. Manninen, R. Stock,
Phys.Rev.C {\bf 69} 024905 (2004).
\bibitem{RHICfits}
J.~Adams {\it et al.}  [STAR Collaboration], Nucl.\ Phys.\  A {\bf 757}, 102
(2005).
\bibitem{FreezeOut}
J. Cleymans, H. Oeschler, K. Redlich, and S. Wheaton, Phys. Rev. C {\bf 73},
034905 (2006);
J.Cleymans, and K.Redlich, Phys.Rev. C {\bf 60}, (1999) 054908;
J.Cleymans, and K.Redlich, Phys.Rev.Lett. {\bf 81} (1998) 5284-5286;
F. Becattini, J. Manninen, and M. Ga\'zdzicki,  Phys. Rev. C {\bf 73}, 044905
(2006);
A. Andronic, P. Braun-Munzinger, J. Stachel, Nucl. Phys. A {\bf 772}, 167
(2006).
\bibitem{ElemetaryFits}
F. Becattini, Z. Phys. C {\bf 69} 485 (1996);
F.~Becattini and U.~W.~Heinz, Z.\ Phys.\  C {\bf 76}, 269 (1997);
F.~Becattini and G.~Passaleva,  Eur.\ Phys.\ J.\  C {\bf 23}, 551 (2002).
\bibitem{SHM}
R. Hagedorn, CERN-TH-7190/94, 1994;
R. Hagedorn, CERN lectures, {\it Thermodynamics of Strong Interaction} (1970);
R.~Stock, Phys.\ Lett.\  B {\bf 456}, 277 (1999).
\bibitem{Meaning}
F.~Becattini, J.\ Phys.\ Conf.\ Ser.\  {\bf 5}, 175 (2005).
\bibitem{QFTMCE}
F.~Becattini and L.~Ferroni,  arXiv:0704.1967 [nucl-th].
\bibitem{OnsetOfDecon}
M.~Gazdzicki, M.~I.~Gorenstein and S.~Mrowczynski, Phys.\ Lett.\ B {\bf 585},
 115 (2004);
M.~I.~Gorenstein, M.~Gazdzicki and O.~S.~Zozulya, Phys.\ Lett.\ B {\bf 585},
237 (2004).
\bibitem{PhaseTrans}
I.N. Mishustin, Phys. Rev. Lett. {\bf 82}, 4779 (1999);
Nucl. Phys. A {\bf 681}, 56-63 (2001);
H. Heiselberg and A.D. Jackson, Phys. Rev. C {\bf 63}, 064904 (2001).
\bibitem{CriticalPoint}
M.A.~Stephanov, K.~Rajagopal, and E.V.~Shuryak, Phys. Rev. Lett. {\bf 81},
4816 (1998);
Phys. Rev. D {\bf 60},114028 (1999);
M.A.~Stephanov, Acta Phys.Polon.B {\bf 35} 2939 (2004).
\bibitem{QGP-HRG-fluc}
S.~Jeon and V.~Koch, Phys.\ Rev.\ Lett.\  {\bf 85}, 2076 (2000);
S. Jeon, V. Koch, Phys.Rev.Lett.85:2076-2079,2000;
M.~Asakawa, U.~W.~Heinz and B.~Muller, Phys.\ Rev.\ Lett.\  {\bf 85}, 2072
(2000).
M. Asakawa, U.W. Heinz,B. Muller, Phys.Rev.Lett.85:2072-2075,2000.
\bibitem{fluc-mult}
S.V.~Afanasev {\it et al}., [NA49 Collaboration],
Phys. Rev. Lett. {\bf 86}, 1965 (2001);
M.M.~Aggarwal {\it et al}., [WA98 Collaboration], Phys. Rev.
C {\bf 65}, 054912 (2002);
J.~Adams {\it et al}., [STAR Collaboration], Phys. Rev. C {\bf 68}, 044905
(2003);
C.~Roland {\it et al}., [NA49 Collaboration], J. Phys. G {\bf 30} S1381
(2004);
Z.W.~Chai {\it et al}., [PHOBOS Collaboration], J. Phys. Conf.Ser. {\bf 27},
128 (2005);
M.~Rybczynski {\it et al.}  [NA49 Collaboration],
J.\ Phys.\ Conf.\ Ser.\  {\bf 5}, 74 (2005).
\bibitem{fluc-pT}
H.~Appelshauser {\it et al.}  [NA49 Collaboration],
Phys.\ Lett.\ B {\bf 459}, 679 (1999);
D.~Adamova  {\it et al}., [CERES Collaboration], Nucl. Phys. A {\bf 727},
97 (2003);
T. Anticic {\it et al}., [NA49 Collaboration], Phys. Rev.
C {\bf 70}, 034902 (2004);
S.S.~Adler {\it et al}., [PHENIX Collaboration], Phys. Rev.
Lett.  {\bf 93}, 092301 (2004);
J.~Adams  {\it et al}., [STAR Collaboration], Phys. Rev.
C {\bf 71}, 064906 (2005).
\bibitem{NA49}
  B.~Lungwitzt {\it et al.} [NA49 Collaboration],
  arXiv:nucl-ex/0610046.
\bibitem{MCEvsData}
V.V. Begun,  M. Ga\'zdzicki, M.I.~Gorenstein, M.~Hauer,
V.P.~Konchakovski, and B. Lungwitz, arXiv:nucl-th/0611075.
\bibitem{Horn}
S.~V.~Afanasiev {\it et al.}  [The NA49 Collaboration],  Phys.\ Rev.\  C {\bf
  66}, 054902 (2002);
M.~Gazdzicki {\it et al.}  [NA49 Collaboration], J.\ Phys.\ G {\bf 30}, S701
(2004)
\bibitem{ScanProgram}
M.~Gazdzicki {\it et al.}  [NA49-future Collaboration], arXiv:nucl-ex/0612007.
\bibitem{non-eq}
C.~M.~Ko, V.~Koch, Z.~w.~Lin, K.~Redlich, M.~A.~Stephanov and X.~N.~Wang,
Phys.\ Rev.\ Lett.\  {\bf 86}, 5438 (2001);
O.~Fochler, S.~Vogel, M.~Bleicher, C.~Greiner, P.~Koch-Steinheimer and Z.~Xu,
Phys.\ Rev.\  C {\bf 74}, 034902 (2006);
G.~Torrieri and J.~Rafelski,  Phys.\ Lett.\  B {\bf 509}, 239 (2001)
\bibitem{CEfluc_1}
V.V. Begun, M. Ga\'zdzicki, M.I.~Gorenstein, and O.S.~Zozulya, Phys. Rev. C
{\bf 70}, 034901 (2004).
\bibitem{MCEfluc_1}
V.V.~Begun, M.I.~Gorenstein, A.P.~Kostyuk, and O.S.~Zozulya, Phys. Rev. C {\bf
  71}, 054904 (2005).
\bibitem{QGas}
V.V. Begun and M.I.~Gorenstein, Phys. Rev. C {\bf 73}, 054904
(2006);
V.V.~Begun and M.I.~Gorenstein,
arXiv:hep-ph/0611043;
V.V.~Begun, M.I.~Gorenstein, A.P.~Kostyuk, and O.S.~Zozulya, J. Phys. G {\bf
  32}, 935 (2006).
\bibitem{res}
V.V. Begun, M.I.~Gorenstein, M.~Hauer, V.P.~Konchakovski, and
O.S.~Zozulya, Phys. Rev. C {\bf 74}, 044903 (2006).
\bibitem{GrandMce}
V.V.~Begun, M.I.~Gorenstein, A.P.~Kostyuk, and O.S.~Zozulya, Phys. Rev. C {\bf
  71}, 054904 (2005);
\bibitem{BGZ}
V.V.~Begun, M.I.~Gorenstein, and O.S.~Zozulya,
Phys. Rev. C {\bf 72}, 014902 (2005).
\bibitem{SPE}
F.~Becattini, A.~Ker\"anen, L.~Ferroni, and T.~Gabbriellini, Phys. Rev. C {\bf
  72}, 064904 (2005).
\bibitem{CETurko}
J.~Cleymans, K.~Redlich, and L.~Turko, Phys. Rev. C {\bf 71}, 047902 (2005);
J.~Cleymans, K.~Redlich and L.~Turko, J.\ Phys.\ G {\bf 31}, 1421 (2005).
\bibitem{GCEfluc}
G.~Torrieri, S.~Jeon and J.~Rafelski, Phys.\ Rev.\  C {\bf 74}, 024901 (2006)
G.~Torrieri, J.\ Phys.\ G {\bf 32}, S195 (2006)
S.~Jeon and V.~Koch, Phys.\ Rev.\ Lett.\  {\bf 83}, 5435 (1999)
\bibitem{MonteCarlo}
F.~Becattini, and L.~Ferroni, Eur. Phys. J. C {\bf 35}, 243-258 (2004)
F.~Becattini, and L.~Ferroni, Eur. Phys. J. C {\bf 38}, 225-246 (2004)
\bibitem{Physics}
Reichl L.E. {\it A modern course in statistical physics} (Wiley, New York
,1998);
L.D. Landau, E.M. Lifschitz, {\it Statistical Physics} (Fizmatlit,
Moscow, 2001);
W. Greiner, L. Neise, H.St\"ocker, {\it Thermodynamics and Statistical
  Mechanics} (Springer, New York, 1997).
\bibitem{CEPF}
A.~Ker\"anen, and F.~Becattini, Phys. Rev. C {\bf 65}, 044901 (2002).
\bibitem{PDG}
Particle Data Group, Phys.Rev. D  \textrm{\bf{66}} (2002).
\bibitem{SHARE}
  G.~Torrieri, S.~Steinke, W.~Broniowski, W.~Florkowski, J.~Letessier, and
  J.~Rafelski, Comput. Phys. Commun.\  {\bf 167}, 229 (2005).
\bibitem{THERMUS} S. Wheaton and J. Cleymans, arXiv:hep-ph/0407174.
\bibitem{Formulas}
I.S. Gradshteyn, I.M. Ryzhik, {\it Table of Integrals, Series, and Products}
(Academic Press, San Diego, 2000);
M. Abramowitz and I.A. Stegun, {\it Handbook of Mathematical Functions with
  Formulas, Graphs, and Mathematical Tables}, New York, Dover (1965).
\bibitem{MATH}
W.~Feller, {\it An Introduction to Probability Theory and Its Applications}
Vol. I (Wiley, New York, 1968);
W.~Feller, {\it An Introduction to Probability Theory and Its Applications}
Vol. II (Wiley, New York, 1970);
B.~Hughes, {\it Random Walks and Random Environments} Vol. I, (Clarendon
Press, Oxford, 1995);
B.~Hughes, {\it Random Walks and Random Environments} Vol. II, (Clarendon
Press, Oxford, 1996).
\end{thebibliography}
\end{document}